\documentclass[useAMS,usenatbib,usegraphicx,a4paper]{mn2e}
\usepackage{deluxetable}
\usepackage{amssymb} 

\newcommand{\Htwo}{\mbox{H$_2$}}
\newcommand{\pccm}{\ensuremath{\rm{cm}^{-3}}}  
\newcommand{\HI}{\mbox{H{\sc i}}}
\newcommand{\HII}{\mbox{H{\sc ii}}}
\newcommand{\Halpha}{\mbox{H$\alpha$}}
\newcommand{\Gnaught}{$G_0$}
\newcommand{\dtg}{\ensuremath{\delta/\delta_0}}
\newcommand{\HIunits}{\ensuremath{10^{21}\ \rm{cm}^{-2}}}
\newcommand{\UVunits}{\ensuremath{10^{-15}\ \rm{ergs\ cm^{-2}\ s^{-1}\ \mathrm{\r{A}}^{-1}}}}
\newcommand{\ecsa}{\ensuremath{\rm{ergs\ cm^{-2}\ s^{-1}\ \AA^{-1}}}}
\newcommand{\hmsdms}[6]{\ensuremath{#1^h#2^m#3^s\phantom{0}#4^\circ#5^\prime#6^{\prime\prime}}}
\providecommand{\note}[1]{\textbf{\textit{(#1)}}}	
\renewcommand{\note}[1]{\textbf{\textit{(#1)}}}	

\providecommand\phn{\phantom{0}}
\providecommand{\nat}{Nature} 
\providecommand{\apj}{ApJ}
\providecommand{\apjl}{ApJL}
\providecommand{\apjs}{ApJS}
\providecommand{\aap}{A{\&}A}
\providecommand{\mnras}{MNRAS}
\providecommand{\aj}{AJ}

\title[Atomic Hydrogen produced in M\,33 Photodissociation Regions]{Atomic Hydrogen produced in M\,33 Photodissociation Regions}
\author[J. S. Heiner, R. J. Allen and P. C. van der Kruit]{J. S. Heiner$^{1,2,3}$\thanks{E-mail: jonathan.heiner.1@ulaval.ca}, R. J. Allen$^2$ and P. C. van der Kruit$^3$\\
  $^1$D\'{e}partement de Physique, Universit\'{e} Laval, Qu\'{e}bec, QC G1V 0A6, Canada\\
	$^2$Space Telescope Science Institute, Baltimore, MD 21218, USA\\
  $^3$Kapteyn Astronomical Institute, University of Groningen, PO Box 800, 9700 AV Groningen, the Netherlands}
\begin{document}
\date{Draft of \today}
\pagerange{\pageref{firstpage}--\pageref{lastpage}} \pubyear{2011}
\maketitle
\label{firstpage}

\begin{abstract}
We derive total (atomic + molecular) hydrogen densities in giant molecular clouds (GMCs) in the nearby spiral galaxy M\,33 using a method that views the atomic hydrogen near regions of recent star formation as the product of photodissociation. Far-UV photons emanating from a nearby OB association produce a layer of atomic hydrogen on the surfaces of nearby GMCs. Our approach provides an estimate of the total hydrogen density in these GMCs from observations of the excess far-UV emission that reaches the GMC from the OB association, and the excess 21-cm radio \HI\ emission produced after these far-UV photons convert \Htwo\ into \HI\ on the GMC surface. The method provides an alternative approach to the use of CO emission as a tracer of \Htwo\ in GMCs, and is especially sensitive to a range of density well below the critical density for CO(1-0) emission.

We describe our ``PDR method'' in more detail and apply it using GALEX far-UV and VLA 21-cm radio data to obtain volume densities in a selection of GMCs in the nearby spiral galaxy M\,33. We have also examined the sensitivity of the method to the linear resolution of the observations used; the results obtained at 20 pc are similar to those for the larger set of data at 80 pc resolution. The cloud densities we derive range from 1 to 500 cm$^{-3}$, with no clear dependence on galactocentric radius; these results are generally similar to those obtained earlier in M\,81, M\,83, and M\,101 using the same method.
\end{abstract}


\section{Introduction}

Molecular hydrogen is the major component of baryonic matter in the Universe, but direct observations of its large-scale distribution and motions within and between galaxies are hampered by the fact that the molecule has no dipole moment. The absorption and emission of radiation from \Htwo\ can occur via the much weaker quadrupole coupling, but the lowest rotational transition has $E/k = 510$K ($\lambda = 28.2 \mu$) and emission therefore requires a significant heat source for excitation. Secondary tracers of \Htwo\ include the rotational lines of the CO molecule and infrared continuum emission from the accompanying dust, but the excitation requirements of the former favor gas at relatively high density ($n \gtrsim 1000$ \pccm) and the thermal radiation properties of the latter favor gas at relatively high temperature ($T_K \gtrsim 20$ K) and depend on grain characteristics. The cool ($\lesssim 10$K), sparse ($n \lesssim 100$ \pccm) component of the ISM occupies a region of parameter space that is difficult to explore, and its contribution to the total gas content of a galaxy remains uncertain.

In nearby galaxies, the most common indirect means of detecting molecular hydrogen is carbon monoxide (CO), tracing dense, relatively cold molecular gas. On galaxy-wide scales, CO may or may not be a good indicator of the total molecular gas content, but on the scale of individual complexes of giant molecular clouds (GMCs) the situation becomes more complicated. See, for example, \citet{2007ApJ...658.1027L} and \citet{2009ApJ...702..352L} on CO-free \Htwo\ envelopes in the Small Magellanic Cloud. \citet{2010A&A...521L..17L} present recent observations of $\rm{\left[CII\right]}$ emission without CO counterparts, tracing warm \Htwo\ gas.

Another way to trace the 'total gas' is through infrared dust emission \citep{1997A&A...328..471I}. Recent results from \textit{Herschel} have greatly expanded the potential application of this approach \citep[e.g.][]{2010A&A...518L..69B}.

Here we use an alternative method that is sensitive to low-density gas, and is relatively insensitive to  temperature. The tracer is atomic hydrogen, considered to be a result of photodissociation of \Htwo\ by far-uv photons emanating from OB associations of young massive stars and impinging on the surfaces of nearby giant molecular clouds (GMCs). A simple model of the photodissociation process in these photodissociation regions (PDRs) permits us to estimate the total volume density (atomic + molecular) of the GMC as a function of its relative dust content, the intensity of the incident far-UV radiation, and the column density of the \HI\ created on the cloud surface. Observations of nearby galaxies by the GALEX satellite provide the far-UV data, the VLA radio telescope provides \HI\ column densities from 21-cm emission, and optical emission lines in the ionized gas near the OB association provide estimates of the local dust/gas ratio. \citet{1986Natur.319..296A} presented the first evidence that dissociation of \Htwo\ into \HI\ occurs on kiloparsec scales in the nearby galaxy M\,83, and in the meantime several nearby galaxies have been studied in more detail including M\,81 \citep{1997ApJ...487..171A,2008ApJ...673..798H}, M\,101 \citep{2000ApJ...538..608S}, and (again) M\,83 \citet{2008A&A...489..533H}. We will refer to this approach as the ``PDR method'', which we describe and apply here to M\,33.

The nearby galaxy M\,33 is an obvious choice to apply the PDR method to and derive total hydrogen volume densities from it. M\,33 is close enough to permit resolving some of the atomic hydrogen structures around sites of recent star formation. Besides utilizing the full resolution available in the FUV and \HI\ images ($\approx 20$pc), we also apply the method at a deliberately reduced resolution ($\approx 80$pc), which has the advantage of revealing more of the larger scale structure of the PDRs and allows us to compare the M\,33 results to our previous M\,81 and M\,83 results. The latter two galaxies are at a distance of 3.6 and 4.5 Mpc respectively, but M\,33 is considerably closer, at 847 kpc \citep{2006ApJS..165..108S}, and therefore offers a superior linear resolution at the same angular resolution. At the same time M\,33 is not too close to lose the larger-scale morphology, while being reasonably face-on. The influence of resolution effects on the PDR method can also be studied this way.

Recent work on M\,33 has revealed the detailed distribution of molecular clouds detectable in CO emission \citep[e.g.][]{2010ApJ...722L.127O,2010A&A...522A...3G,2007ApJ...661..830R}. Metallicities in M\,33 have been investigated with increasing spatial resolution \citep[][]{2008ApJ...675.1213R,2010A&A...512A..63M}, and \citet{2010A&A...518L..69B} treat the dust and gas this galaxy. To this expansive host of results, we add independent total hydrogen volume density measurements using the PDR method.

In \citet{2009ApJ...700..545H} we presented the first results of applying the PDR method to M\,33 at high resolution. In \citet{2010ApJ...719.1244H} we showed how the PDR method at reduced resolution can be used to measure the power-law slope of the volume Schmidt law of star formation as originally formulated by Schmidt (using volume density instead of surface density). In this paper we present our full PDR method results of M\,33 that were used in the previous two papers, at high resolution as well as reduced resolution. The aim of this work is to compare the densities with our previous M\,81 and M\,83 results and test the consistency of the PDR method at two widely different linear resolutions.

The data used as input for the PDR method and the method itself are detailed in Section \ref{sec:datamethod}. The total hydrogen volume densities in candidate PDRs in M\,33 are presented in the next section: the densities from the reduced resolution analysis in Section \ref{sec:equiv} and the full resolution results in Section \ref{sec:full}. We briefly compare the results of the PDR method in M\,33 to those in M\,81 and M\,83 and summarize our conclusions in Section \ref{sec:conclusions}.

\section{Data and method}
\label{sec:datamethod}

\begin{figure}
  \centering
  \includegraphics[width=0.45\textwidth]{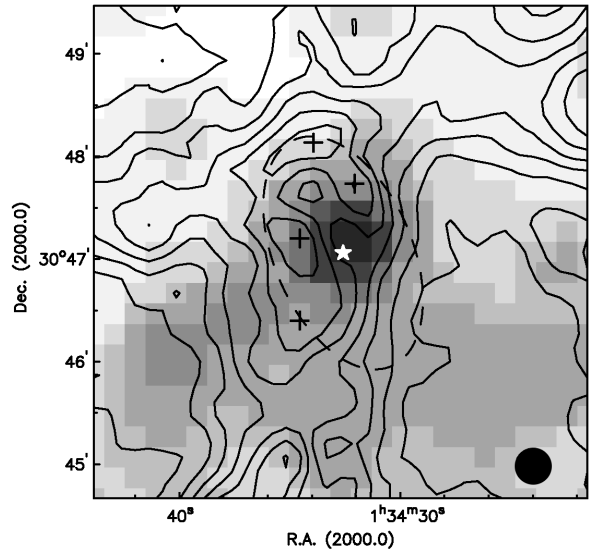}
  \caption{\label{fig:ngc604_20} Example: NGC 604 seen as a large candidate PDR. Gray scale levels (FUV counts per second per pixel) are 0.25, 0.5, 0.75, 1, 2.5, 5, 7.5, 10, 25, 50. The fitted centre of the UV emission is marked with a white star. The \HI\ contour levels are (in \HIunits) 0.5, 1, 1.5, 2, 2.5. Measured \HI\ patches are indicated with black crosses. The dashed ellipse is an example of how $\rho_{\HI}$ was measured - namely de-projected using M\,33's position angle and inclination. Its major axis has a radius of about 280 parsec. A 20-arcsec beam is drawn in the lower right.}
\end{figure}

\begin{figure*}
  \centering
  \includegraphics[width=0.45\linewidth]{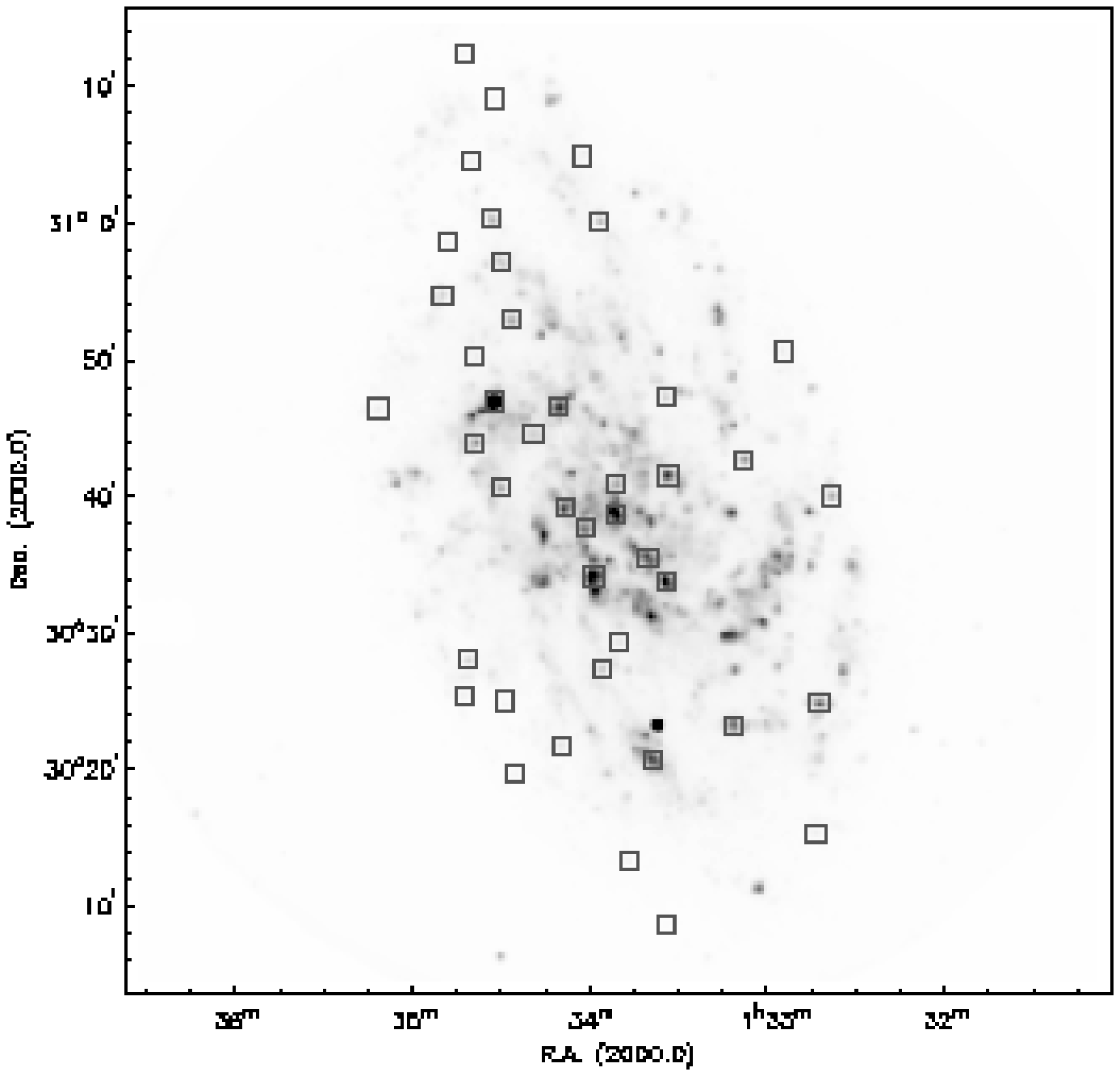}
  \includegraphics[width=0.45\linewidth]{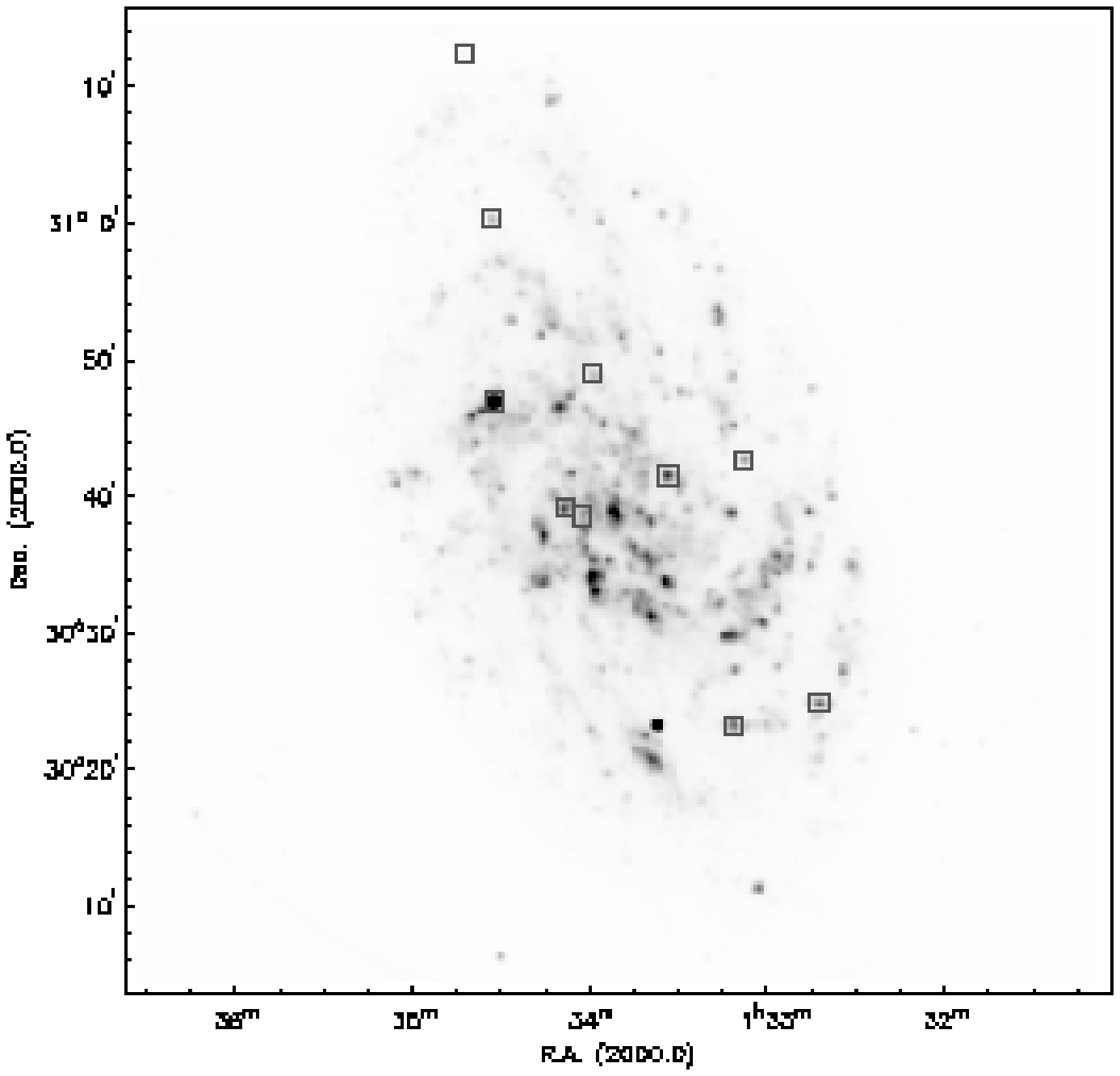}
  \caption{\label{fig:locplot} Candidate PDR locations are indicated with boxes on this \textit{GALEX} FUV image (smoothed to 30-arcsec). The candidate PDRs were selected on the basis of their FUV emission. Left panel: reduced resolution regions. Right panel: full resolution regions.}
\end{figure*}

\begin{table*}
\begin{tabular}{crrrrr}
\hline\hline
           &             &              & Radius & $F_{FUV}~^a$  & Aperture \\
Source no. & R.A. (2000) & Decl. (2000) & (kpc)  &            & (arcsec) \\
\hline
\phn1& 1 33 51.371 &   30 38 52.01        &  0.23 & 305.35 &  96 \\ 
\phn2& 1 33 51.052 &   30 41  \phn1.18    &  0.40 &  26.08 &  36 \\
\phn3& 1 33 40.271 &   30 35 34.72        &  1.16 & 137.22 &  72 \\
\phn4& 1 34 \phn1.540 &   30 37 43.83     &  1.27 &  71.03 &  60 \\
\phn5& 1 34 \phn8.658 &   30 39 15.42     &  1.63 &  96.25 &  72 \\
\phn6& 1 33 34.135 &   30 33 53.24        &  1.71 & 151.01 &  72 \\
\phn7& 1 33 33.401 &   30 41 37.88        &  1.88 & 188.70 & 156 \\
\phn8& 1 33 58.709 &   30 34 14.98        &  1.92 & 518.50 & 156 \\
\phn9& 1 34 10.340 &   30 46 39.52        &  2.06 & 195.81 & 108 \\
10& 1 34 19.129 &   30 44 37.26        &  2.35 &  20.08 &  48 \\
\hline                                       \multicolumn{6}{l}{$^a$ \UVunits}
\end{tabular}
\caption{\label{tab:sources} Locations and FUV fluxes of candidate PDRs (example, full table available in the \note{online version of this journal}). }
\end{table*}                              

\subsection{The PDR method}
Determining the total hydrogen volume density in candidate PDRs requires measurements of the local radiation field, which we estimate from FUV photometry derived from the publicly available \textit{GALEX} M\,33 image. It also requires identifying potential GMCs, traced by 'patches' of atomic hydrogen: distinct features of \HI\ found near OB star clusters. We focus on FUV emission above the diffuse background radiation field and we assume that this emission causes the distinct \HI\ features above the \HI\ background levels. In this manner we attempt to obtain clear and simple components to our method and keep the geometry of the candidate PDRs surrounding the OB clusters simple.

The 21-cm 20-arcsec and 5-arcsec M\,33 images were provided by David Thilker and Rob Braun (2007, private communication). The linear resolution for the reduced resolution comparison is about 80 pc at the distance of M\,33, from a combination of data from the VLA (B, C and D configurations) and the GBT \citep[for a complete explanation of the procedure, see][]{2009ApJ...695..937B}\footnote{Project codes for the radio data: (VLA) AT206, AT268, and (GBT) GBT02A-038}, and the full resolution analysis was carried out at a 5-arcsec resolution. The images are a mosaic of radio pointings that cover the extent of the \textit{GALEX} image. At the 5-arcsec resolution, typical noise levels are $2 \times 10^{20} \rm{cm}^{-2}$.

\begin{figure*}
  \centering
  \begin{tabular*}{\textwidth}{r p{1.5cm} l} 
    \includegraphics[width=0.35\textwidth]{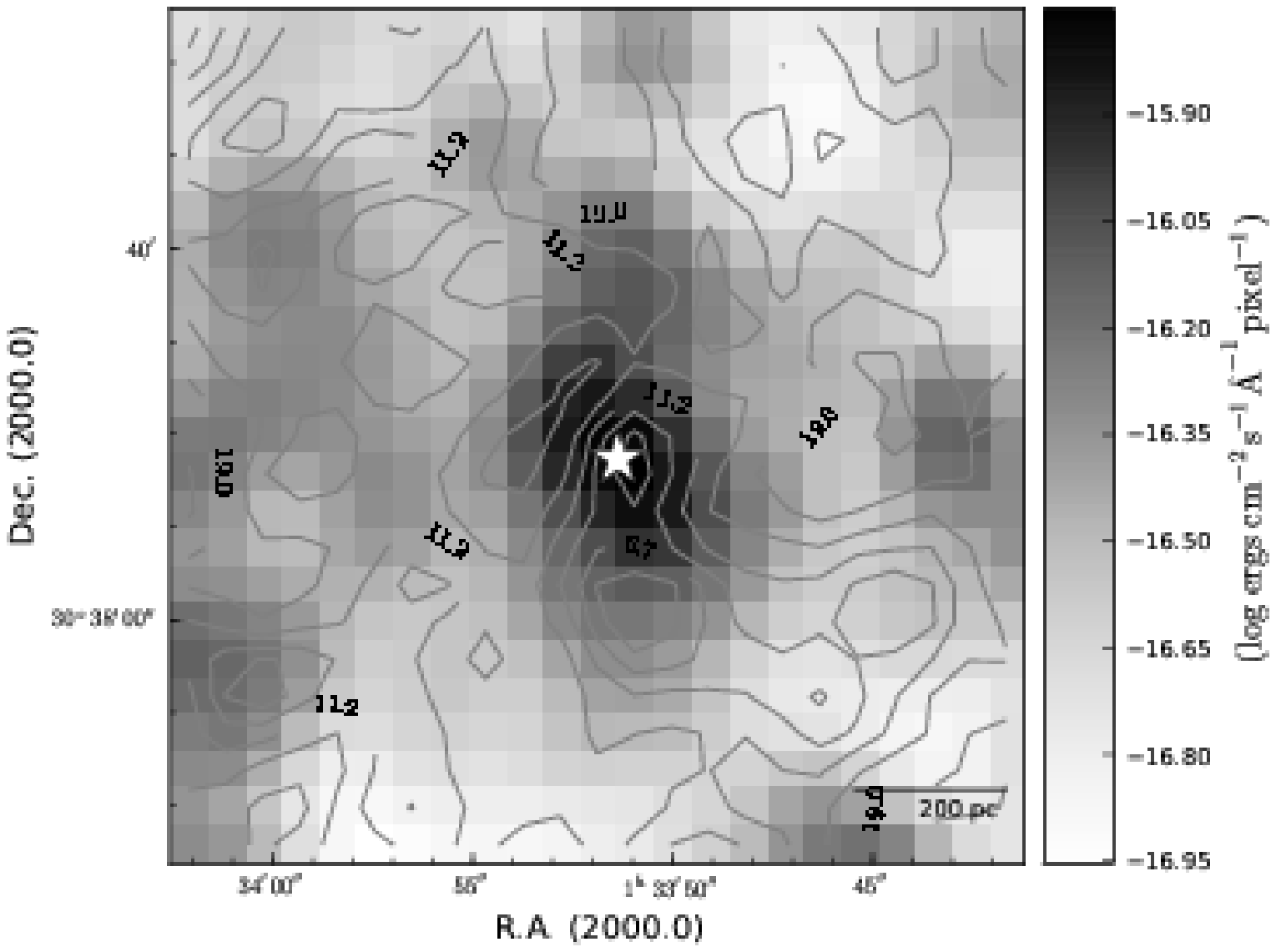}
    & \centering{\scriptsize{\hspace{0.5cm}1\hspace{0.5cm}2}} &
    \includegraphics[width=0.35\textwidth]{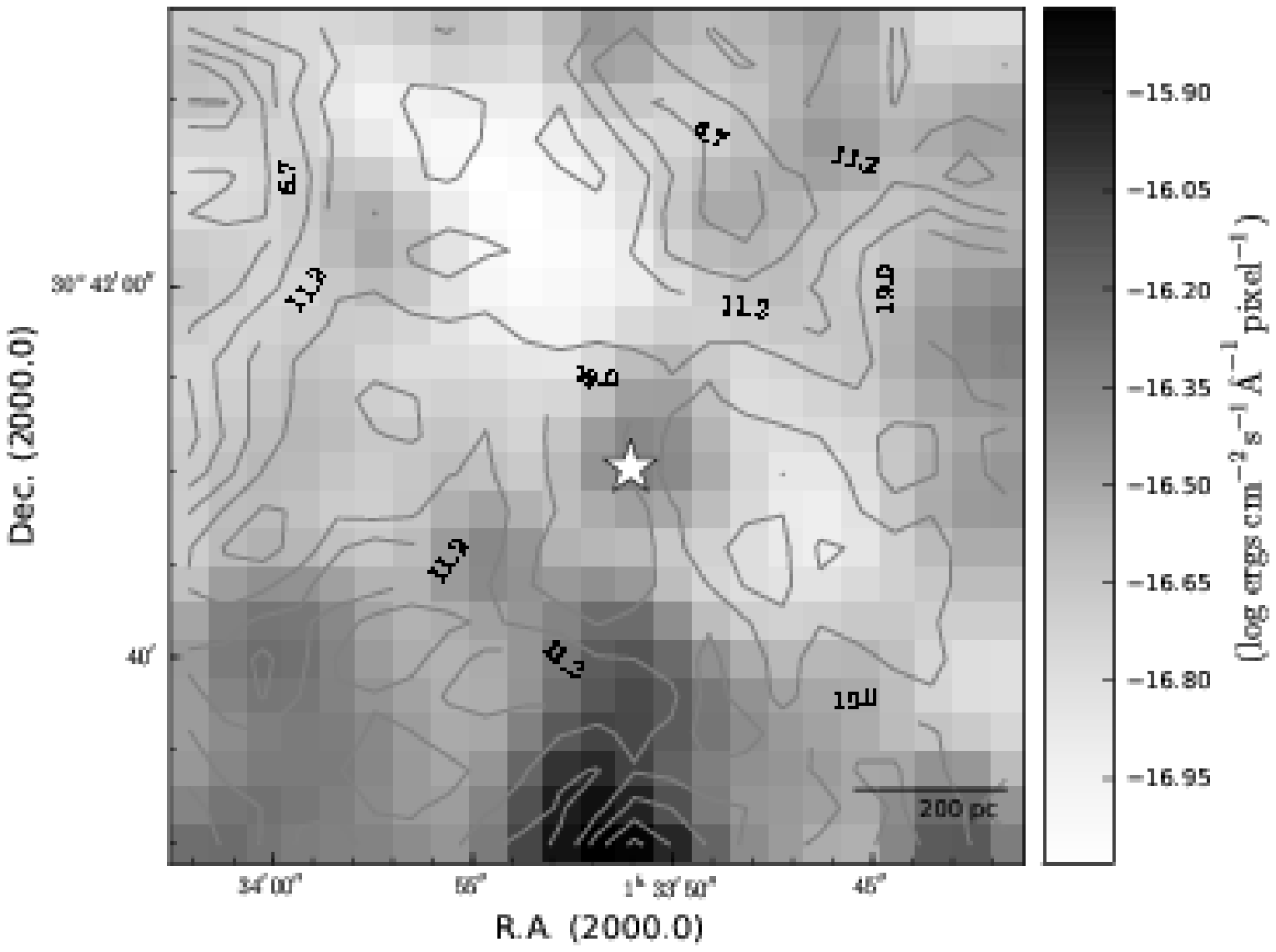} \\
  \end{tabular*}
	\caption[]{\label{fig:example_overlay} M\,33 candidate PDRs at reduced resolution (the full set is available in the \note{online version of this journal}). The \HI\ column density contours are plotted against a background of the FUV flux in gray scale. The fitted position of the central UV complex is marked by a star. A distance scale is indicated, but see Figure \ref{fig:ngc604_20} about how we de-project the distances. The integrated FUV fluxes can be found in Table \ref{tab:sources_full}.}
\end{figure*}

\begin{table*}
\begin{tabular}{ccccccc}
\hline\hline
Source no. & $\rho_{\HI}$ (pc) & $N_{\HI}$ (\HIunits) & \Gnaught & $G/G_{bg}$ & $n$ ($\rm{cm}^{-3}$) & $^a$\\
\hline
1a       &  121   &  1.91    &  6.53    &  8.01      &  487        &  0.60  \\
1b       &  161   &  2.02    &  3.67    &  4.51      &  249        &  0.59  \\
1c       &  403   &  2.17    &  0.59    &  0.72      &  35         &  0.58  \\
2a       &  40    &  2.11    &  5.02    &  8.18      &  323        &  0.97  \\
2b       &  202   &  2.37    &  0.20    &  0.33      &  10         &  0.61  \\
2c       &  363   &  1.57    &  0.06    &  0.10      &  6          &  0.52  \\
2d       &  363   &  2.17    &  0.06    &  0.10      &  4          &  0.58  \\
3a       &  242   &  1.82    &  0.73    &  0.81      &  67         &  0.54  \\
3b       &  242   &  1.51    &  0.73    &  0.81      &  89         &  0.51  \\
3c       &  403   &  1.99    &  0.26    &  0.29      &  21         &  0.55  \\
\hline
\multicolumn{7}{l}{$^a$ Fractional error}
\end{tabular}
\caption{\label{tab:results} Results (example, full table available in the \note{online version of this journal}).}
\end{table*}

To calculate linear distances within M\,33, necessary to compute incident UV fluxes, we assume this galaxy to be located at a distance of 847 kpc \citep{2006ApJS..165..108S}.  \citep[The same parameters were used in][]{2009ApJ...700..545H}. To de-project distances, we adopt a position angle of $23^\circ$ and an inclination of $56^\circ$ for its disc, from \citet{1989AJ.....97...97Z}. Its $R_{25}$ is 28.8\arcmin, or 7.1 kpc \citep{1992MNRAS.259..121V}, which we used to normalize our galactocentric radius measurements. The radius measurements are needed to calculate the dust-to-gas ratio with the assumption of a galactic metallicity gradient. A uniform foreground extinction of $A_{FUV} = 0.33 \rm{mag}$ is assumed to correct the measured UV flux, using E(B-V) from \citet{1998ApJ...500..525S} and the expression for $A_{FUV}$ from \citet{2007ApJS..173..185G}. As in our previous papers, we ignore internal extinction, under the assumption that it is uniform, or at least acting in all directions equally in a spherical morphology, and that the distance through the disc is comparable to the separation between the UV source and the \HI\ patch. In that idealized case the internal extinction can be ignored, as it works equally between the UV source and the observer as between the UV source and the \HI\ patch. The adoption of a single value for the internal extinction affects the calculated values of the total hydrogen volume densities equally. For example, if an internal extinction of 1 mag were adopted, all values of $n$ would need to be multiplied by 2.5. We also use the GALEX FUV emission at 1500\AA\ as a proxy for the dissociating radiation at 1000\AA, assuming a locally flat spectral energy distribution \citep{1988ApJ...334..771V}.

The photodissocation rate is regulated by the dust content, through obscuration of FUV radiation and catalyzation of the formation of molecular hydrogen. We derive the dust-to-gas ratio $\delta/\delta_0$, which is normalized to the solar neighborhood value, from the metallicity 12 + log(O/H) after \citet{1990A&A...236..237I}. Under this assumption, the dust-to-gas ratio is directly proportional to the metallicity log(O/H). The metallicities in M\,33 were measured by \citet{2008ApJ...675.1213R} and \citet{2007A&A...470..865M}. We use a solar metallicity of 8.69 from \citet{2001ApJ...556L..63A} to obtain the dust-to-gas ratio scaled to the solar neighborhood. 
The dependence of the dust-to-gas ratio on the galactocentric radios $R$ (in kpc) is:
\begin{equation}
  \log{\delta/\delta_0} = -0.027 R - 0.33,
  \label{eqn:dtg}
\end{equation}
using \citet{2008ApJ...675.1213R}. More recently, \citet{2010A&A...512A..63M} report a metallicity slope of 0.037 dex kpc$^{-1}$. However, we will keep the slope of 0.027 for consistency with our previous results.

With the full resolution data local metallicity measurements are sometimes available for our candidate PDRs; in these cases we adopt an equivalent expression to that used in \citet{2009ApJ...700..545H}, independent of the galactocentric radius:
\begin{equation}
  \log{\delta/\delta_0} = (12 + \log{(O/H)}) - 8.69.
  \label{eqn:dtgindiv}
\end{equation}

The foreground extinction that was used to correct the far-UV flux is 0.33 mag, based on \citet{1998ApJ...500..525S} and \citet{2007ApJS..173..185G}. The equation describing the photodissociation process, as derived from \citet{1988ApJ...332..400S} and \citet{2004ASSL..319..731A} is:
\begin{equation}
  N_{\HI} = \frac{7.8 \times 10^{20}}{\dtg} \ln\left[1+\frac{106G_0}{n}\left(\frac{\delta}{\delta_0}\right)^{-1/2}\right]~\rm{cm^{-2}}.
  \label{eqn:n}
\end{equation}
In this form it is a clear expression of the atomic hydrogen column density that is created by photodissociation on the surface of GMCs.

\citet{2009ApJ...694..978H} derived a somewhat improved version of this equation, requiring all derived values of $n$ in this paper to be multiplied by $(\delta/\delta_0)^{0.2}$, which is well within our current estimated levels of uncertainty in the case of M\,33. We maintain the use of Equation \ref{eqn:n} at this time for consistency with our results in \citet{2009ApJ...700..545H} and \citet{2010ApJ...719.1244H}.

We considered the overall \HI\ distribution across the disc of M\,33 and adopted a background \HI\ column density of $1.2 \times 10^{20}~\rm{cm^{-2}}$ inside 1.1 $R_{25}$ and $0.5 \times 10^{20}~\rm{cm^{-2}}$ outside of this radius. We do this in an attempt to isolate the additional \HI\ produced by photodissociation under the influence of the nearby OB associations. We checked the reduced resolution background levels of the \HI\ column densities throughout M33 and chose to characterize the trend with these two values. They have a very small impact on the final results as long as they are only a fraction of the $7.8 \times 10^{20}$ scaling in the exponential of the model when computing the total hydrogen volume density $n$.
We measured the distances between the central UV source and nearby \HI\ 'patches', which are defined as a local maximum in the \HI\ column density, and recorded their column densities out to a separation of 400 pc, the approximate maximum size of the candidate PDRs. This value is somewhat arbitrary and based on the \HI\ morphology surrounding the central UV source. In \citet{2008ApJ...673..798H} we used 600 pc, although almost all measured values were below 400 pc. In \citet{2008A&A...489..533H} we used 480 pc. Erroneously including \HI\ patches that are not produced under the influence of the assumed central UV source results in low computed total hydrogen volume densities.

We censored background-subtracted column densities below $0.5 \times 10^{20}~\rm{cm^{-2}}$. As an example of the resolution of the data and the scale of the candidate PDRs, we show NGC 604 in Figure \ref{fig:ngc604_20}. The \HI\ contours illustrate the morphology of atomic hydrogen surrounding the central concentration of O and B stars in NGC 604, the brightest \HII\ region in M\,33. While this region has been studied in much more detail (at a much higher resolution), it is important in the context of large-scale PDRs to view this region in its entirety and to get global gas density estimates from it.
In the full resolution \HI\ data we measured individual local \HI\ column density backgrounds based on the average \HI\ column density near the candidate PDRs. These background levels will be listed for every region in the full resolution results. 

\begin{figure}
  \centering
  \includegraphics[width=\linewidth]{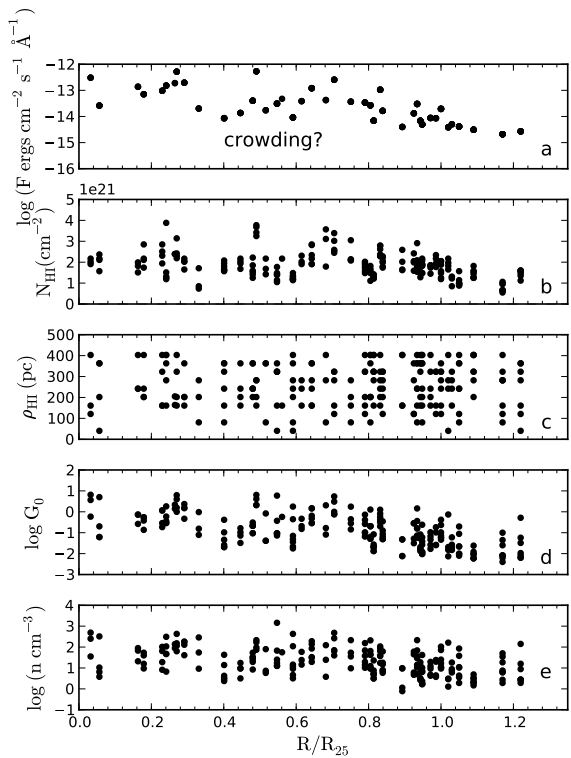}
	\caption{\label{fig:multi} Combined results are plotted here as a function of the normalized galactocentric radius. (a) FUV fluxes of the candidate PDRs, not corrected for extinction. (b) \HI\ column densities associated with the central UV sources. (c) $\rho_{\HI}$, the separation between the central UV source and each \HI\ patch. (d) Incident flux \Gnaught\ on each \HI\ patch. (e) Total hydrogen volume densities $n = n_{\HI} + 2n_{\Htwo}$.}
\end{figure}

\subsection{Improvements to the PDR method}

The full resolution images allow for a slight improvement to the PDR method. Resolved UV sources are measured independently (although they are still expected to harbor a certain amount of O, B stars each) and their incident flux \Gnaught\ on individual \HI\ patches is summed to derive a cumulative \Gnaught. We did not consider whether one UV source blocks another in the direction of an \HI\ patch, since we could not de-project the three dimensional structure of these OB clusters from the FUV image. It can however be assumed that this is not a big issue if the extinction close to these sources is low.

\begin{figure}
  \includegraphics[clip=,width=\linewidth]{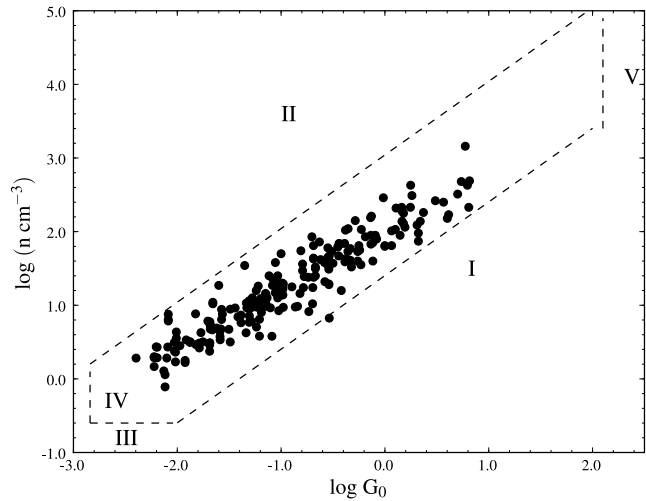}
  \caption{\label{fig:G0_n} Observational limits are marked in this plot of $G_0$ versus $n$. The roman numerals are explained in the text.}
\end{figure}

\begin{table}
  \centering
  \begin{tabular}{ll}
    \hline\hline
    Name & Reduced resolution no.\\
		\hline
    CPSDP Z204  & 5 \\
    CPSDP 0087g & near 5 \\
    BCLMP 0695	& none \\
    BCLMP 0650  & 24 \\
    BCLMP 0288  & 19 \\
    BCLMP 0269	& 27 \\
    NGC 604	& 14 \\
    NGC 595	& 7  \\	
    BCLMP 0256	& 20 \\
    Region 42	& 42 \\
		\hline
  \end{tabular}
  \caption{\label{tab:correspondence} Candidate PDRs, full resolution}
\end{table}

Another improvement to the method is the use of locally measured metallicity values where available (Equation \ref{eqn:dtgindiv}). Otherwise, we use the single metallicity gradient equation for M\,33 (Equation \ref{eqn:dtg}). As in M\,81 and M\,83, we did not aim for a complete sample of candidate PDRs in M\,33, but rather we selected those regions that stood out in their FUV emission and had a morphology that seemed to indicate the presence of large-scale PDRs. The latter merely means that we preferred regions with a relatively simple apparent \HI\ morphology. We selected regions with a progressively more (apparent) complex \HI\ structure. Additionally, when a larger, resolved \HI\ structure appeared to be present in a candidate PDR, we measured its average column density and calculated a global total hydrogen volume density. This is similar to the method used in M\,81 and M\,83, except that in M\,33 the structures are resolved. We explicitly assume that a large-scale PDR with a shell of \HI\ is observed in these cases, with a radius of up to a few hundred parsec. NGC 595 is an example where this morphology is particularly obvious. 

In addition to the UV-selected candidate PDRs, we included a few regions that had either individual CO detections associated to it, from \citet{2003ApJS..149..343E}, or individual metallicity measurements.
Almost all of these candidate PDRs, primarily selected by FUV emission, are previously catalogued \HII\ regions. The ones starting with BCLMP derive their names from \citet{1974A&A....37...33B}. The ones starting with CPSDP come from \citet{1987A&A...174...28C}. Our sample includes one region that has no name at this time, which we will call ``Region 42'', consistent with the numbering of the reduced resolution results.

\begin{figure*}
  \centering
  \includegraphics[width=0.4\textwidth]{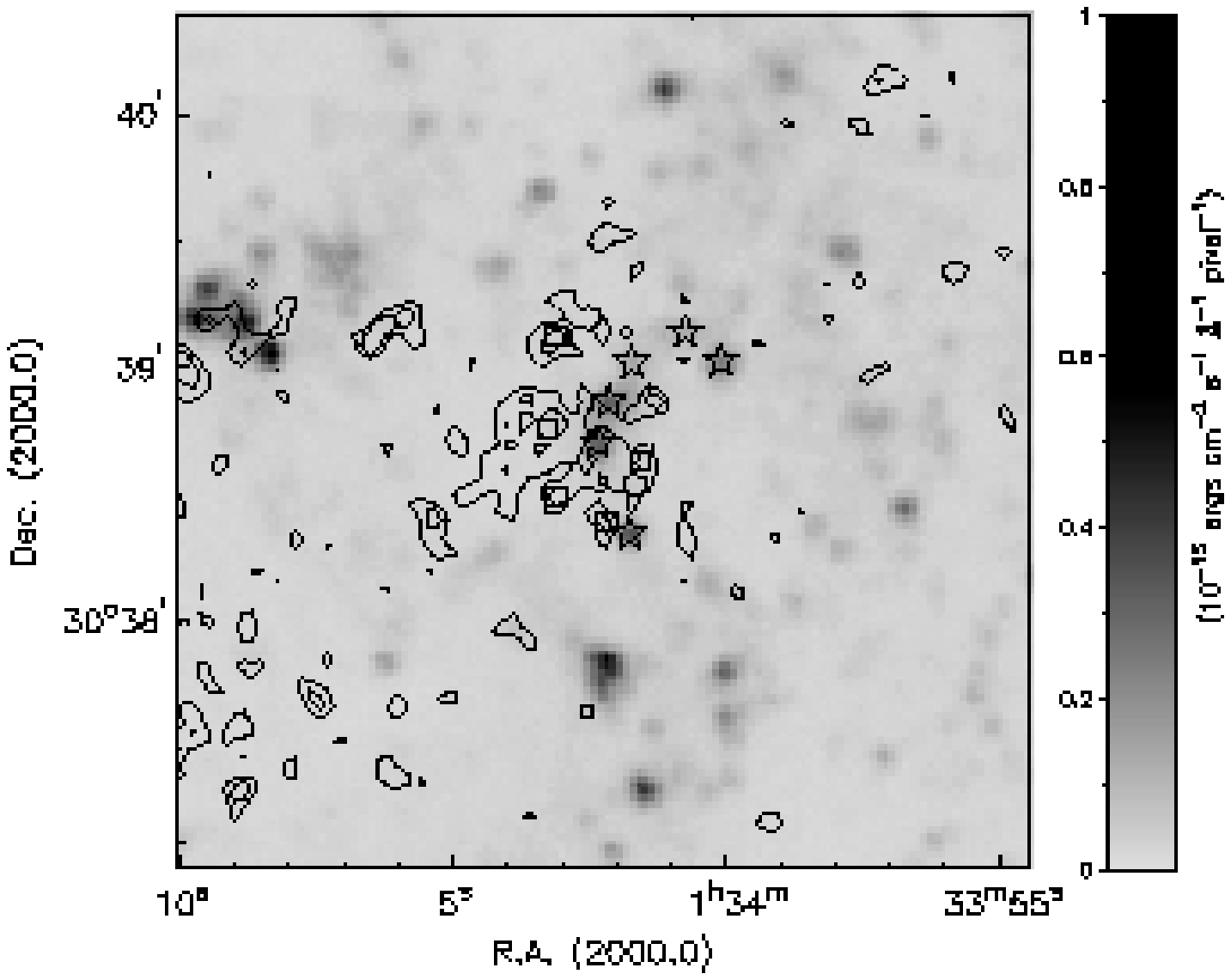}
  \includegraphics[width=0.34\textwidth]{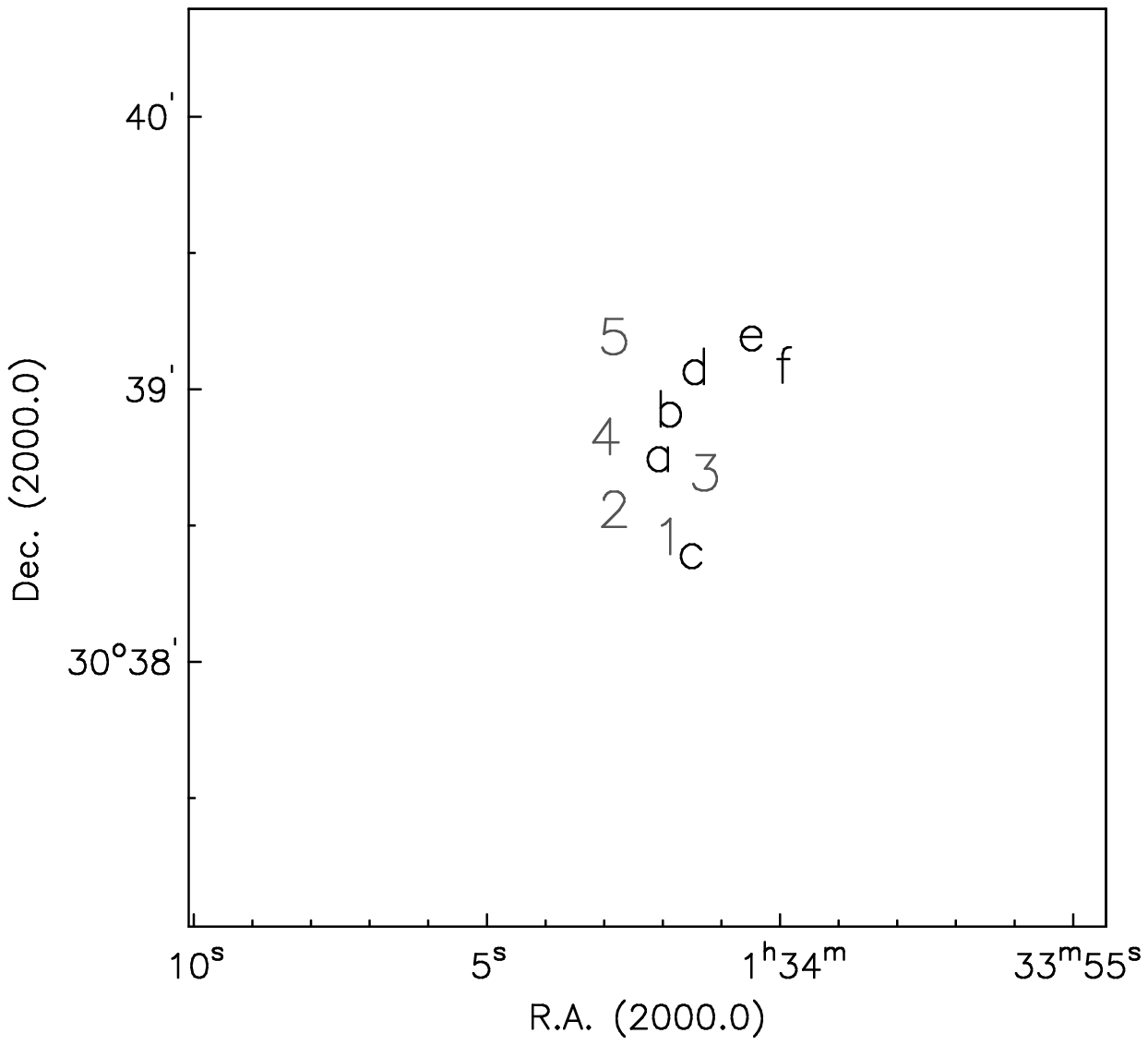}
	\caption{\label{fig:cpsdp0087g_example} Left panel: CPSDP 0087g region at full resolution. The FUV image is overlaid with \HI\ contours (color version available in the \note{online version of this journal}. The \HI\ contours are 3 and 4 $\times 10^{21} \rm{cm}^{-2}$ only, to emphasize the morphology of the \HI\ peaks. Black stars mark the locations where the FUV fluxes were measured. The orange boxes are the locations of the measured \HI\ patches. Green diamonds denote the location of CO detections by \citet{2003ApJS..149..343E}. Part of CPSDP Z204 can be seen in the eastern part of the image. Right panel: Finding chart of the same region with letters (UV sources) and numbers (HI patches).}
\end{figure*}
\begin{table*}
  \begin{tabular}{lp{6.5cm}}
	\hline
	\hline
    $R_{gal}$ (kpc) 		& 1.05 \\
    \dtg			& 0.65 \\
    FUV sources (J2000)         & \hmsdms{1}{34}{2.363}{30}{38}{42.30}$^a$,
                                  \hmsdms{1}{34}{2.175}{30}{38}{52.13}$^b$,
                                  \hmsdms{1}{34}{1.792}{30}{38}{20.99}$^c$,
                                  \hmsdms{1}{34}{1.754}{30}{39}{01.48}$^d$,
                                  \hmsdms{1}{34}{0.781}{30}{39}{08.93}$^e$,
                                  \hmsdms{1}{34}{0.127}{30}{39}{01.85}$^f$\\
    FUV fluxes ($10^{-15}$ \ecsa)	& $4.55^a$, $4.68^b$, $4.55^c$, $1.56^d$, $1.65^e$, $4.72^f$ \\
    $N_{bg} (\HIunits)$		& 0.96 \\
    $N_{HI} (\HIunits)$		& $3.37^{1:abcd}$, $4.10^{2:abc}$, $3.67^{3:abcdef}$, $4.63^{4:abcd}$, $3.64^{5:abcd}$ \\
    $G_0$ (cumulative)		& $1.69^1$, $0.32^2$, $0.97^3$, $0.48^4$, $0.37^{5}$ \\
    $G/G_{bg}$ range		& $0.01-1.77^1$, $0.03-0.17^2$, $0.01-0.23^3$, $0.01-0.16^4$, $0.02-0.09^{5}$\\
    n (derived, in \pccm)	& $15^1$, $1^2$, $6^3$, $1^4$, $3^{5}$ \\
    Fractional error range	& $0.35 - 0.45$ \\
		\hline
	\end{tabular}
  \caption{\label{tab:cpsdp0087g_example} Detailed measurements of CPSDP 0087g}
\end{table*}

\section{Results}
\label{sec:results}

\subsection{Source selection}

The candidate PDRs that were selected for our analysis are shown in Figure \ref{fig:locplot}.
We selected 42 prominent FUV sources in M\,33 at reduced resolution, while attempting to cover the full range of galactocentric radii available to us.
Each UV-selected region has multiple measured \HI\ patches, providing a range of total hydrogen volume density measurements per candidate PDR. Note that we did not aim to identify all candidate PDRs, as source completeness was not our goal. 

At the full resolution, we selected 10 regions. At this level of detail, the larger scale (a few hundred parsec) morphology of \HI\ is resolved into smaller scale (several tens of parsec) \HI\ features.
It is important to see if the results of the PDR method change when applying it at full resolution. We also wanted to take advantage of the availability of individual dust-to-gas ratio measurements. Finally, some regions also had measurements of CO available, so the resulting cloud densities can be compared to CO results. At the same time, we tried to preserve the spread in galactocentric radius coverage.

The full resolution \HI\ morphology is considerably more detailed, and identifying clear \HI\ patches is not easy as in the reduced resolution case. Therefore, we selected a limited subset of the regions featured in the reduced resolution analysis and studied them more extensively. Table \ref{tab:correspondence} shows how the selected regions correspond to the reduced resolution regions. The full resolution regions are chosen with several criteria in mind: apparent presence of a large scale \HI\ (partial) shell, known \HII\ regions (e.g. NGC 604), availability of individual metallicity data, and availability of CO measurements, while still keeping a representative spread in galactocentric radius. The final sample included 10 regions. 8 of these are also present in the reduced resolution sample, one is an \HII\ region close to a reduced resolution region and one does not appear in the reduced resolution analysis. These 10 regions feature a rich and detailed morphology, while at the same time the resulting total hydrogen volume densities are shown not to differ significantly from the reduced resolution analysis result. (See Section \ref{sec:reseffects}.)

\subsection{Cloud densities in M\,33, reduced density}
\label{sec:equiv}

Two example overlays of M\,33 candidate PDRs (locations are marked in Figure \ref{fig:locplot}) at reduced resolution are shown in Figure \ref{fig:example_overlay}. A full set of overlay plots can be found in the \note{online version of this journal}.

The FUV fluxes of the central sources in our candidate PDRs are displayed in Figure \ref{fig:multi}a, and listed in Table \ref{tab:sources} along with their locations and the aperture used to measure the FUV flux and the galactocentric radius of each source. 

A gradual decline in maximum values can be seen going outwards. A similar decline can be seen in the minimum values, but we suspect that crowding effects are possible here, leading us to miss the fainter sources. We are assuming that the maximum values are real and not a selection effect, whereas a sensitivity limit has been reached at larger galactocentric radius. The fluxes show the same spread as we reported for M\,83 \citep{2008A&A...489..533H} - namely 2 dex or less at any given galactocentric radius. Since associations of O and B stars are measured, the number of those stars per cluster could decrease with radius, causing the general decline. On an absolute scale the difference in flux values is caused only by the different distances of M\,83 and M\,33 (the luminosities of the UV sources are in the same range).

\begin{figure*}
  \centering
  \includegraphics[width=0.45\textwidth]{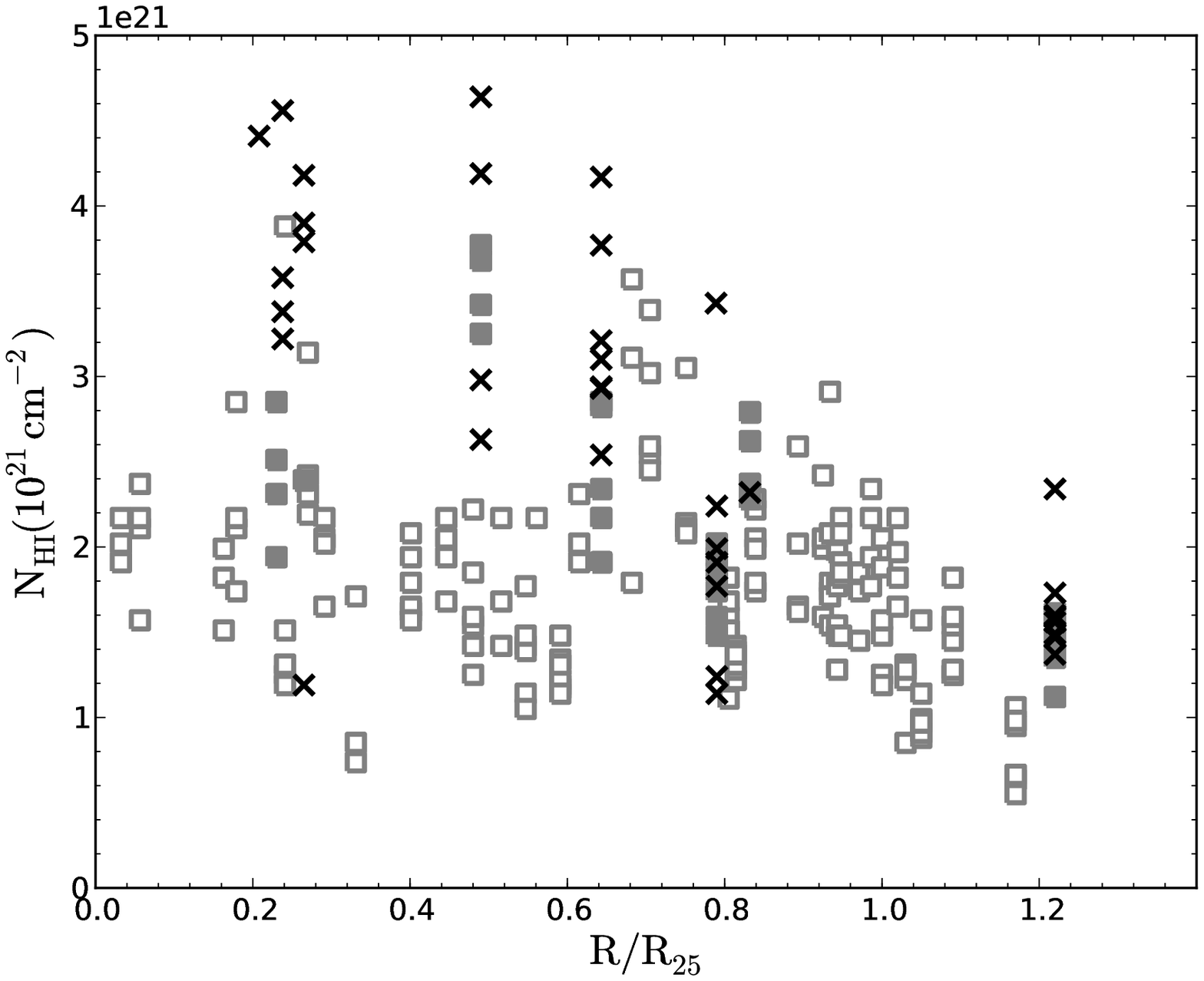}
  \includegraphics[width=0.45\textwidth]{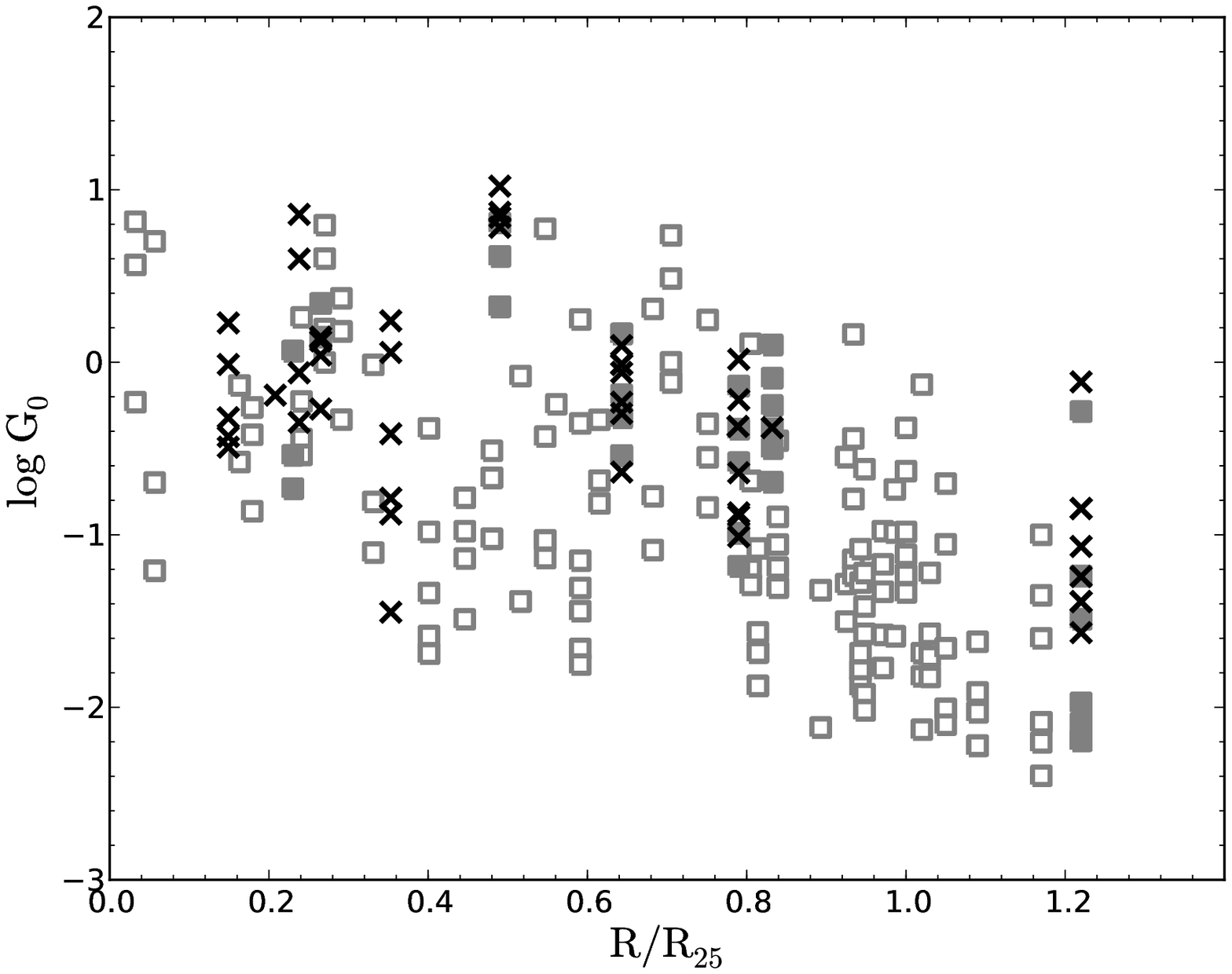}
  \includegraphics[width=0.45\textwidth]{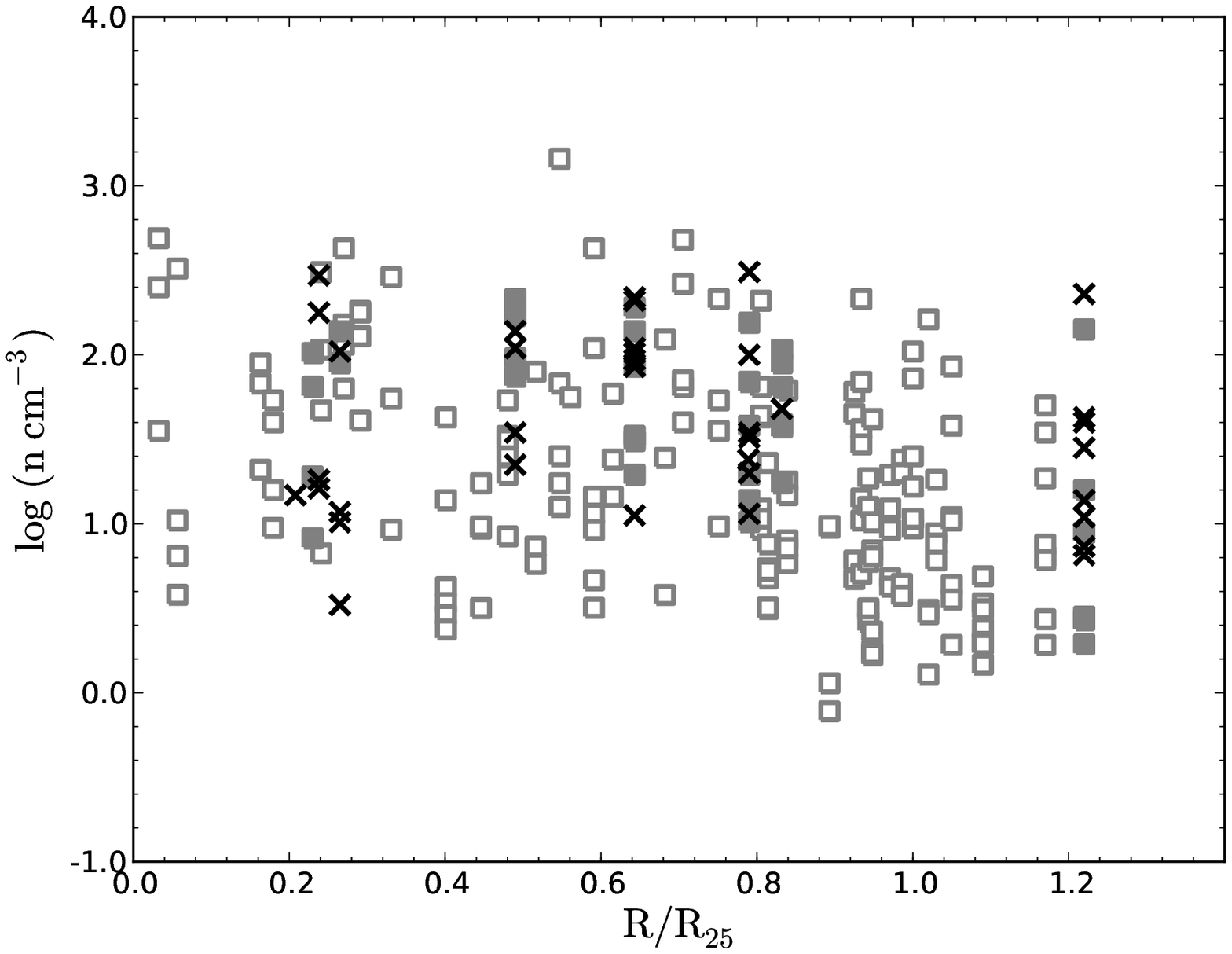}
  \caption{\label{fig:consolidated} The results of the 9 out of 10 candidate PDRs that have detailed results (not the larger scale measurements) are shown here (as crosses): the \HI\ column densities, the incident flux \Gnaught, and the total hydrogen volume densities. The reduced resolution results are plotted as gray boxes. The filled boxes are the regions with full resolution result counterparts. Region BCLMP 0288 is not included since it does not feature a discernible detailed morphology.}
\end{figure*}

\begin{figure*}
  \centering
	\includegraphics[width=0.45\textwidth]{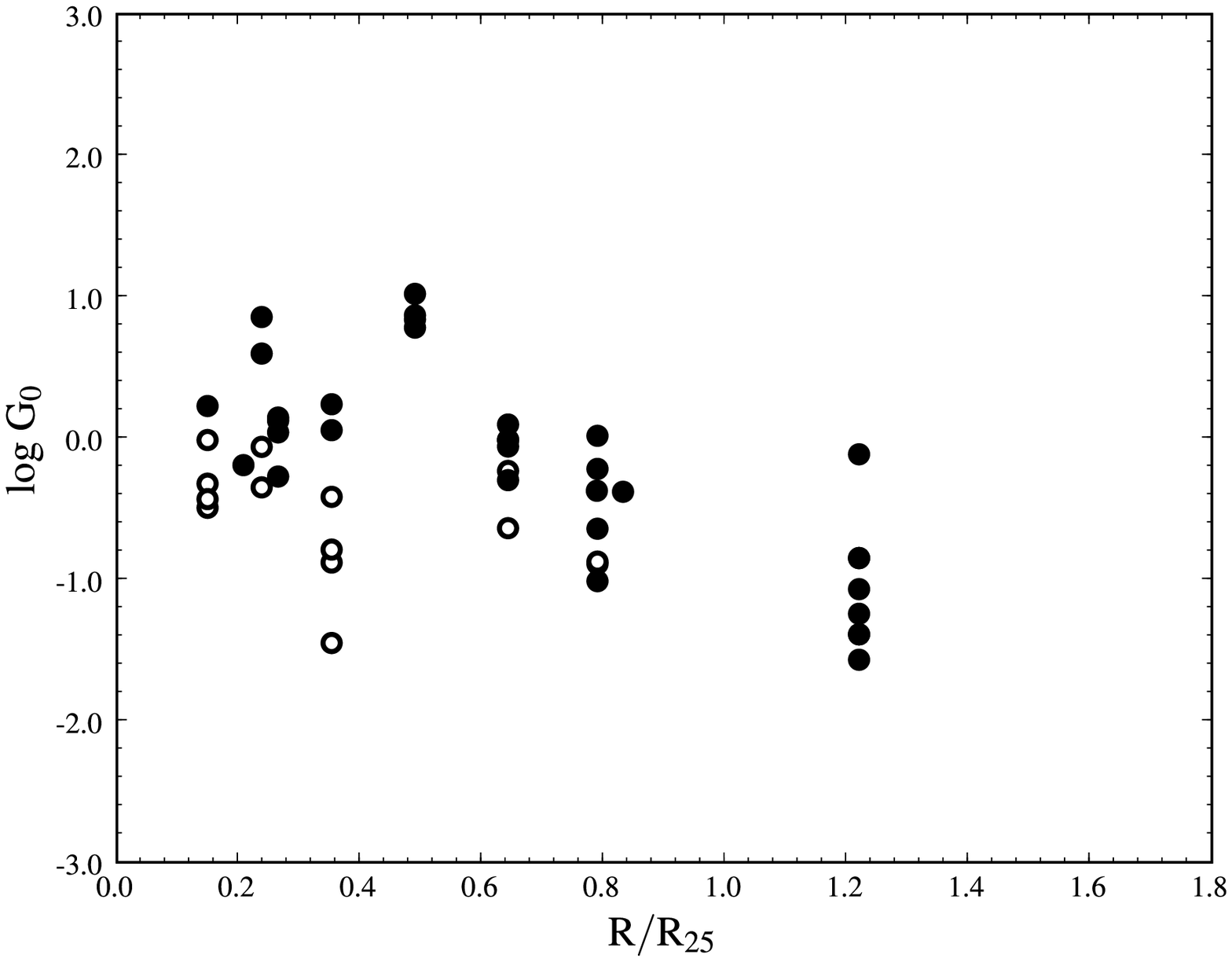}
  \includegraphics[width=0.45\textwidth]{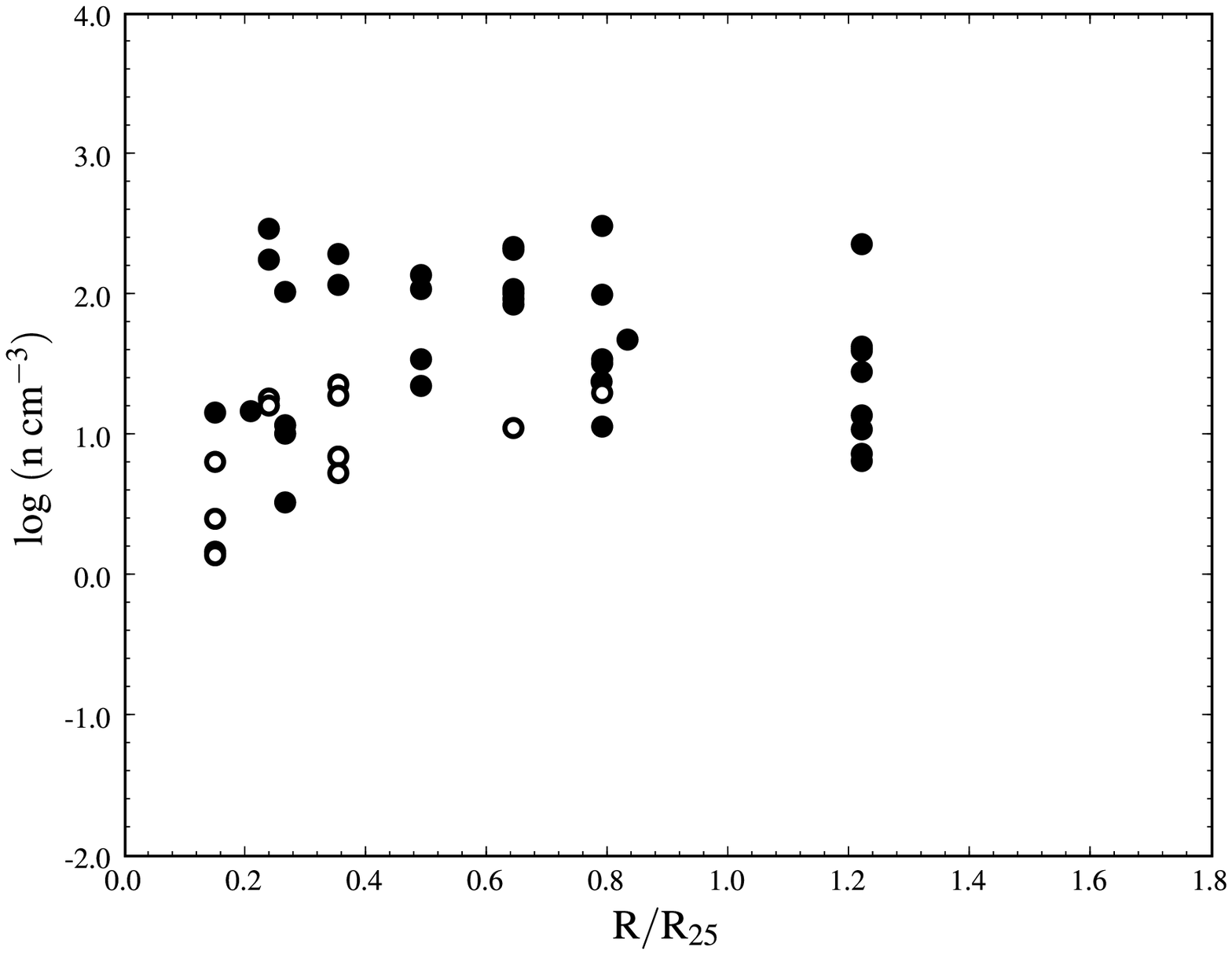}
  \includegraphics[width=0.45\textwidth]{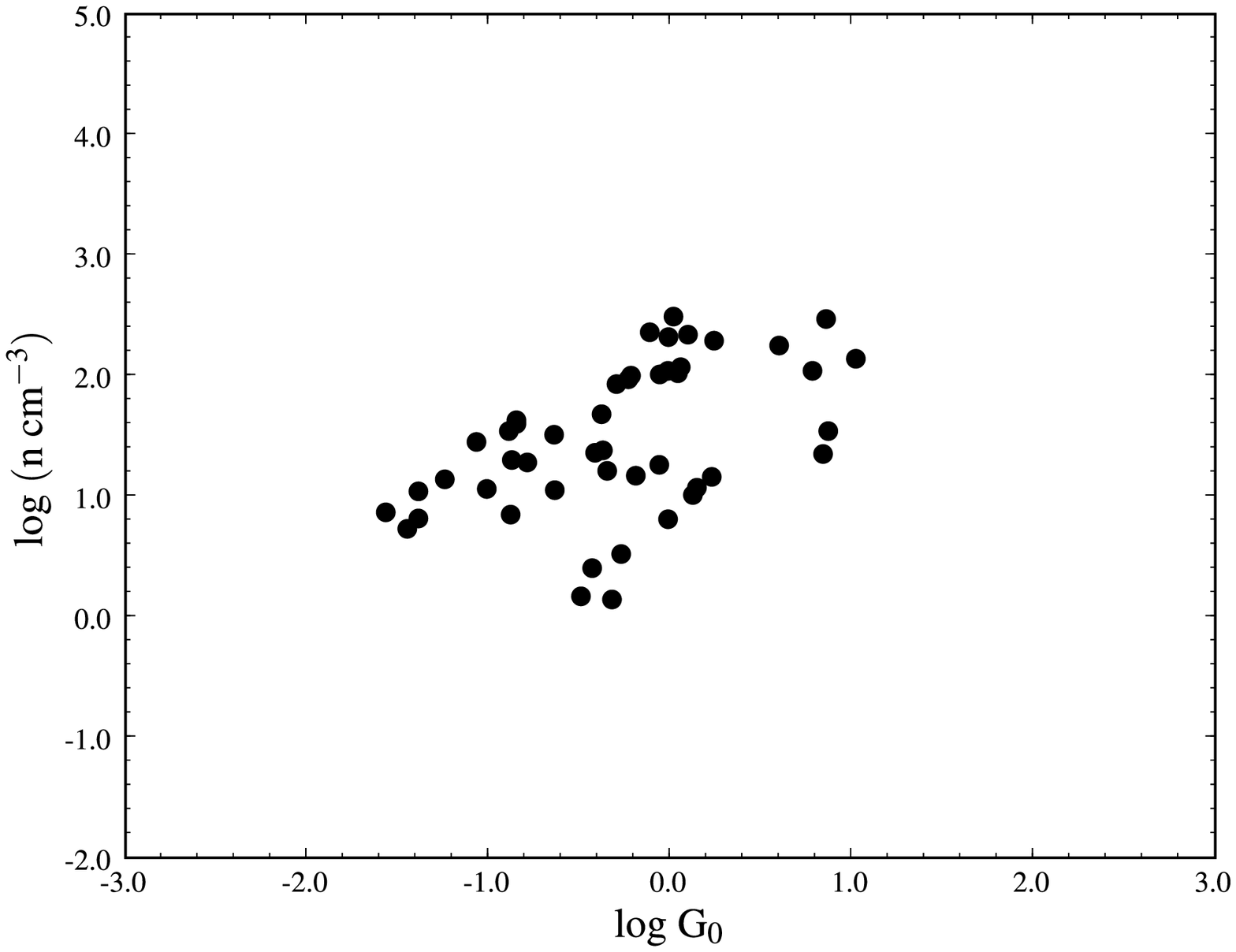}
	\caption{\label{fig:contrast} The results of the 9 out of 10 regions that have detailed results (not the larger scale measurements) are shown here: the incident flux \Gnaught, the total hydrogen volume densities and a plot of $n$ vs. \Gnaught. The open circles indicate \HI\ patches with a source contrast below 0.5. Region BCLMP 0288 is not included since it does not feature a discernible detailed morphology.}
\end{figure*}

The measurements of the candidate PDRs are shown in Figure \ref{fig:multi} and in Table \ref{tab:results}.
The figure and table show the measured \HI\ column densities of candidate PDRs, the separation between the OB clusters and associated \HI\ patches, the calculated incident flux \Gnaught,and finally the derived total hydrogen volume densities.

Figure \ref{fig:multi}b shows the \HI\ column densities of the patches found near each of the UV sources. The patches are local maxima in the column densities that are assumed to be caused by \HI\ produced in a PDR. Each UV source can have multiple \HI\ patches associated to it. The measured \HI\ columns are fairly flat out to a radius of 0.8 $R_{25}$ (with some spread), after which they start to decline. The maximum measured column densities are affected by beam smoothing and related to the linear resolution.

Using the measured separation between the central UV source in each candidate PDR and its associated \HI\ patches ($\rho_{\HI}$, Figure \ref{fig:multi}c), the incident flux \Gnaught\ on every patch is calculated, and plotted in Figure \ref{fig:multi}d. The distribution of \Gnaught\ values is mostly flat out to a radius of 0.8 $R_{25}$, and after that the values start to decline due to the decreasing measured FUV fluxes. The distribution of measured separations $\rho_{\HI}$ does not change with galactocentric radius and shows no preferred value.

Finally we calculate the total hydrogen volume densities $n$, which are plotted in Figure \ref{fig:multi}e, using Equation \ref{eqn:n}. Values range from 1 to 500 \pccm\ with values going up to 2500 \pccm\ and do not change significantly with galactocentric radius. This corresponds to \Htwo\ densities of up to 250 \pccm\ in GMCs across M\,33. A hint of an upturn at large galactocentric radius may be caused by overestimating the dust content and is not significant.

It is also instructive to plot the relation between $G_0$ and $n$, as displayed in Figure \ref{fig:G0_n}, because of the selection effects, marked by dashed lines:

\makeatletter
\renewcommand{\theenumi}{\Roman{enumi}}
\renewcommand{\labelenumi}{\theenumi.}
\makeatother
\begin{enumerate}
  \item \HI\ column density upper limit of $5 \times 10^{21}\ \rm{cm}^{-2}$ at a characteristic \dtg\ of 0.33.
  \item \HI\ column density lower limit of $6 \times 10^{19}\ \rm{cm}^{-2}$ at a characteristic \dtg\ of 0.25. Note however that this lower limit can be higher in practice, as it also depends on our ability to discern \HI\ patches against the general \HI\ background.
  \item Lowest observable $n$ of $0.2\ \rm{cm}^{-3}$ related to the radio beam size and the \HI\ lower limit.
  \item Minimum usable $G_0$ of $1.4 \times 10^{-3}$, depending on the maximum accepted size of candidate PDRs and the lowest accepted UV flux.
  \item Maximum $G_0$ obtainable by the PDR method considering our data of M\,33 of $1.2 \times 10^{2}$.
\end{enumerate}

Clearly, the selection effects limit the possible values of \Gnaught\ and $n$.

\begin{figure}
  \centering
  \includegraphics[width=\linewidth]{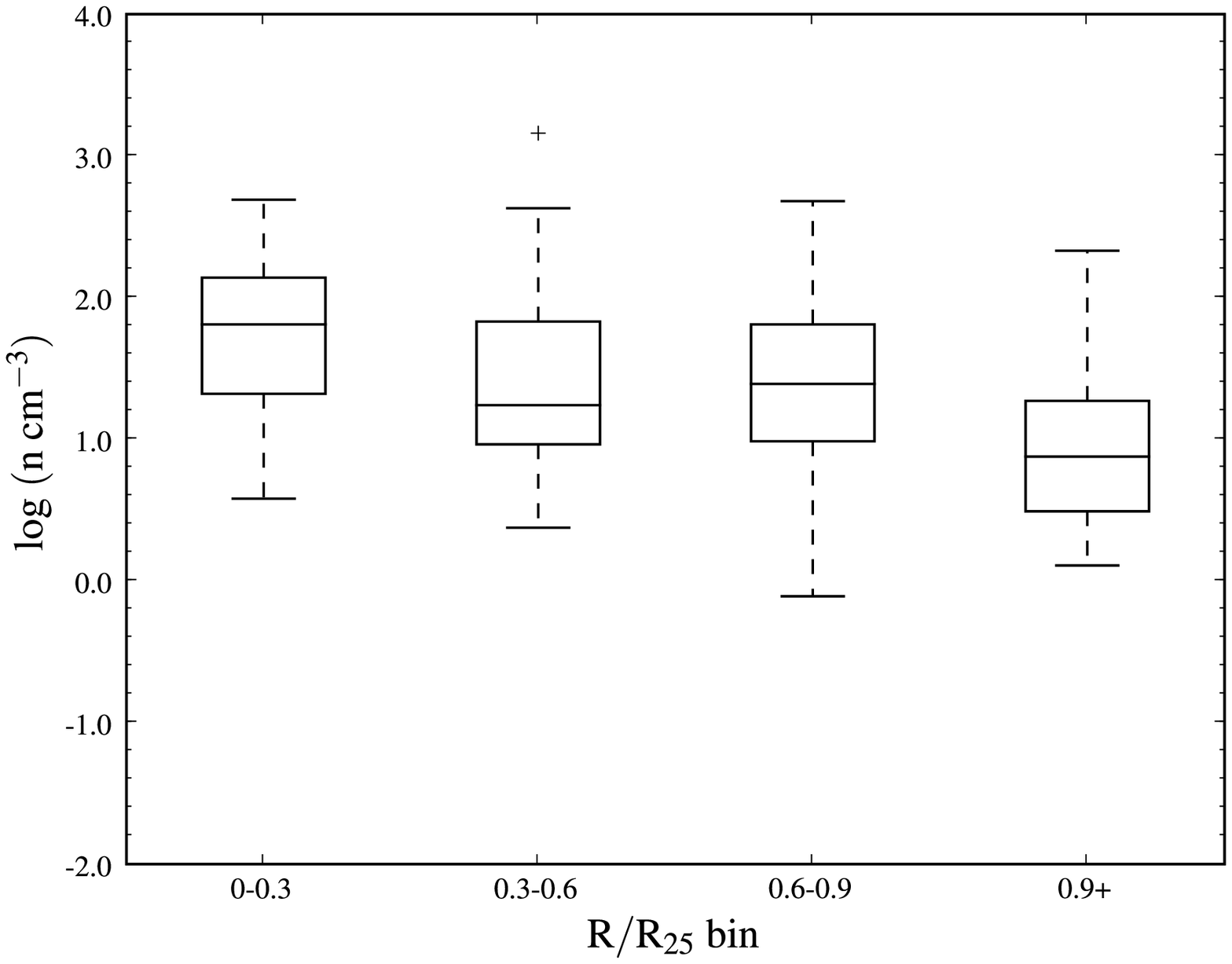}
  \includegraphics[width=\linewidth]{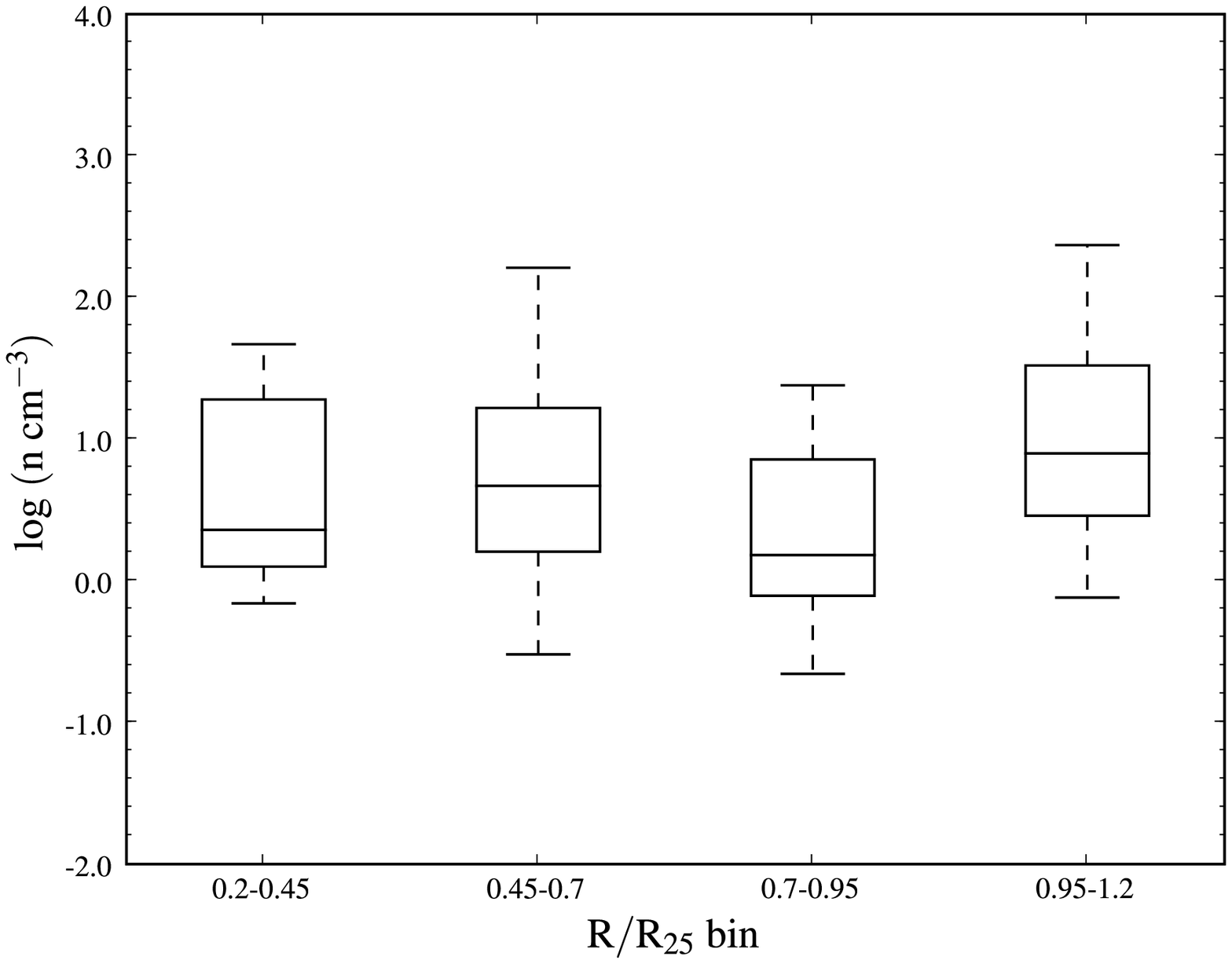}
  \includegraphics[width=\linewidth]{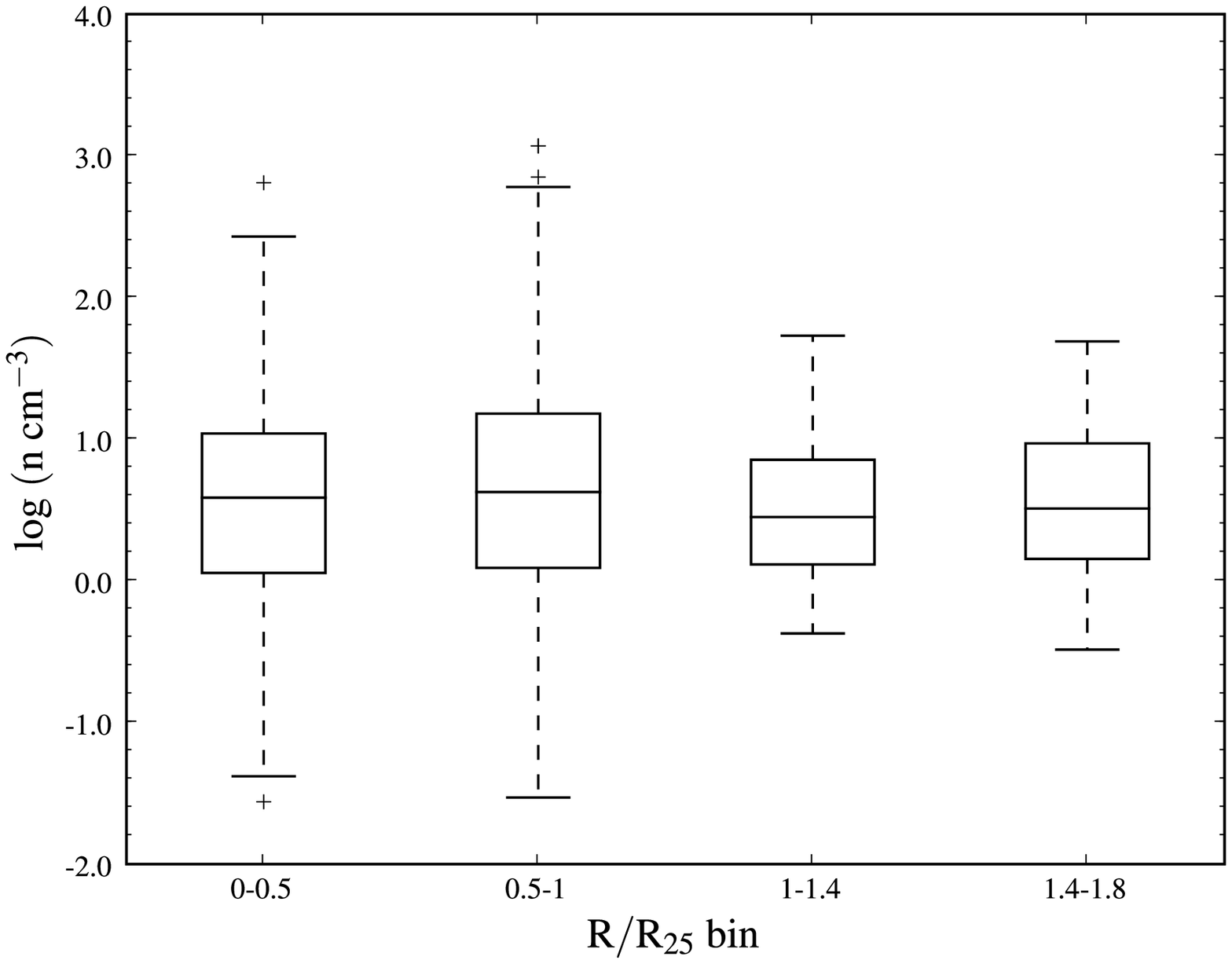}
  \caption{\label{fig:boxplots} Boxplots showing the range of total hydrogen volume densities in the galaxies M\,33, M\,81 and M\,83 respectively (from left to right). The bins have been chosen per galaxy to divide the available range of galactocentric radii in four parts. The M\,83 results, previously published in \citet{2008A&A...489..533H}, show a greater spread in values within $R_{25}$ than the other results. The M\,33 results are narrower in spread and tend to be higher in total hydrogen volume densities.}
\end{figure}

\subsection{Cloud densities in M\,33, full resolution}
\label{sec:full}

We will now present the cloud densities in candidate PDRs as derived with the PDR method from the full resolution ($\sim$ 20 pc linear resolution) data. These are mostly a subset of the regions selected for the reduced resolution analysis. The corresponding regions are listed in Table \ref{tab:correspondence}.
We will give a short description of each region, accompanied by an overlay plot, a finding chart and a data table. The color plots, finding charts and data tables are available in \note{the online version of this journal}. An example plot, finding chart and table are shown for CPSDP 0087g in Figure \ref{fig:cpsdp0087g_example} and Table \ref{tab:cpsdp0087g_example}.
The results for the regions NGC 604 and CPSDP Z204 are included for completeness, but were presented previously in \citet{2009ApJ...700..545H}.
Finally, the results are gathered in a couple of consolidated plots. An overview plot of the location of the regions that were investigated is shown in Figure \ref{fig:locplot}.

\begin{figure*}
  \centering
  \includegraphics[width=0.4\textwidth]{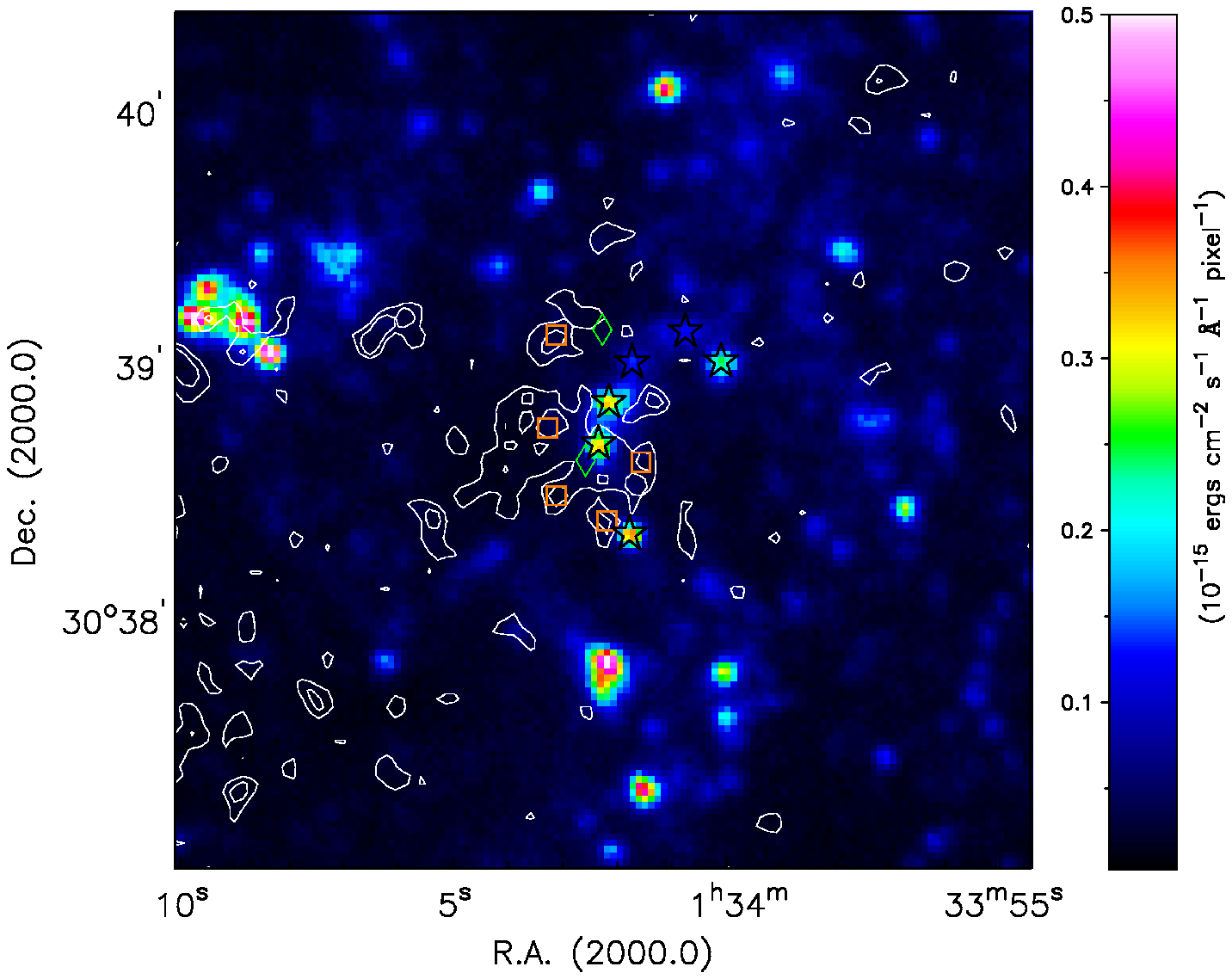}
  \includegraphics[width=0.34\textwidth]{fig6_1r}
  \caption{\label{fig:cpsdp0087g} Left panel: CPSDP 0087g region at full resolution. The FUV color image is overlaid with \HI\ contours. The \HI\ contours are 3 and 4 $\times 10^{21} \rm{cm}^{-2}$ only, to emphasize the morphology of the \HI\ peaks. Black stars mark the locations where the FUV fluxes were measured. The orange boxes are the locations of the measured \HI\ patches. Green diamonds denote the location of CO detections by \citet{2003ApJS..149..343E}. Part of CPSDP Z204 can be seen in the eastern part of the image. Right panel: Finding chart of the same region with letters (UV sources) and numbers (HI patches).}
\end{figure*}

CPSDP 0087g: We measured fluxes in a string of UV sources, that seems to be surrounded by \HI\ emission. (Figure \ref{fig:cpsdp0087g}, Table \ref{tab:cpsdp0087g}.)
The \HI\ morphology is clumpy and is mostly located to the east of the UV sources.
This region has CO detections, that are not directly coinciding with measured \HI\ patches. CO detection close to one of the UV sources is likely to arise from high temperatures. The metallicity in this region is relatively high, higher than, for example, CPSDP Z204 nearby. This depresses the total hydrogen volume densities in the PDR model.

\begin{figure*}
  \centering
  \includegraphics[width=0.4\textwidth]{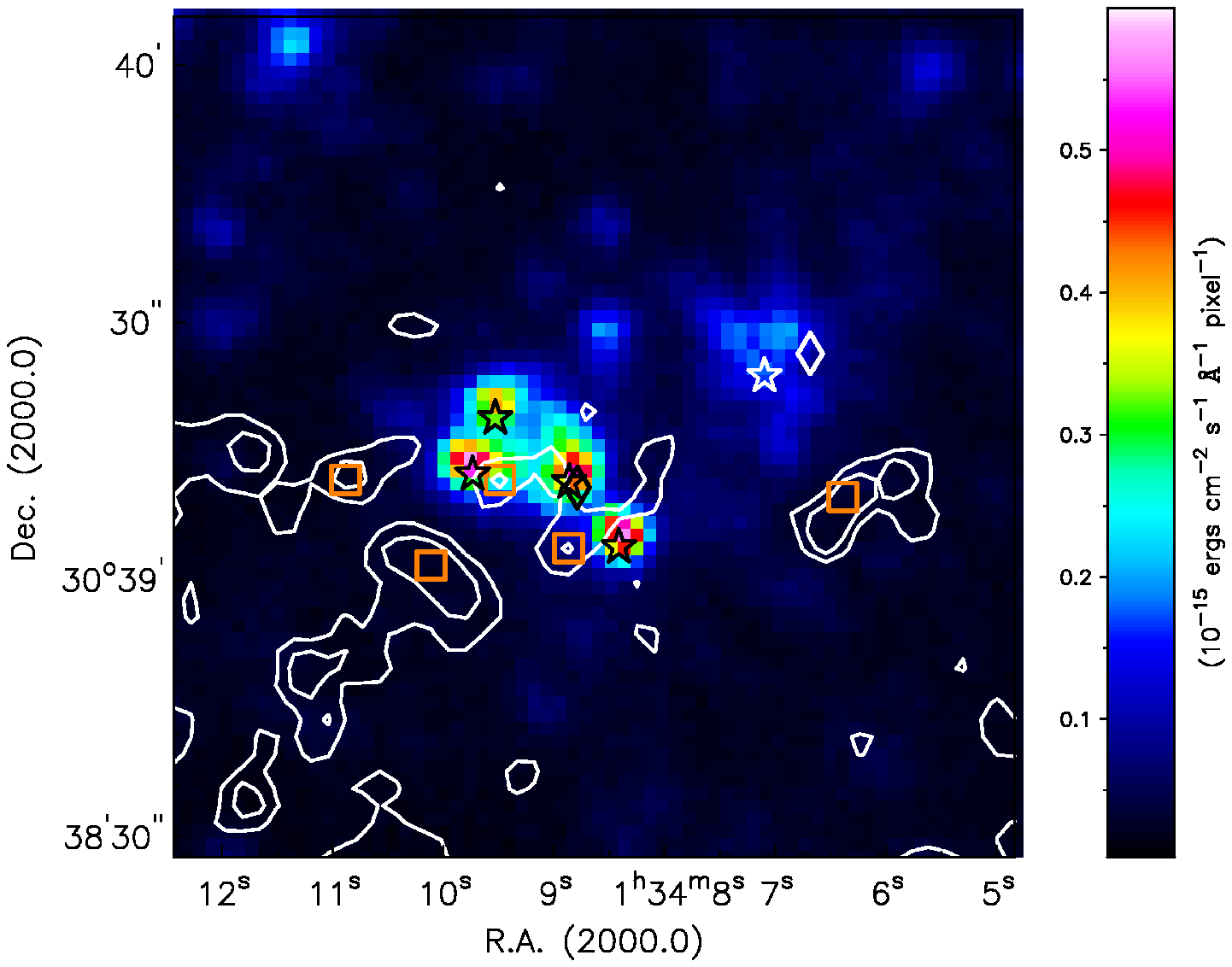}
  \includegraphics[width=0.32\textwidth]{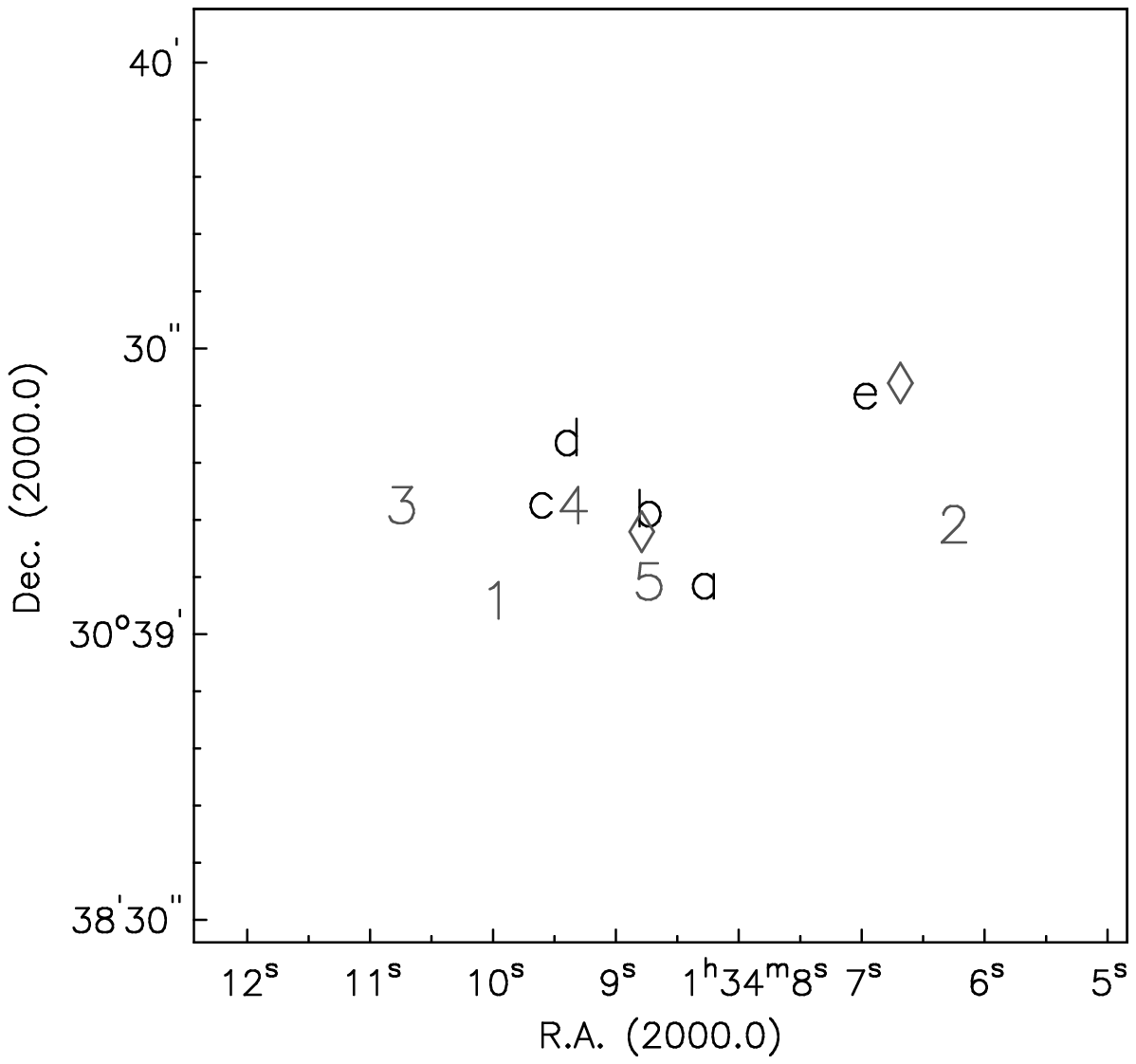}
  \caption{\label{fig:cpsdpz204} Left panel: CPSDP Z204 region at full resolution. The FUV color image is overlaid with \HI\ contours. The \HI\ contours are 3 and 4 $\times 10^{21} \rm{cm}^{-2}$ only, to emphasize the morphology of the \HI\ peaks. Black stars mark the locations where the FUV fluxes were measured. The orange boxes are the locations of the measured \HI\ patches. Diamonds denote the location of CO detections by \citet{2003ApJS..149..343E}. Right panel: Finding chart of the same region with letters (UV sources) and numbers (HI patches), as well as the CO detection diamonds.}
\end{figure*}

CPSDP Z204 is about 1.6 kpc away from the centre of M\,33 and is shown in Figure \ref{fig:cpsdpz204} and Table \ref{tab:cpsdpz204}. It was described in more detail in \citet{2009ApJ...700..545H} and shows a morphology of \HI\ patches surrounding a complex of UV sources, making these \HI\ clumps candidate PDRs. The detection of CO is a further confirmation of the presence of GMCs in the area.

\begin{figure*}
  \centering
  \includegraphics[width=0.4\textwidth]{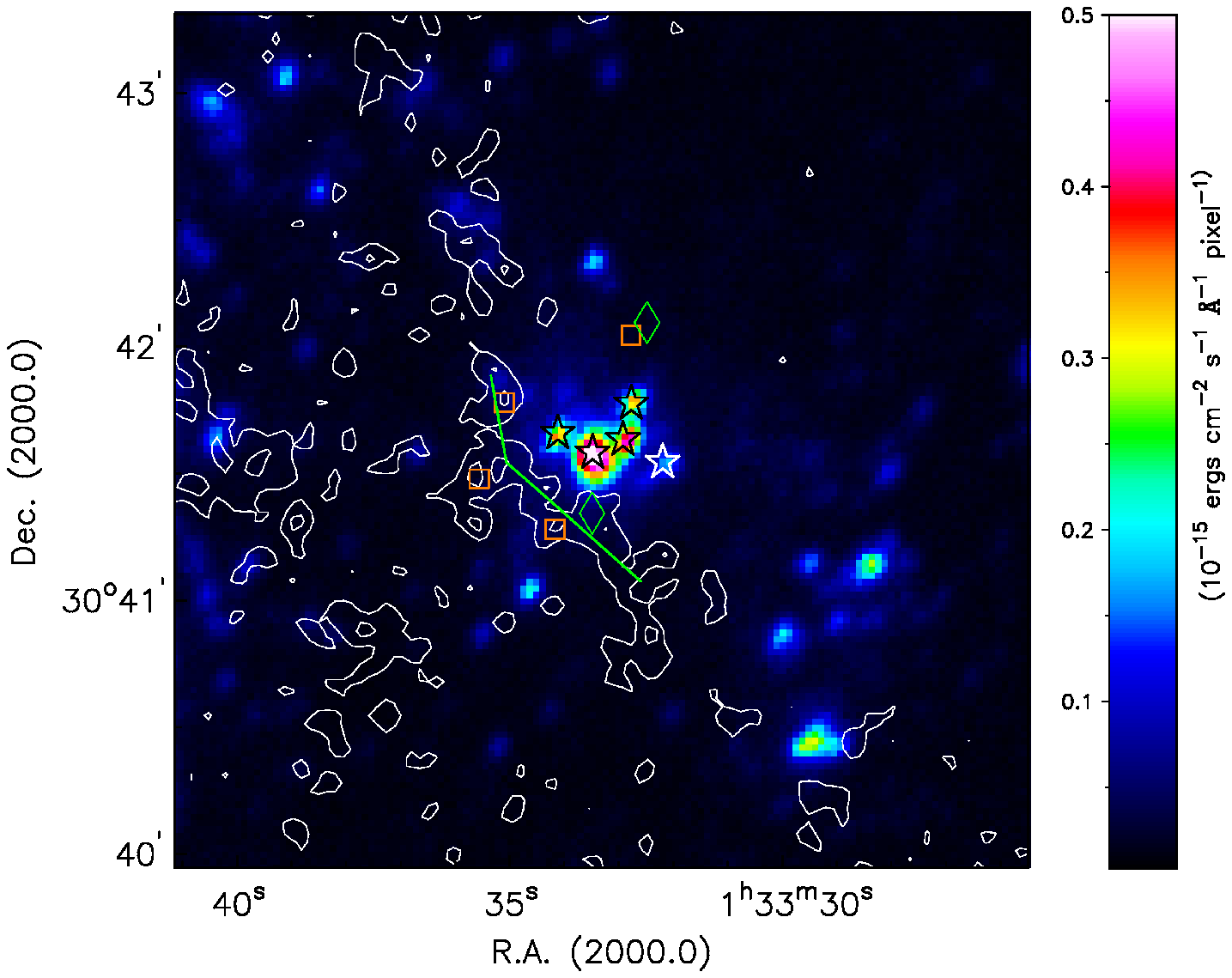}
	\includegraphics[width=0.32\textwidth]{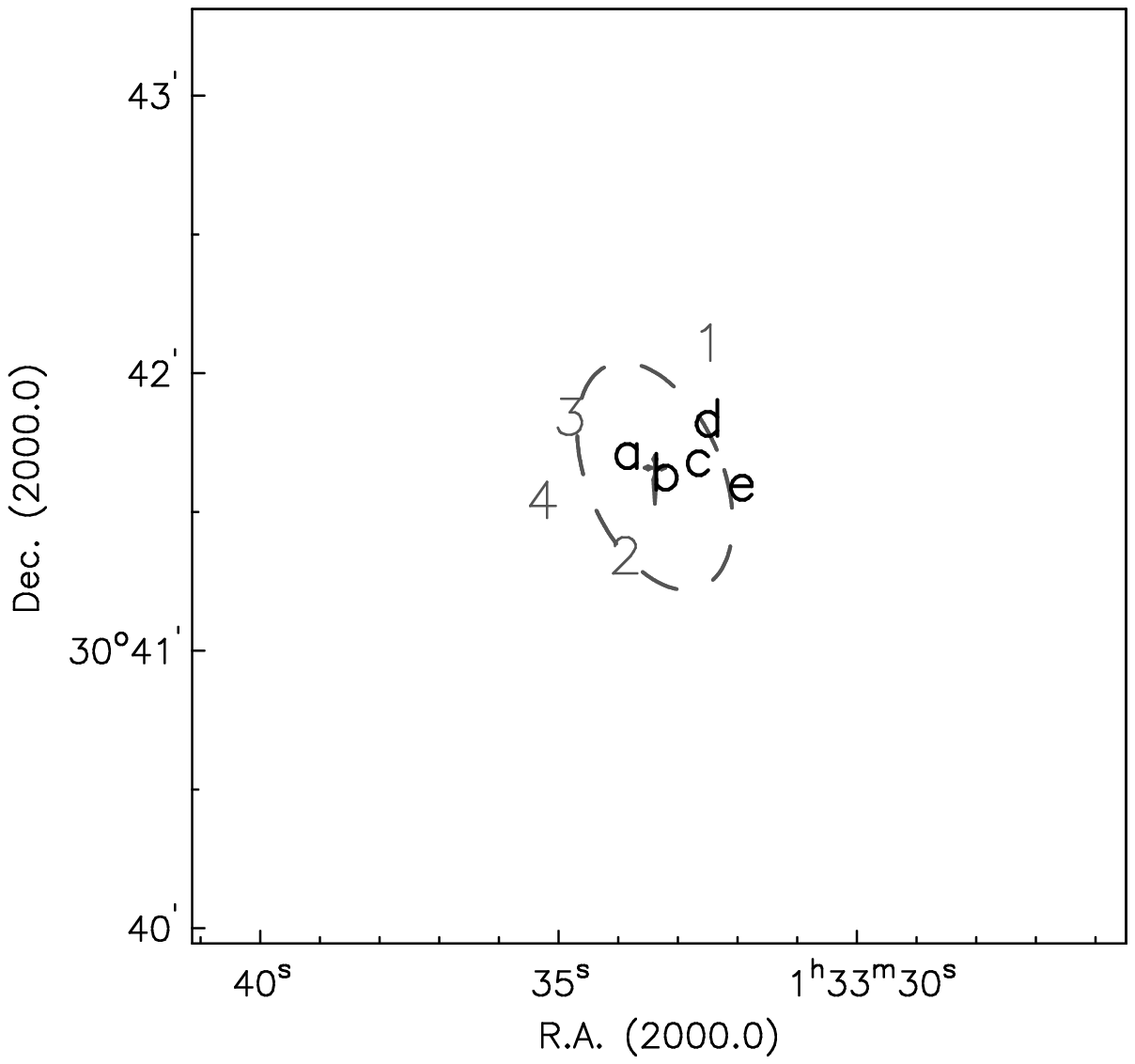}
  \caption{\label{fig:ngc595} Left panel: NGC 595 region at full resolution. The FUV flux range is indicated in the color bar. The HI contours are 3, and 4.5 $\times 10^{21} \rm{cm}^{-2}$. The green line indicates the path along which the HI columns were averaged for the large scale measurement. The green diamonds mark the location of CO detections. Right panel: Finding chart of the same region with letters (UV sources) and numbers (HI patches).}
\end{figure*}

NGC 595 is M\,33's second brightest \HII\ regions and has been studied extensively. \citet{1992ApJ...385..512W} studied the molecular content of this region and found it to have less molecular gas mass than NGC 604 by about an order of magnitude (half a million solar masses). They noted a significant atomic mass component and point to photodissociation as the likely cause of this \HI. \citet{2009ApJ...700.1847L} derive an overal electron density of the nebula of 162 \pccm\ from \Halpha, [O III] and [S II] kinematic observations.
The finding chart in Figure \ref{fig:ngc595} shows that \HI\ patch no. 1, that was selected because it has CO associated to it, has a density of 105 \pccm\ (Table \ref{tab:ngc595}). However, patch no. 2 is associated with CO emission as well, but has a lower density (only 12 \pccm). The CO emission here may be the result of a higher temperature of the gas rather than a higher density of the GMCs. A larger scale measurements (green polygon in the overlay), yields a slightly higher density of 45 \pccm.

\begin{figure*}
  \centering
  \includegraphics[width=0.4\textwidth]{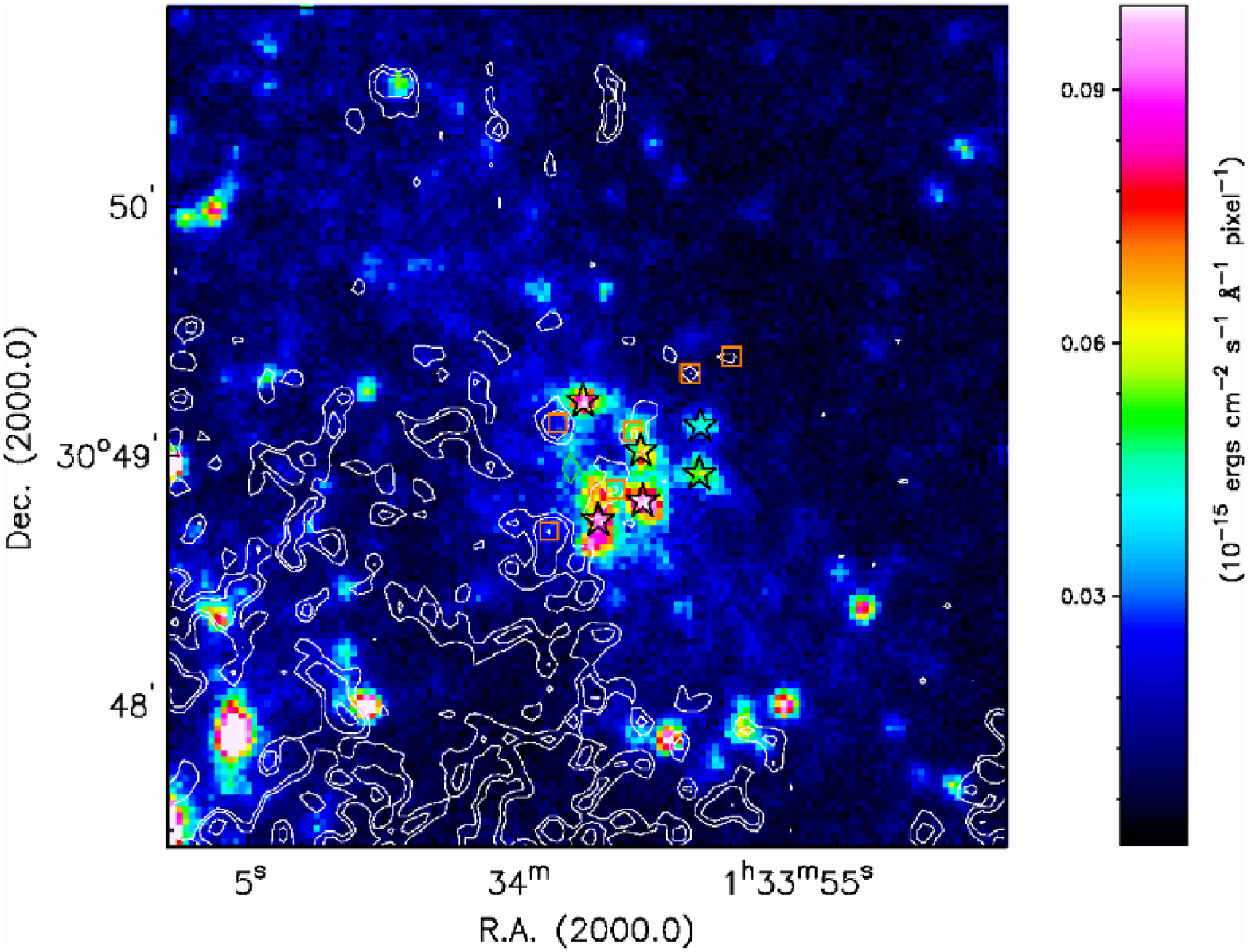}
	\includegraphics[width=0.32\textwidth]{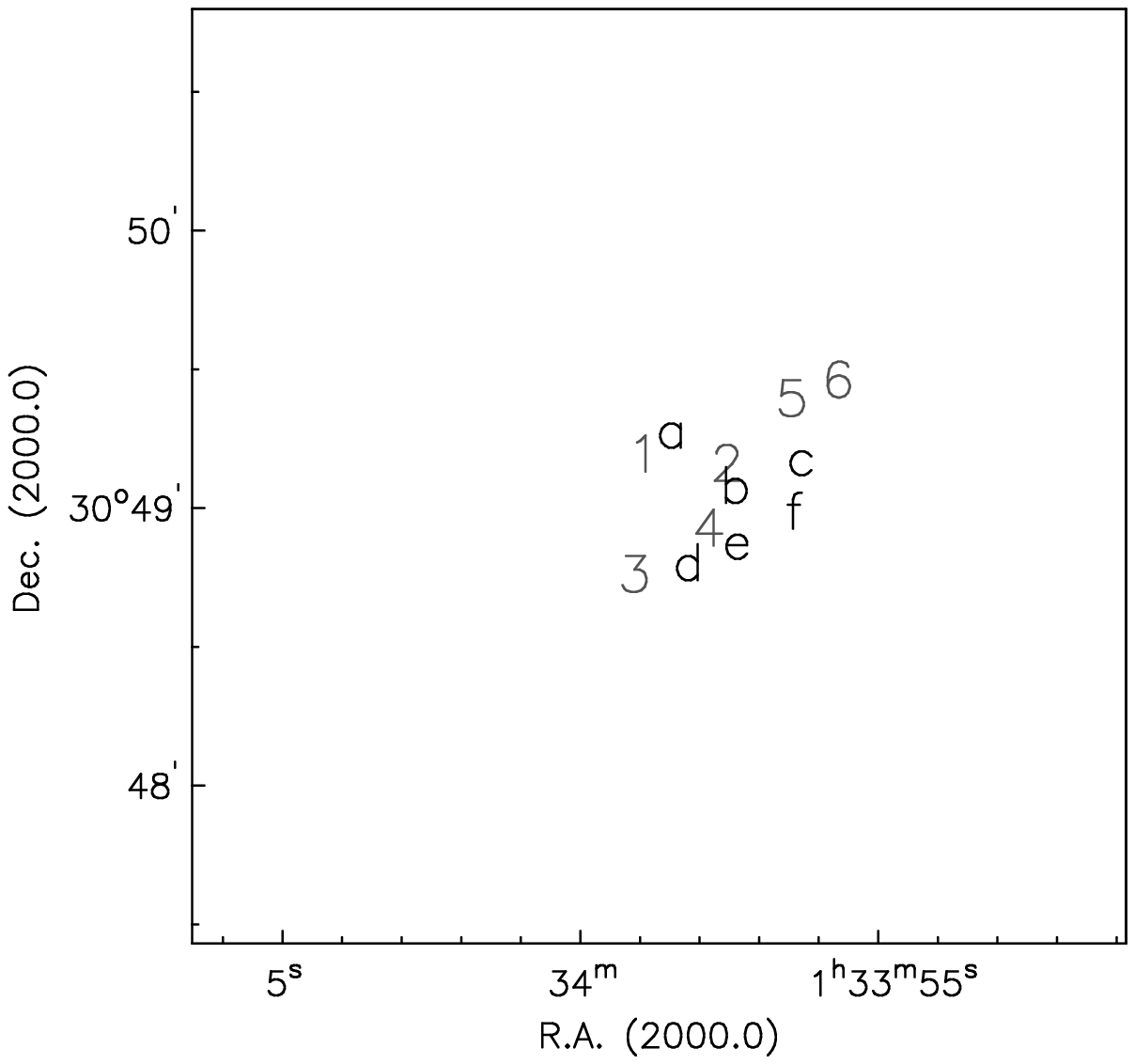}
  \caption{\label{fig:bclmp0695} Same as Figure \ref{fig:cpsdpz204}, but for BCLMP 0695. The \HI\ contours are 2, 2.5 and 3.5 $\times 10^{21} \rm{cm}^{-2}$. A CO detection (diamond) lies between \HI\ patches 1 and 4.}
\end{figure*}

BCLMP 0695: No clear large-scale structure can be discerned in this \HII\ region, although the distribution of atomic hydrogen seems to suggest some filamentary structures. The UV fluxes are comparable to the lower fluxes in NGC 595, and the number of sources points to a lot of recent star formation (Figure \ref{fig:bclmp0695}). The UV sources surround the single detection of CO in this region, and so do the \HI\ patches. The closest ones are no. 1 (23 \pccm) and no. 4 (116 \pccm). We would also expect CO emission at no. 2, since it has a density of 195 \pccm, but it is also close enough to the UV source that any CO would have been broken up. In that case it would be of interest to look for ionized carbon atoms.

\begin{figure*}
  \centering
  \includegraphics[width=0.4\textwidth]{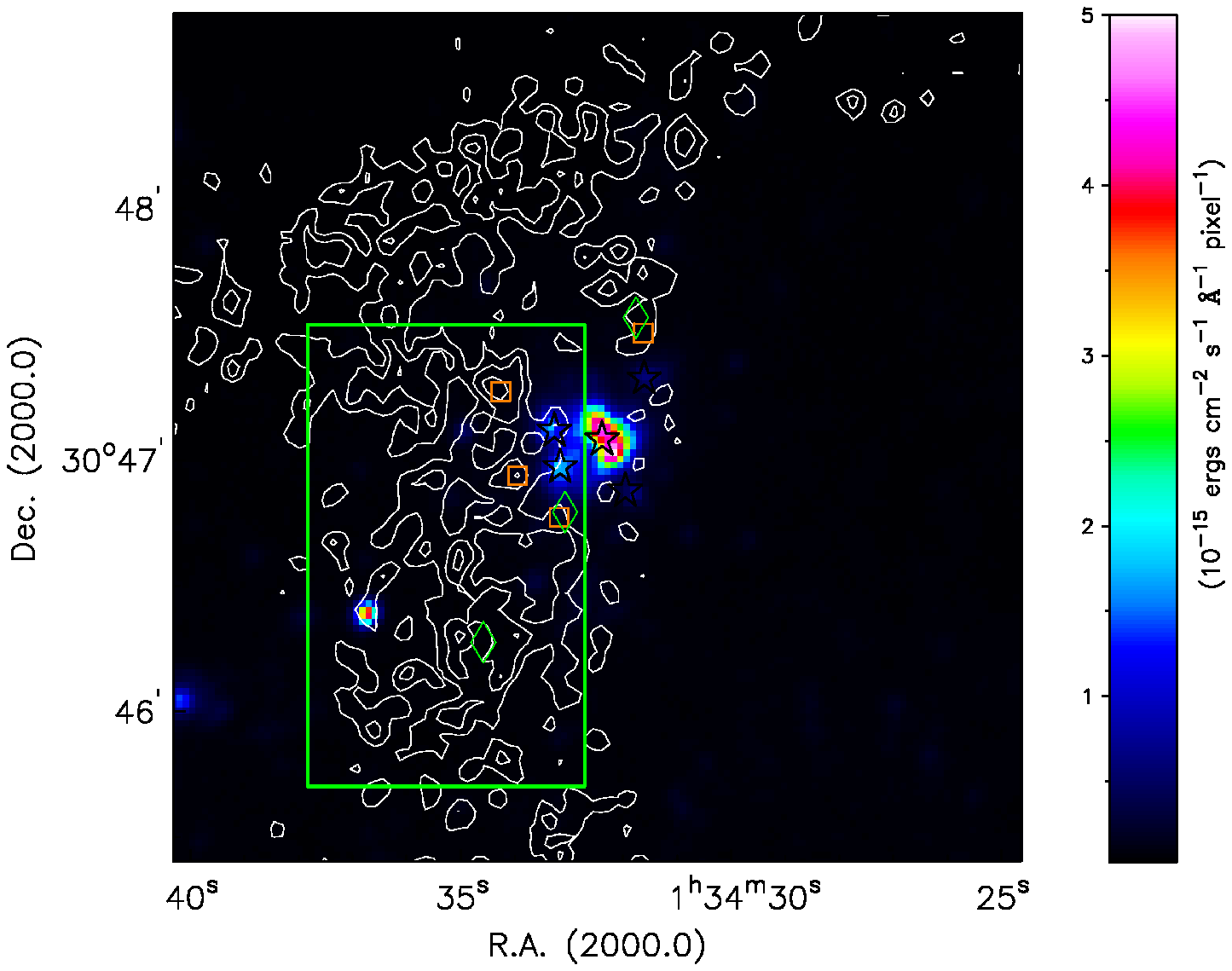}
	\includegraphics[width=0.32\textwidth]{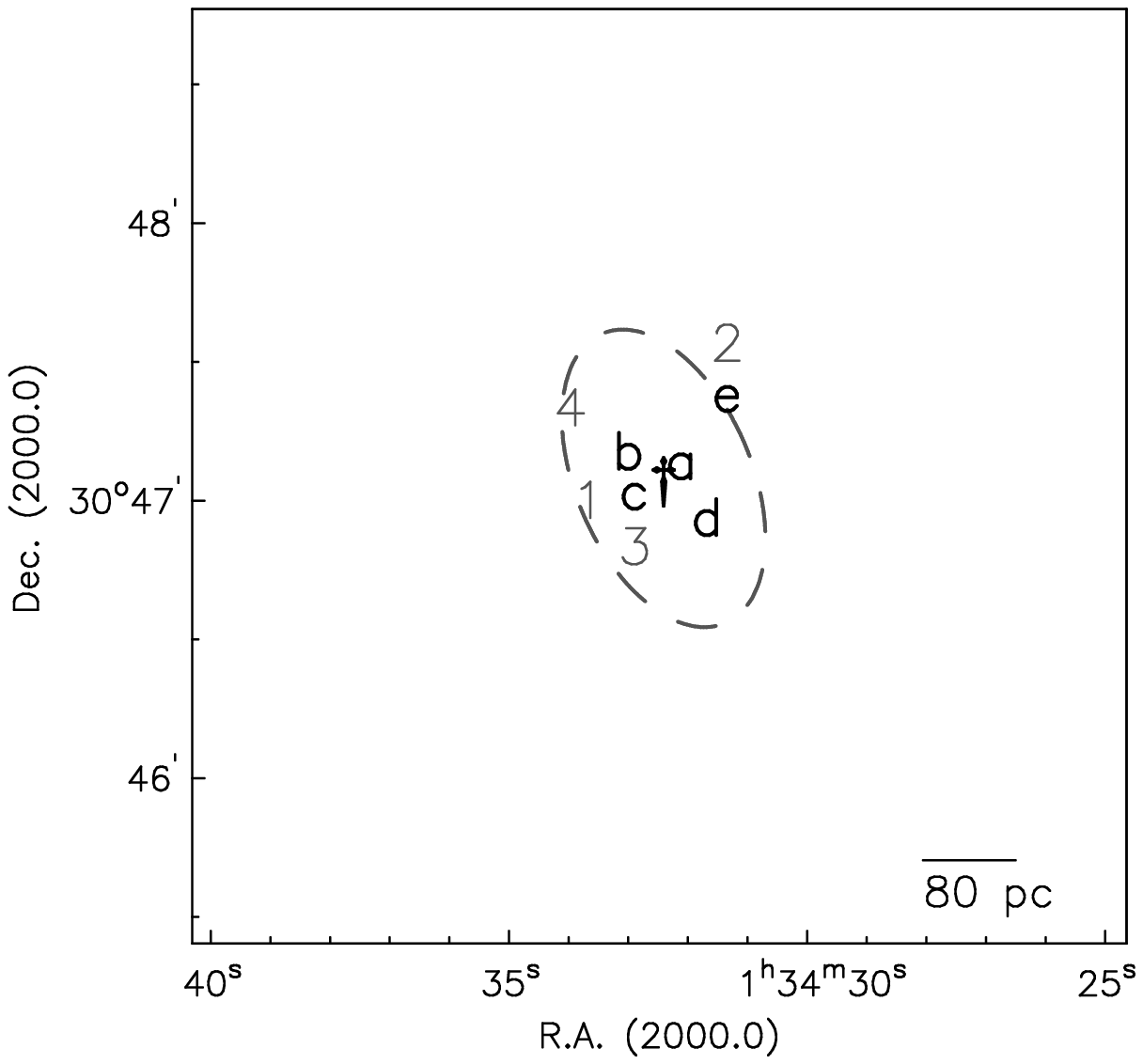}
  \caption{\label{fig:ngc604} NGC 604 region at full resolution. The FUV flux range is indicated in the color bar. The HI contours are 3, 4 and 5 $\times 10^{21} \rm{cm}^{-2}$. The green rectangle indicates the region where the HI columns were averaged for our measurement. The orange boxes are the locations of the measured \HI\ patches. See Table \ref{tab:ngc604}, for the measurements corresponding to the finding chart.}
\end{figure*}

NGC 604 is the largest and most luminous \HII\ region in M\,33.
Our measurements of NGC 604 are listed in Table \ref{tab:ngc604}. A detailed view of NGC 604 is shown in Figure \ref{fig:ngc604}.
The central cluster of OB stars is situated on the edge of an \HI\ arm. We attempted to get a global measurement using an average \HI\ column density measured on the \HI\ arm. \HI\ column densities to the west of NGC 604 drop to below the sensitivity limit, although a diffuse component shows on single dish GBT data (Thilker, 2008, private communication). Since the UV emission does not particularly follow the \HI\ arm, this faint \HI\ emission deserves further attention.

\begin{figure*}
  \centering
  \includegraphics[width=0.4\textwidth]{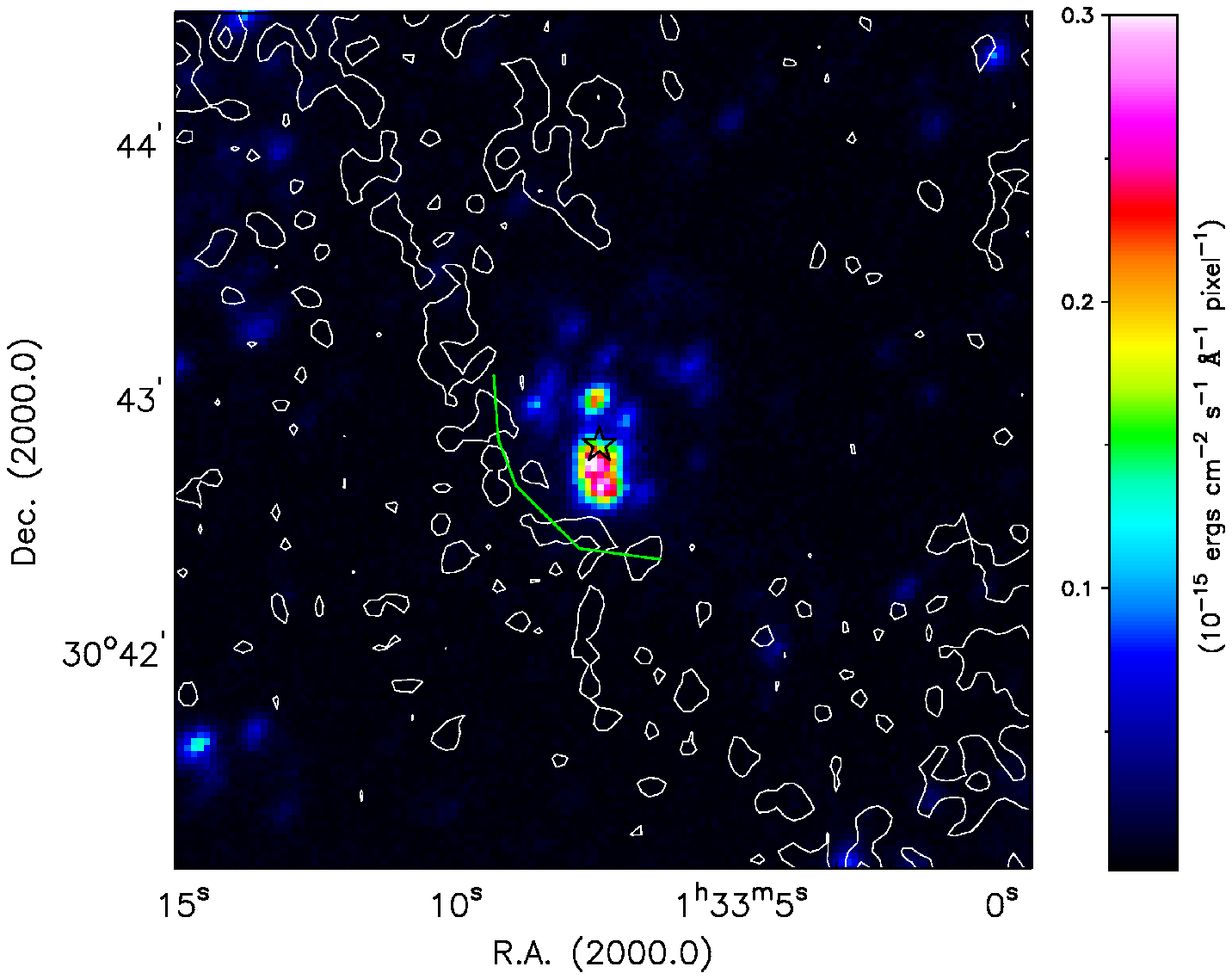}
  \includegraphics[width=0.32\textwidth]{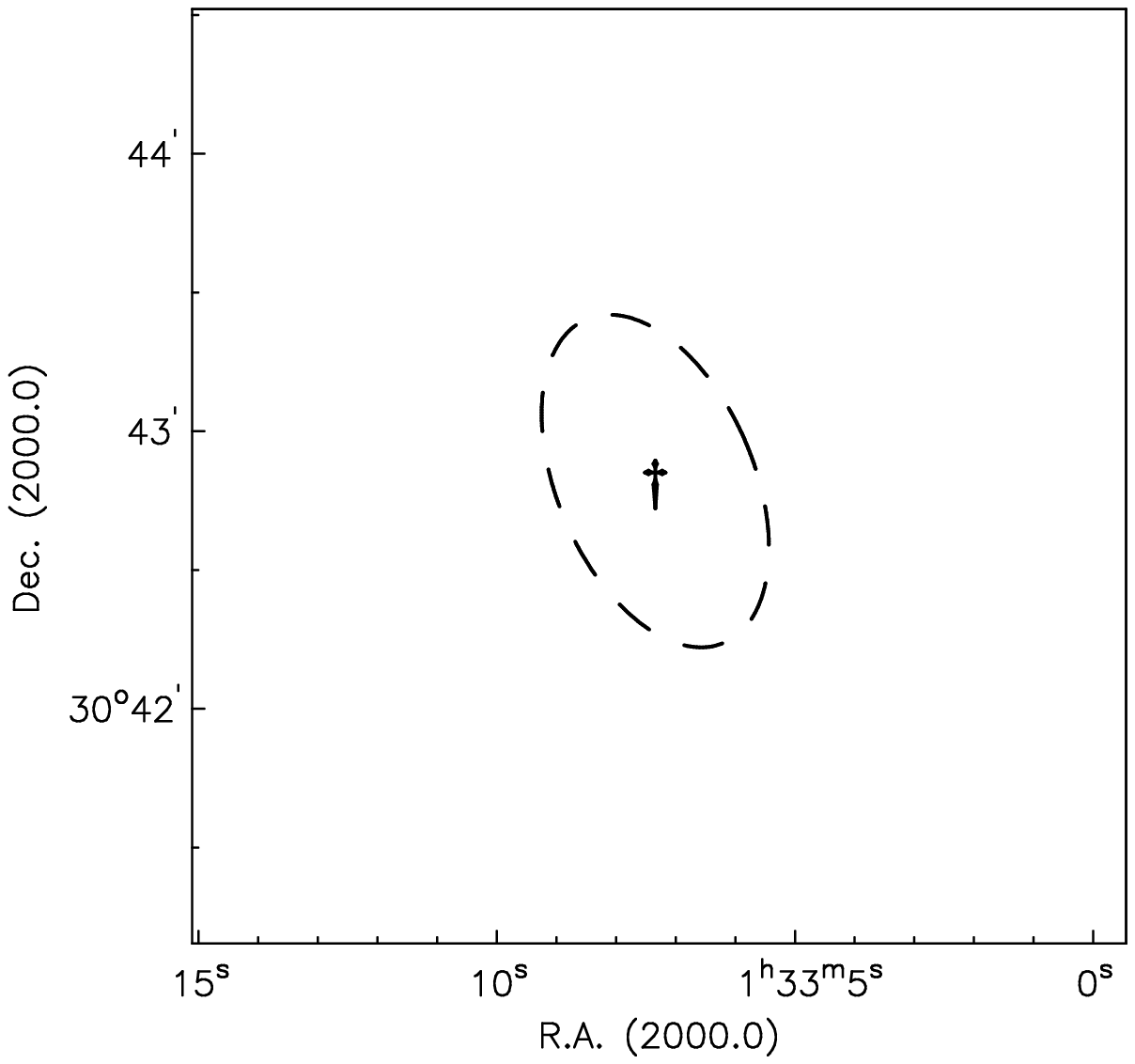}
  \caption{\label{fig:bclmp0288} Same as Figure \ref{fig:cpsdpz204}, but for BCLMP 0288. The \HI\ contours are 2.5 and 4 $\times 10^{21} \rm{cm}^{-2}$. This region did not show a detailed morphology, so only a large-scale measurement was made.}
\end{figure*}

Region BCLMP 0288 only features a morphology that suggests a large-scale PDR, which makes it an interesting target in its own right. The central cluster is embedded in \HI\ emission that doesn't show discernible structure on a smaller scale. The highest \HI\ columns go around BCLMP 0228 on the eastern side and we used an average along that ridge for our measurements (Figure \ref{fig:bclmp0288}, Table \ref{tab:bclmp0288}).

\begin{figure*}
  \centering
  \includegraphics[width=0.4\textwidth]{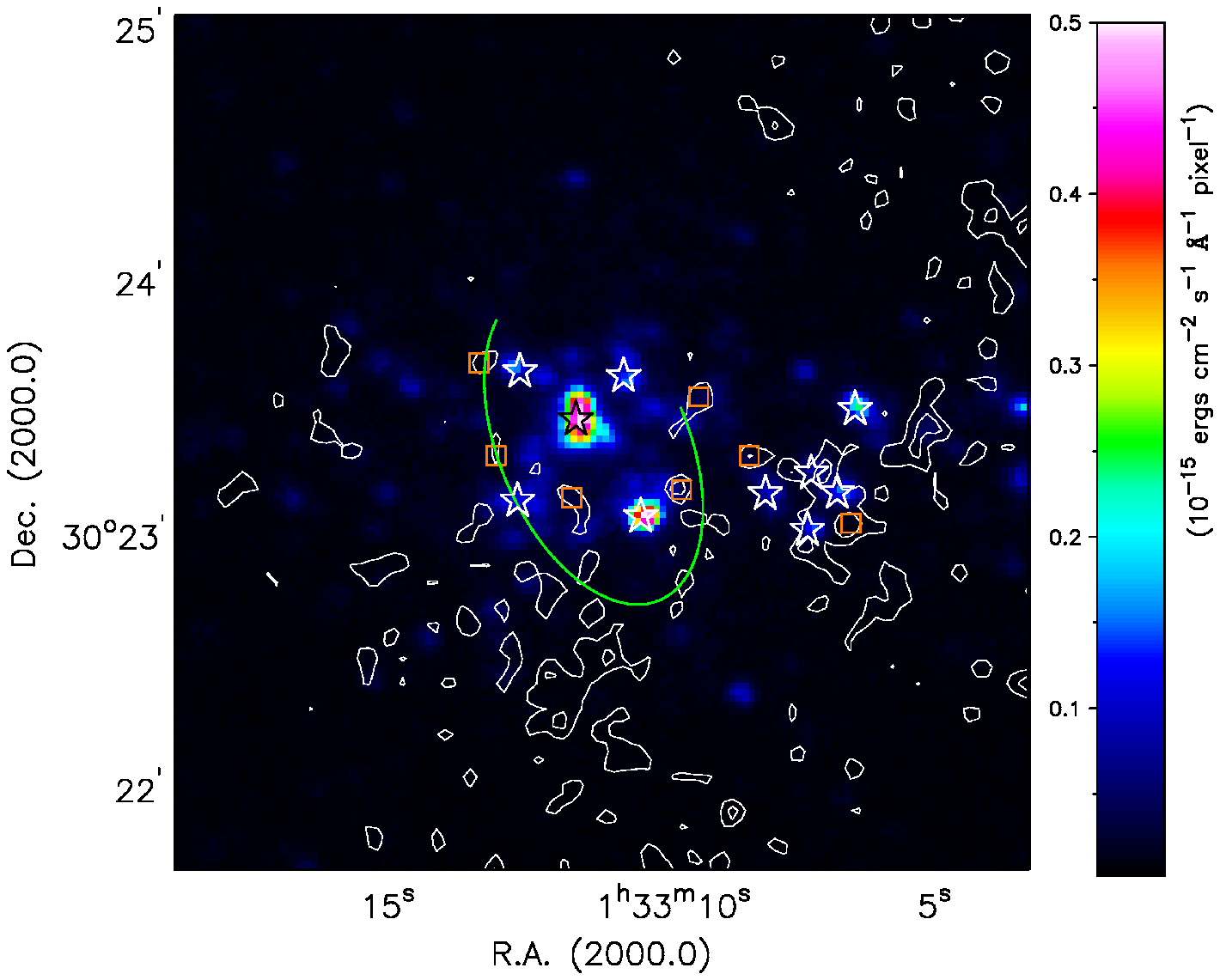}
	\includegraphics[width=0.32\textwidth]{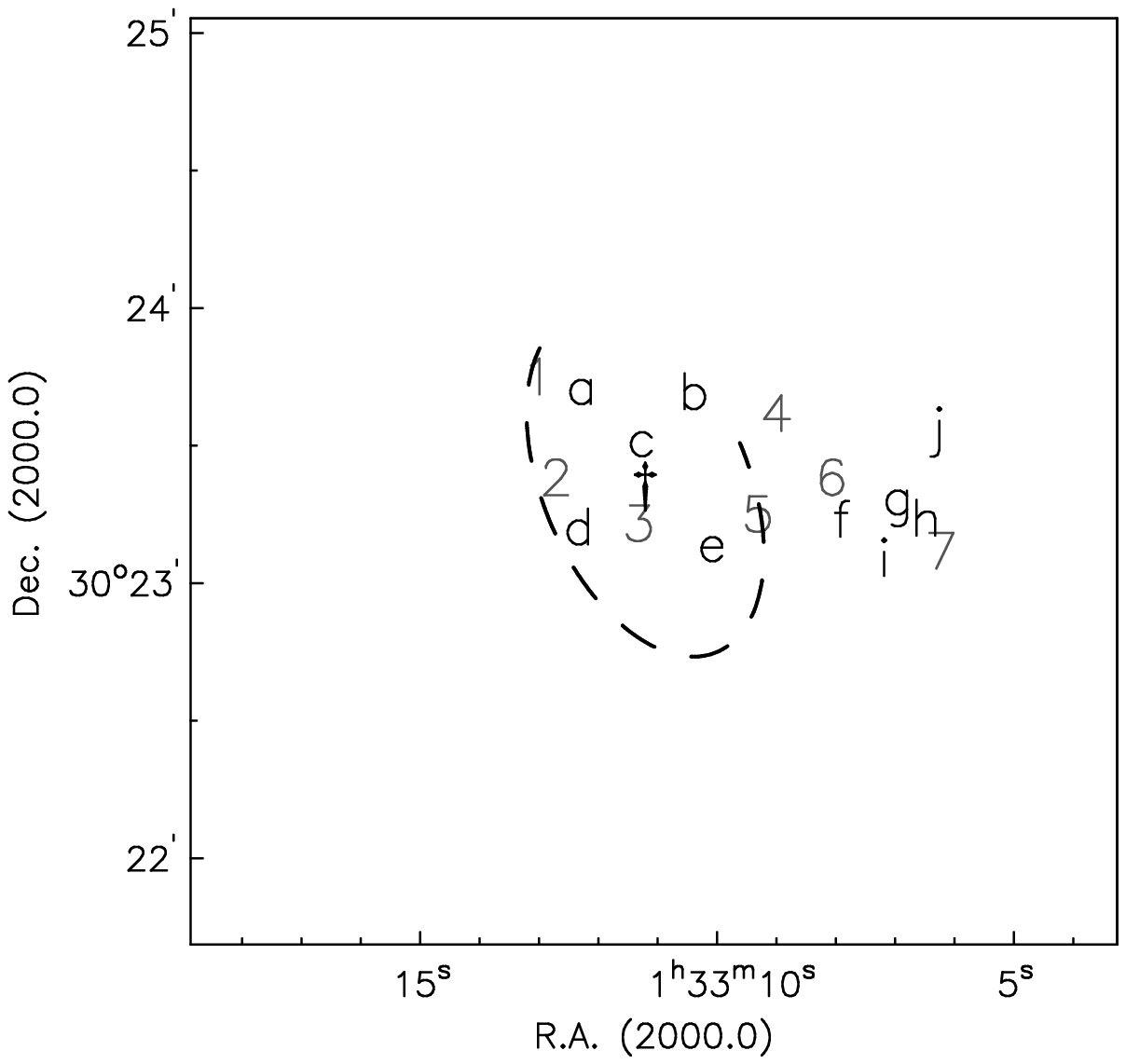}
  \caption{\label{fig:bclmp0256} Same as Figure \ref{fig:cpsdpz204}, but for BCLMP 0256. The \HI\ contours are 3.5 and 4.5 $\times 10^{21}~\rm{cm}^{-2}$, to reduce confusion from the presence of generous amounts of \HI. The partial ellipse traces the general \HI\ distribution peaks.}
\end{figure*}

BCLMP 0256: This \HII\ region contains three individual metallicity measurements from \citet{2008ApJ...675.1213R}. We adopted the nearest measurement to each \HI\ patch, as can be seen in Table \ref{tab:bclmp0256}. A partial \HI\ ridge surrounding BCLMP 0256 on the basis of surrounding \HI\ patches is shown, among others, in Figure \ref{fig:bclmp0256}. Additional UV sources were measured towards the west, where the extra metallicity measurements were available.

\begin{figure*}
  \centering
  \includegraphics[width=0.4\textwidth]{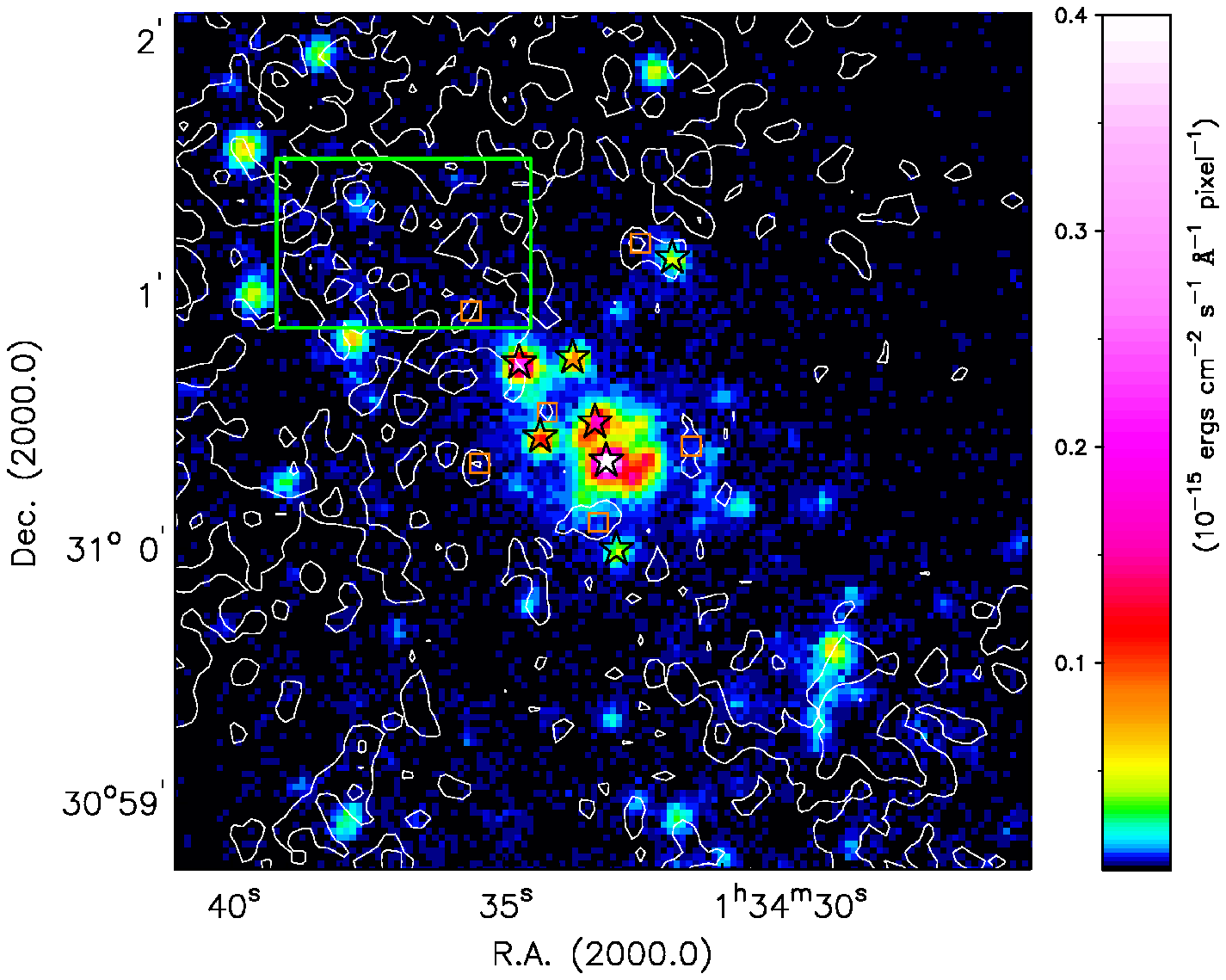}
	\includegraphics[width=0.32\textwidth]{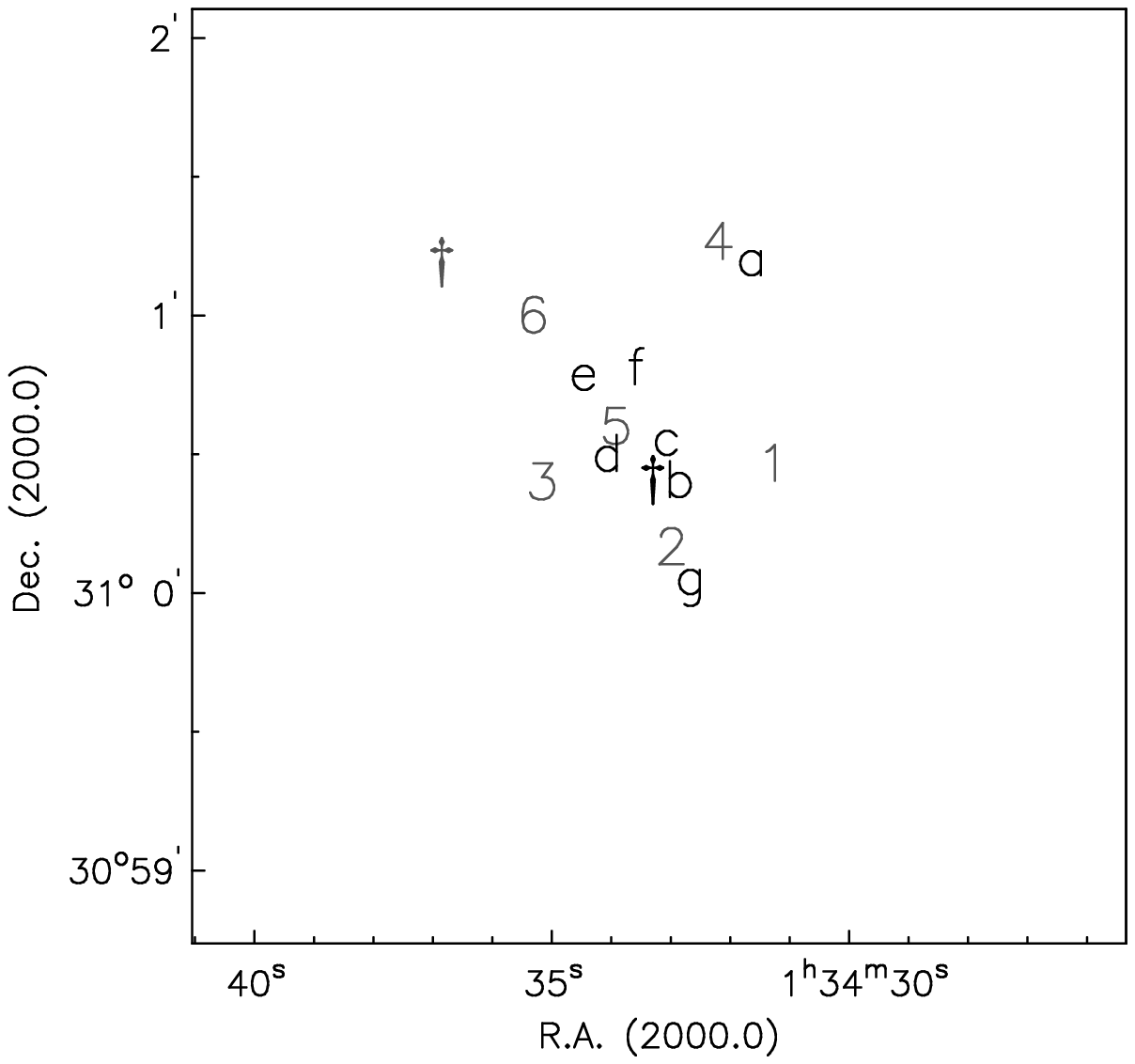}
  \caption{\label{fig:bclmp0650} Same as Figure \ref{fig:cpsdpz204}, but for BCLMP 0650. The \HI\ contours are 1.5 and 2.5 $\times 10^{21}~\rm{cm}^{-2}$. A global \HI\ measurement was attempted in the rectangular region, centered on the $\dagger$.}
\end{figure*}

The region BCLMP 0650 (Figure \ref{fig:bclmp0650}) shows a rich morphology of a large-scale PDR and intermediate size PDRs, of which we have analysed 5. While the large-scale \HI\ shell is not very dominant, we still attempted to measure its average column density in a rectangular area, finding a total hydrogen density of 18 \pccm.

\begin{figure*}
  \centering
  \includegraphics[width=0.4\textwidth]{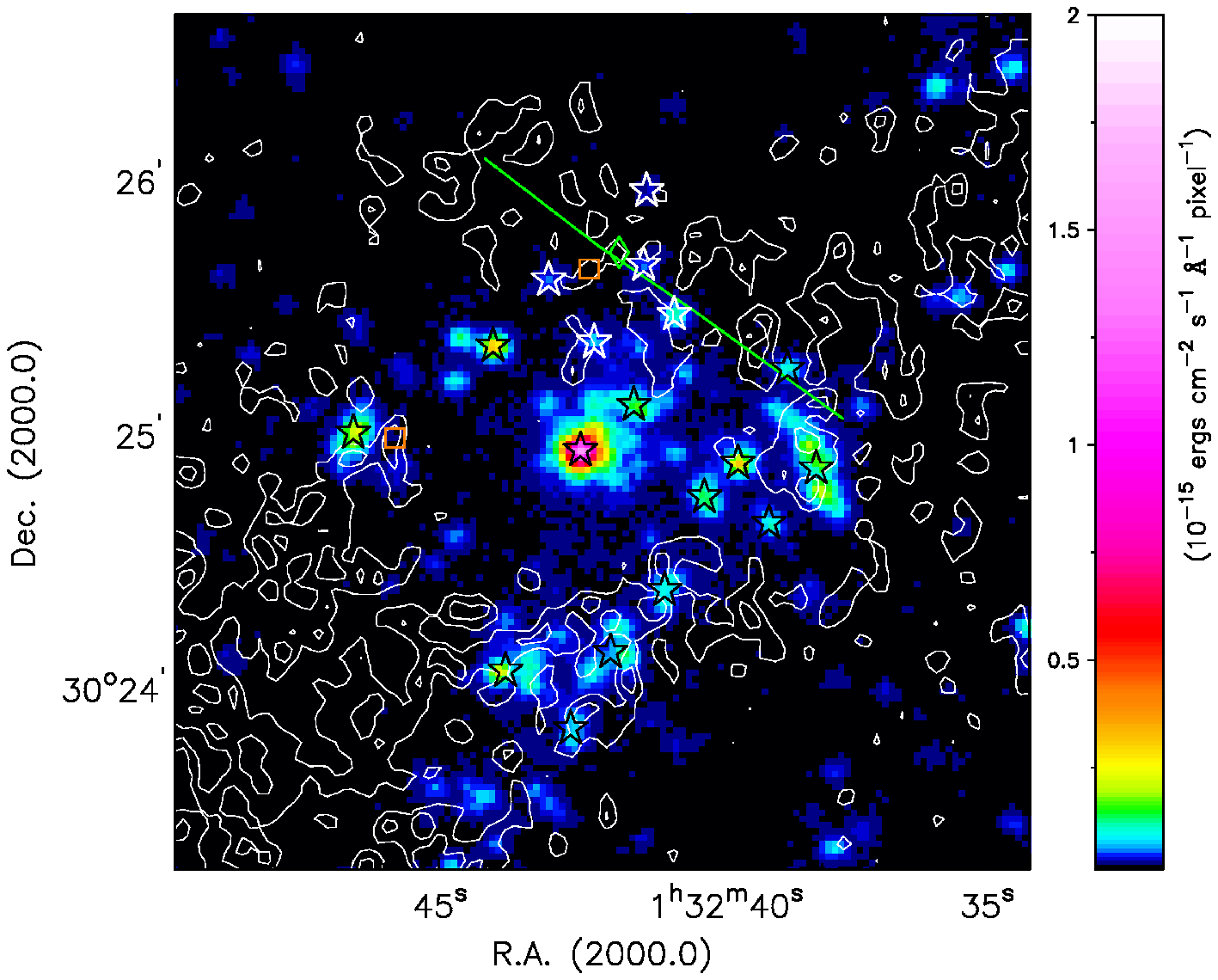}
	\includegraphics[width=0.32\textwidth]{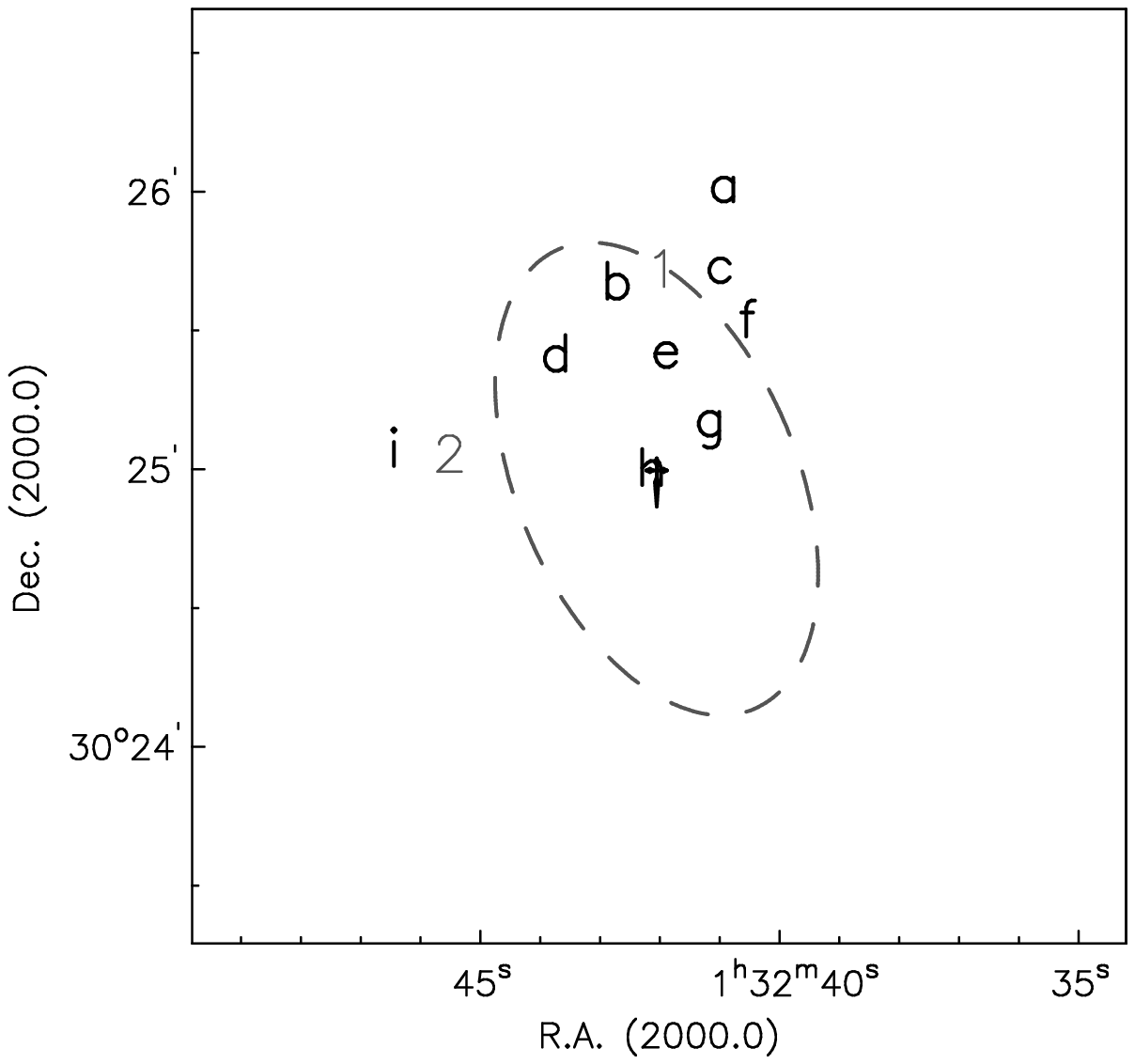}
  \caption{\label{fig:bclmp0269} Same as Figure \ref{fig:cpsdpz204}, but for BCLMP 0269. The \HI\ contours are 3 and 4 $\times 10^{21}~\rm{cm}^{-2}$. The right panel only contains the UV sources that were considered to be significantly impinging on the \HI\ patches. A more global \HI\ measurement was taken along the green line. A CO detection can also be found along this line.}
\end{figure*}

BCLMP 0269: This region features a lot of recent star formation, scattered over a relatively large area. The distribution of the atomic hydrogen surrounding the young star clusters is reminiscent of a large-scale shell. The average \HI\ column density of this shell is measured  along a straight line in Figure \ref{fig:bclmp0269}). The region includes a CO measurement by \citet{2003ApJ...599..258R}, that we connect to an intermediate size PDR. We also include a similarly sized PDR on the eastern part of BCLMP 0269, which looks like our typical candidate PDR.

\begin{figure*}
  \centering
  \includegraphics[width=0.4\textwidth]{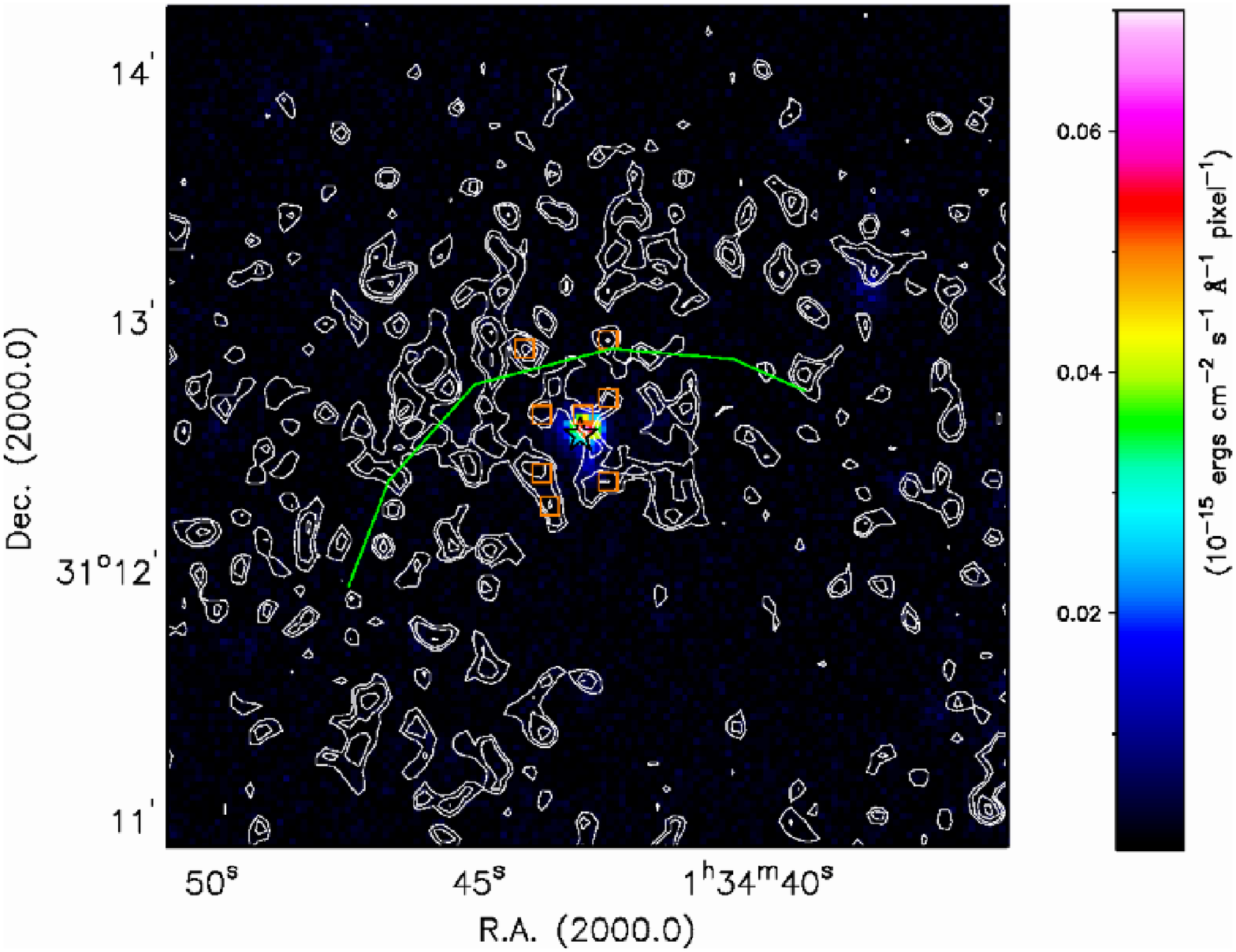}
	\includegraphics[width=0.32\textwidth]{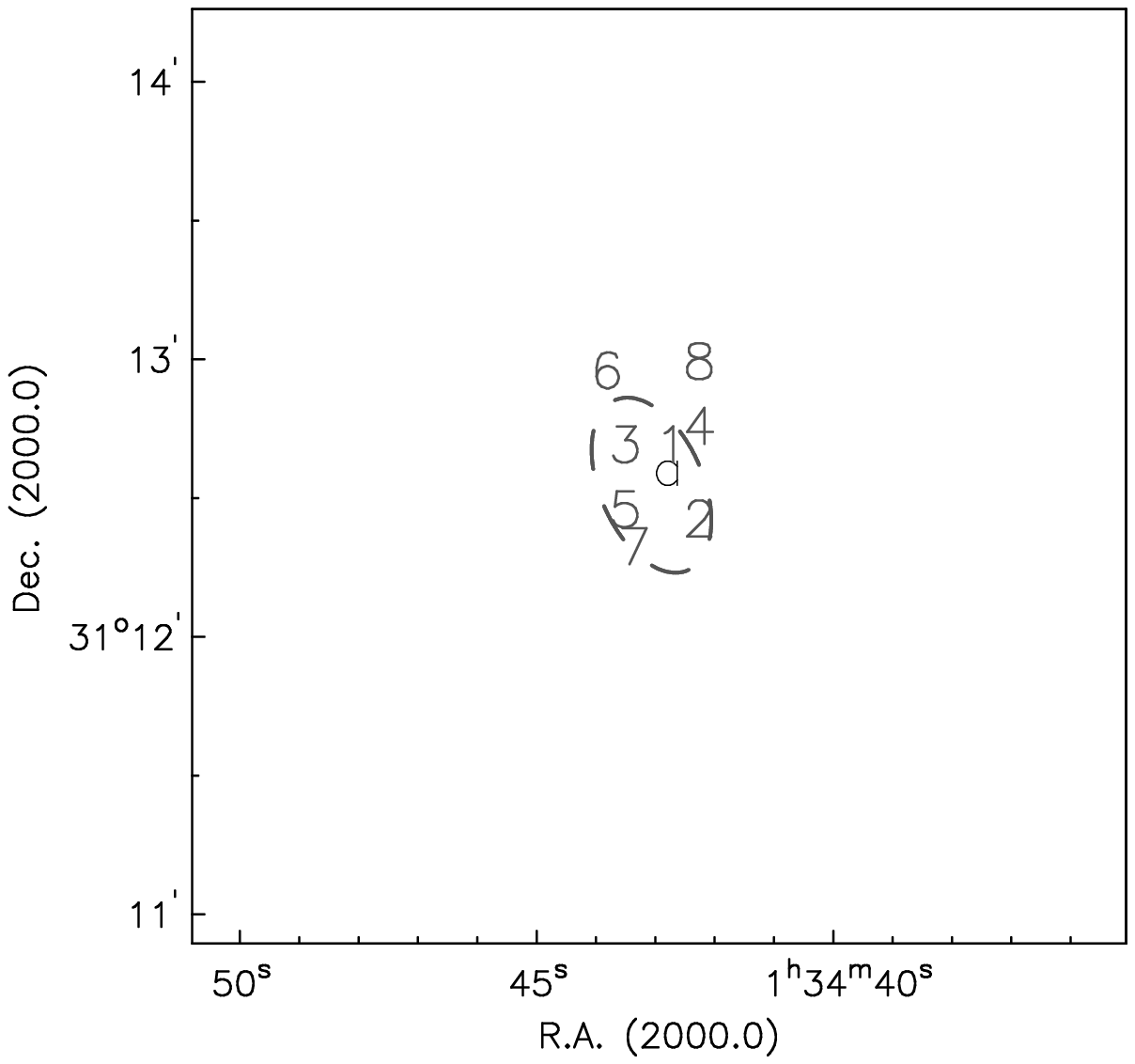}
  \caption{\label{fig:r43} Our Region 42 at full resolution. Left panel: The FUV color image is overlaid with \HI\ contours. The \HI\ contours are 1.7, 2, 2.5, 3 and 3.5 $\times 10^{21} \rm{cm}^{-2}$.
  The central FUV source is marked with a black star. Orange boxes mark the locations of \HI\ patches that were measured. The green spline tracks the path along which we averaged \HI\ column densities to get a measure of the potential large scale PDR. Right panel: Finding chart of the same region with 'a' for the UV source and numbers (HI patches). The ellipse indicates the fitted radius to the green spline in the left panel (the corresponding measurements are in {\it italic} in Table \ref{tab:r43}).}
\end{figure*}

Our Region 42 lies relatively far from the centre of M\,33, at about 8.6 kpc (Figure \ref{fig:r43}, Table \ref{tab:r43}). It features a single UV complex at its centre, surrounded by various \HI\ patches that can be used to probe the underlying GMCs in this candidate PDR. These patches are expected to indicate medium scale PDRs. The total hydrogen volume densities found in this way range from 6 \pccm\ to about 220 \pccm.
Additionally, we attempted to probe the large-scale PDR by averaging the \HI\ column densities along an arc of \HI\ surrounding the central FUV source. This measurements yields a total hydrogen volume density of $n = 70\ \pccm$. As in M\,81 and M\,83, our method points to the presence of GMCs in the outer regions of these galaxies.

\subsection{Combined plots}
\label{sec:comparison}

The measurements of the regions with candidate PDRs presented here are consolidated in Figure \ref{fig:consolidated}. The results at full resolution are indicated with black crosses. The gray boxes in the background are the results at reduced resolution for reference. The filled boxes correspond to the regions investigated at full resolution. Note that one of the regions, BCLMP 0288, is not included since we only have a larger-scale measurement there (see Section \ref{sec:full}).

The \HI\ plot is a sampling of \HI\ peaks that decline with galactocentric radius. Close to the centre of M\,33 the values are high enough to raise worries about the \HI\ being optically thick. The values of \Gnaught\ decline as well and generally span a range of 0.1 to 10. The total hydrogen volume densities are remarkably constant in range and values, ranging from approximately 10 to 300 \pccm. The values in the region closest to the centre, CPSDP 0087g, are lower. We noted previously that this region has a particularly high dust content, depressing the densities that we obtain. The high dust content may also indicate heavy internal extinction for which we did not correct. 

As an additional indication of the likelihood of the \HI\ patches being produced by photodissociation, we looked at the source contrast $G/G_{bg}$. The same results are plotted without the reduced resolution results in the background. The open circles in Figure \ref{fig:contrast} indicate a corresponding source contrast of less than 0.5. These \HI\ patches generally correspond to a lower value of $n$, and a lower value of \Gnaught. Finally, plotting \Gnaught\ and $n$ does not show a clear correlation, although if an actual correlation were present, it would be more pronounced without the CPSDP 0087g sources. 

\section{Discussion and Conclusions}
\label{sec:conclusions}

\subsection{Resolution effects}
\label{sec:reseffects}

Despite the analysis at different resolutions, the resulting cloud densities do not differ from each other, as they are in the same range. This is shown in Figure \ref{fig:consolidated}, where the densities are shown for those regions that are included at both resolutions. The full resolution data offers the advantage of (partially) resolving the parent GMCs connected to the OB associations with the opportunity to study the behavior of the \HI\ in more detail \citep{2009ApJ...700..545H}. On the other hand, there is more confusion in the distribution of the \HI, where the spatial distribution at the reduced resolution makes it easier to identify the \HI\ patches. With the reduced resolution data presented here, we were able to study the volume Schmidt Law of star formation \citet{2009ApJ...700..545H}.

Either way, the resulting cloud densities do not differ significantly, and selecting more regions for full resolution study is not expected to lead to more insight at this time.

\subsection{Performance of the PDR method across M\,33, M\,81, M\,83}

The M\,81 and M\,83 results were presented in \citet{2008ApJ...673..798H,2008A&A...489..533H}. Both galaxies show no clear drop in total hydrogen volume densities out to larger galactocentric radii. In the case of M\,83 we were able to identify candidate PDRs out to relatively large radii. In the view of photodissociated atomic hydrogen, there is no reason to assume that individual GMC densities change further away from the galactic centre. In fact, \citet{2010ApJ...720L..31B} confirm that far-UV and \HI\ emission continue to be correlated spatially out to 3 times $R_{25}$ even. At these distances we would expect PDRs and photodissociated \HI\ to occur just as they would more towards the galactic centre.

We generated boxplots for M\,81 and reproduced boxplots for M\,83 for comparison to the M\,33 results at equivalent resolution. These are shown in Figure \ref{fig:boxplots}. In a box-and-whisker plot, or boxplot, the boxes span the lower to upper quartiles of the data (50\% of the datapoints combined), with a line indicating the median value. The so-called 'whiskers' show the range of the data (1.5 times the inner quartile range of points), with outliers plotted individually (+).

The M\,81 results were recalculated to match certain parameters used in M\,83 and M\,33 for consistency, like the solar metallicty and the foreground extinction. A solar metallicity of 8.68 was adopted \citep{2001ApJ...556L..63A}, and the M\,81 foreground extinction was taken to be 0.63 mag, where 1.37 mag was used in \citet{2008ApJ...673..798H}. As a result of this, all total hydrogen volume densities derived for M\,81 used here are a factor 2 lower than calculated previously.

Finally, the results presented and summarized here are in many ways similar to those of M\,101 presented by \citet{2000ApJ...538..608S}, but cannot be compared directly to our results. They found total hydrogen volume densities in the range of 30 to 1000 \pccm. They used UIT data, while newer \textit{GALEX} data has better sensitivity and the UV flux normalization can be expected to be slightly different due to the respective UIT and \textit{GALEX} filter function. The method that was applied to M\,101 used a different way of measuring the \HI\ column density, since averages of concentric ellipses around the central UV sources were used instead of individual \HI\ patch measurements. While averaging the column densities is expected to yield a lower overall column densities, they did not subtract a local or global \HI\ background level. These two differences may balance each other out. Another difference is their assumptions about an internal extinction correction, which we have chosen not to apply due to geometrical considerations \citep{2008ApJ...673..798H}.

The first observation that can be made, is that there seems to be no fundamental difference between the three galaxies. In all cases we find a range of values that does not change significantly with galactocentric radius. The lowest cloud densities that we find should not constitute an actual physical limit, since these values are either due to a large separation $\rho_{\HI}$ (where it becomes questionable whether the \HI\ is really connected to the central UV source), or possibly pushing the method to the limits of its applicability. 
The largest cloud densities are a better indication, since the responsible \HI\ patches are generally closer to the central UV source and their connection is therefore more convincing. The largest densities in M\,81 are lower than those of M\,33 and M\,83. We already speculated in \citet{2008ApJ...673..798H} that this is consistent with the fainter levels of CO emission in M\,81. In the case of M\,83 they appear to be lower outside the optical disc, but this feature does not show in M\,33.

\subsection{Conclusions}

In \citet{2009ApJ...700..545H} we presented our first findings of total hydrogen volume densities in candidate PDRs in M\,33 using the PDR method. We used reduced resolution results to recover the volume Schmidt law of star formation in M\,33 in \citet{2010ApJ...719.1244H}. In this paper we present the complete results of applying the PDR method to M\,33. We summarize our findings as follows.

\renewcommand{\labelenumi}{\arabic{enumi}.}
\begin{enumerate}
	\item We presented total hydrogen volume densities of candidate PDRs in M\,33 obtained with the PDR method, and compared them with those found in M\,81 and M\,83. These PDRs occur throughout the optical discs of these galaxies and beyond. UV emission can be matched to the presence of atomic hydrogen out to the detection limit of 21-cm images. 
	\item We find total hydrogen volume densities of candidate PDRs in M\,33 ranging from 1 to 500 cm$^{-3}$.
	\item The densities in candidate PDRs of M\,81 and M\,83 are in the same range, although the spread in the M\,83 results is higher, indicating the presence of higher density GMCs in M\,83. The M\,33 densities go up to the same level as in M\,83, although the spread is narrower. The PDR method therefore yields similar results nearby (M\,33) as well as further away (M\,83).
	\item The cloud densities presented here at two different resolutions show that the PDR method yields consistent results that are not significantly affected by resolution effects.
\end{enumerate}

\section*{Acknowledgments}
	JSH acknowledges support by a CRAQ postdoctoral fellowship. Significant parts of this work were carried out at the Kapteyn Astronomical Institute and the Space Telescope Science Institute funded by the STScI Director's Discretionary Research Fund. Data analysis for this work was done primarily using the Groningen Image Processing System GIPSY \citep{1973A&A....27...77E,1992ASPC...25..131V,2001ASPC..238..358V}. We thank the anonymous referee for useful comments that improved this paper.

\bibliographystyle{mn2e} 

\label{lastpage}

\clearpage

Online materials

\clearpage



\begin{deluxetable}{crrrrr}
\centering
\tablecolumns{6}
\tablewidth{0pt}
\tablehead{
  \colhead{} &
  \colhead{} &
  \colhead{} &
  \colhead{Radius} &
  \colhead{$F_{FUV}$\tablenotemark{a}}  &
  \colhead{Aperture} \\
  \colhead{Source no.} &
  \colhead{R.A. (2000)} &
  \colhead{DEC. (2000)} &
  \colhead{(kpc)}  &
  \colhead{} &
  \colhead{(arcsec)} \\
}
\startdata
\phn1& 1 33 51.371 &   30 38 52.01        &  0.23 & 305.35 &  96 \\ 
\phn2& 1 33 51.052 &   30 41  \phn1.18    &  0.40 &  26.08 &  36 \\
\phn3& 1 33 40.271 &   30 35 34.72        &  1.16 & 137.22 &  72 \\
\phn4& 1 34 \phn1.540 &   30 37 43.83     &  1.27 &  71.03 &  60 \\
\phn5& 1 34 \phn8.658 &   30 39 15.42     &  1.63 &  96.25 &  72 \\
\phn6& 1 33 34.135 &   30 33 53.24        &  1.71 & 151.01 &  72 \\
\phn7& 1 33 33.401 &   30 41 37.88        &  1.88 & 188.70 & 156 \\
\phn8& 1 33 58.709 &   30 34 14.98        &  1.92 & 518.50 & 156 \\
\phn9& 1 34 10.340 &   30 46 39.52        &  2.06 & 195.81 & 108 \\
10& 1 34 19.129 &   30 44 37.26        &  2.35 &  20.08 &  48 \\
11& 1 33 49.897 &   30 29 29.97        &  2.84 &   8.61 &  36 \\
12& 1 33 34.036 &   30 47 26.00        &  3.16 &  13.63 &  48 \\
13& 1 34 30.170 &   30 40 45.20        &  3.40 &  39.95 &  48 \\
14& 1 34 32.634 &   30 47 \phn4.24     &  3.47 & 533.69 &  96 \\
15& 1 33 56.051 &   30 27 29.73        &  3.66 &  17.29 &  48 \\
16& 1 34 26.824 &   30 53 \phn1.06     &  3.87 &  30.87 &  60 \\
17& 1 34 39.056 &   30 44 \phn2.28     &  3.97 &  47.50 &  60 \\
18& 1 34 39.500 &   30 50 23.27        &  4.19 &   9.20 &  60 \\
19& 1 33 \phn7.349 &   30 42 49.34     &  4.36 &  38.62 &  60 \\
20& 1 33 11.217 &   30 23 21.93        &  4.55 & 121.16 & 108 \\
21& 1 34 30.256 &   30 57 16.33        &  4.84 &  42.43 & 108 \\
22& 1 33 38.929 &   30 20 52.34        &  4.99 & 254.41 & 120 \\
23& 1 34 50.221 &   30 54 47.24        &  5.32 &  36.57 &  84 \\
24& 1 34 33.331 &   31 \phn0 25.62     &  5.60 &  34.03 & 108 \\
25& 1 33 56.608 &   31 \phn0 17.50     &  5.71 &  26.68 & 108 \\
26& 1 34 48.391 &   30 58 39.46        &  5.77 &   6.93 &  96 \\
27& 1 32 42.053 &   30 24 57.91        &  5.90 & 104.96 &  96 \\
28& 1 34 \phn9.835 &   30 21 50.99     &  5.95 &  16.44 &  84 \\
29& 1 34 28.980 &   30 25 \phn8.31     &  6.33 &   3.97 &  60 \\
30& 1 34 41.170 &   30 28 \phn7.71     &  6.55 &  13.20 &  48 \\
31& 1 32 37.749 &   30 40 \phn8.36     &  6.62 &  30.09 &  84 \\
32& 1 34 40.370 &   31 \phn4 33.49     &  6.68 &   6.86 &  84 \\
33& 1 35 12.068 &   30 46 31.11        &  6.72 &   4.99 &  48 \\
34& 1 34 \phn2.713 & 31 \phn5 \phn1.82 &  6.88 &   8.76 &  72 \\
35& 1 32 54.210 &   30 50 43.38        &  6.99 &   8.58 &  72 \\
36& 1 32 43.372 &   30 15 28.60        &  7.09 &  19.50 & 180 \\
37& 1 34 42.536 &   30 25 32.17        &  7.25 &   3.86 &  48 \\
38& 1 33 46.842 &   30 13 27.45        &  7.30 &   5.01 &  72 \\
39& 1 34 25.430 &   30 19 46.10        &  7.46 &   4.12 &  72 \\
40& 1 34 32.897 &   31 \phn9 12.91     &  7.72 &   3.12 &  60 \\
41& 1 33 33.651 &   30 \phn8 43.02     &  8.29 &   2.09 &  72 \\
42& 1 34 43.114 &   31 12 32.42        &  8.63 &   2.69 &  60 \\
\enddata
\tablenotetext{a}{\ensuremath{10^{-15}\ \rm{ergs\ cm^{-2}\ s^{-1}\ \mbox{\AA}^{-1}}}}
\tablecaption{\label{tab:sources_full} Locations and FUV fluxes of candidate PDRs}
\end{deluxetable} 
\clearpage

\begin{deluxetable}{ccccccc}
\centering
\tablecolumns{7}
\tablewidth{0pt}
\tablehead{
  \colhead{Source no.} &
  \colhead{$\rho_{\HI}$ (pc)} &
  \colhead{$N_{\HI}$ (\HIunits)} &
  \colhead{\Gnaught} & 
  \colhead{$G/G_{bg}$} &
  \colhead{$n$ ($\rm{cm}^{-3}$)} &
  \colhead{\tablenotemark{a}}\\ 
}
\startdata
1a...... &  121   &  1.91    &  6.53    &  8.01      &  487        &  0.60  \\
1b...... &  161   &  2.02    &  3.67    &  4.51      &  249        &  0.59  \\
1c...... &  403   &  2.17    &  0.59    &  0.72      &  35         &  0.58  \\
2a...... &  40    &  2.11    &  5.02    &  8.18      &  323        &  0.97  \\
2b...... &  202   &  2.37    &  0.20    &  0.33      &  10         &  0.61  \\
2c...... &  363   &  1.57    &  0.06    &  0.10      &  6          &  0.52  \\
2d...... &  363   &  2.17    &  0.06    &  0.10      &  4          &  0.58  \\
3a...... &  242   &  1.82    &  0.73    &  0.81      &  67         &  0.54  \\
3b...... &  242   &  1.51    &  0.73    &  0.81      &  89         &  0.51  \\
3c...... &  403   &  1.99    &  0.26    &  0.29      &  21         &  0.55  \\
4a...... &  202   &  1.74    &  0.55    &  0.69      &  54         &  0.54  \\
4b...... &  202   &  2.11    &  0.55    &  0.69      &  40         &  0.57  \\
4c...... &  242   &  2.85    &  0.38    &  0.48      &  16         &  0.64  \\
4d...... &  403   &  2.17    &  0.14    &  0.17      &  9          &  0.56  \\
5a...... &  161   &  1.94    &  1.16    &  1.84      &  101        &  0.57  \\
5b...... &  161   &  2.51    &  1.16    &  1.84      &  65         &  0.62  \\
5c...... &  323   &  2.31    &  0.29    &  0.46      &  19         &  0.57  \\
5d...... &  403   &  2.85    &  0.19    &  0.29      &  8          &  0.62  \\
6a...... &  161   &  1.25    &  1.82    &  3.21      &  309        &  0.51  \\
6b...... &  282   &  1.19    &  0.59    &  1.05      &  108        &  0.48  \\
6c...... &  363   &  1.51    &  0.36    &  0.64      &  47         &  0.50  \\
6d...... &  403   &  1.31    &  0.29    &  0.51      &  46         &  0.48  \\
6e...... &  403   &  3.88    &  0.29    &  0.51      &  7          &  0.73  \\
7a...... &  164   &  2.39    &  2.19    &  6.06      &  140        &  0.60  \\
7b...... &  205   &  2.39    &  1.40    &  3.88      &  90         &  0.59  \\
8a...... &  161   &  2.31    &  6.23    &  11.83     &  424        &  0.59  \\
8b...... &  202   &  3.14    &  3.99    &  7.57      &  152        &  0.66  \\
8c...... &  323   &  2.19    &  1.56    &  2.96      &  116        &  0.56  \\
8d...... &  403   &  2.42    &  1.00    &  1.89      &  62         &  0.58  \\
9a...... &  161   &  2.17    &  2.35    &  6.08      &  182        &  0.58  \\
9b...... &  202   &  2.05    &  1.51    &  3.89      &  128        &  0.56  \\
9c...... &  202   &  1.65    &  1.51    &  3.89      &  179        &  0.53  \\
9d...... &  363   &  2.02    &  0.47    &  1.20      &  40         &  0.54  \\
10a..... &  81    &  0.85    &  0.97    &  2.76      &  291        &  0.59  \\
10b..... &  202   &  0.74    &  0.15    &  0.44      &  55         &  0.46  \\
10c..... &  282   &  1.71    &  0.08    &  0.23      &  9          &  0.52  \\
11a..... &  81    &  1.94    &  0.41    &  2.31      &  42         &  0.66  \\
11b..... &  161   &  1.62    &  0.10    &  0.58      &  14         &  0.54  \\
11c..... &  242   &  2.08    &  0.05    &  0.26      &  4          &  0.56  \\
11d..... &  323   &  1.65    &  0.03    &  0.14      &  3          &  0.52  \\
11e..... &  363   &  1.57    &  0.02    &  0.11      &  3          &  0.51  \\
11f..... &  363   &  1.79    &  0.02    &  0.11      &  2          &  0.53  \\
12a..... &  161   &  1.94    &  0.16    &  0.81      &  18         &  0.56  \\
12b..... &  202   &  2.17    &  0.10    &  0.52      &  9          &  0.56  \\
12c..... &  242   &  1.68    &  0.07    &  0.36      &  10         &  0.52  \\
12d..... &  363   &  2.05    &  0.03    &  0.16      &  3          &  0.54  \\
13a..... &  202   &  1.42    &  0.31    &  0.88      &  53         &  0.50  \\
13b..... &  242   &  1.54    &  0.21    &  0.61      &  33         &  0.50  \\
13c..... &  242   &  1.85    &  0.21    &  0.61      &  25         &  0.52  \\
13d..... &  242   &  1.59    &  0.21    &  0.61      &  32         &  0.50  \\
13e..... &  363   &  1.25    &  0.09    &  0.27      &  20         &  0.47  \\
13f..... &  363   &  2.22    &  0.09    &  0.27      &  8          &  0.54  \\
14a..... &  161   &  3.77    &  6.42    &  12.80     &  214        &  0.70  \\
14b..... &  202   &  3.42    &  4.11    &  8.19      &  168        &  0.66  \\
14c..... &  282   &  3.68    &  2.10    &  4.18      &  74         &  0.67  \\
14d..... &  282   &  3.25    &  2.10    &  4.18      &  95         &  0.63  \\
15a..... &  81    &  2.17    &  0.83    &  3.63      &  79         &  0.66  \\
15b..... &  363   &  1.42    &  0.04    &  0.18      &  7          &  0.48  \\
15c..... &  363   &  1.68    &  0.04    &  0.18      &  6          &  0.50  \\
16a..... &  40    &  1.14    &  5.94    &  45.90     &  1460       &  0.91  \\
16b..... &  161   &  1.42    &  0.37    &  2.87      &  68         &  0.51  \\
16c..... &  323   &  1.39    &  0.09    &  0.72      &  18         &  0.48  \\
16d..... &  323   &  1.05    &  0.09    &  0.72      &  25         &  0.46  \\
16e..... &  323   &  1.77    &  0.09    &  0.72      &  12         &  0.51  \\
16f..... &  363   &  1.48    &  0.07    &  0.57      &  13         &  0.48  \\
17a..... &  161   &  2.17    &  0.57    &  1.46      &  57         &  0.56  \\
18a..... &  40    &  1.19    &  1.77    &  13.71     &  426        &  0.92  \\
18b..... &  81    &  1.17    &  0.44    &  3.43      &  109        &  0.61  \\
18c..... &  202   &  1.34    &  0.07    &  0.55      &  15         &  0.50  \\
18d..... &  242   &  1.22    &  0.05    &  0.38      &  11         &  0.49  \\
18e..... &  282   &  1.14    &  0.04    &  0.28      &  9          &  0.48  \\
18f..... &  363   &  1.31    &  0.02    &  0.17      &  5          &  0.48  \\
18g..... &  403   &  1.48    &  0.02    &  0.14      &  3          &  0.49  \\
19a..... &  161   &  1.91    &  0.46    &  4.44      &  59         &  0.54  \\
19b..... &  242   &  2.02    &  0.21    &  1.98      &  24         &  0.53  \\
19c..... &  282   &  2.31    &  0.15    &  1.45      &  14         &  0.54  \\
20a..... &  161   &  2.34    &  1.46    &  11.50     &  139        &  0.57  \\
20b..... &  161   &  1.91    &  1.46    &  11.50     &  190        &  0.54  \\
20c..... &  242   &  1.91    &  0.65    &  5.11      &  84         &  0.52  \\
20d..... &  282   &  2.82    &  0.48    &  3.75      &  33         &  0.58  \\
20e..... &  363   &  2.17    &  0.29    &  2.27      &  31         &  0.53  \\
20f..... &  363   &  2.85    &  0.29    &  2.27      &  20         &  0.58  \\
21a..... &  81    &  3.11    &  2.04    &  13.85     &  123        &  0.71  \\
21b..... &  282   &  1.79    &  0.17    &  1.13      &  25         &  0.50  \\
21c..... &  403   &  3.57    &  0.08    &  0.55      &  4          &  0.63  \\
22a..... &  121   &  2.54    &  5.44    &  16.64     &  479        &  0.60  \\
22b..... &  161   &  2.59    &  3.06    &  9.36      &  261        &  0.58  \\
22c..... &  282   &  3.02    &  1.00    &  3.06      &  65         &  0.59  \\
22d..... &  323   &  3.39    &  0.76    &  2.34      &  40         &  0.61  \\
22e..... &  323   &  2.45    &  0.76    &  2.34      &  72         &  0.54  \\
23a..... &  81    &  2.14    &  1.76    &  23.08     &  212        &  0.65  \\
23b..... &  161   &  2.11    &  0.44    &  5.77      &  54         &  0.54  \\
23c..... &  202   &  2.08    &  0.28    &  3.69      &  36         &  0.53  \\
23d..... &  282   &  3.05    &  0.14    &  1.88      &  10         &  0.59  \\
24a..... &  121   &  1.48    &  0.73    &  11.97     &  154        &  0.53  \\
24b..... &  161   &  1.74    &  0.41    &  6.73      &  69         &  0.52  \\
24c..... &  202   &  1.94    &  0.26    &  4.31      &  38         &  0.52  \\
24d..... &  323   &  2.02    &  0.10    &  1.68      &  14         &  0.51  \\
24e..... &  323   &  1.59    &  0.10    &  1.68      &  20         &  0.48  \\
24f..... &  403   &  1.85    &  0.07    &  1.08      &  10         &  0.49  \\
25a..... &  81    &  1.82    &  1.28    &  11.12     &  207        &  0.63  \\
25b..... &  202   &  1.51    &  0.21    &  1.78      &  43         &  0.49  \\
25c..... &  202   &  1.48    &  0.21    &  1.78      &  44         &  0.49  \\
25d..... &  202   &  1.11    &  0.21    &  1.78      &  64         &  0.47  \\
25e..... &  363   &  1.59    &  0.06    &  0.55      &  12         &  0.48  \\
25f..... &  403   &  1.68    &  0.05    &  0.44      &  9          &  0.48  \\
25g..... &  403   &  1.51    &  0.05    &  0.44      &  11         &  0.47  \\
26a..... &  161   &  1.22    &  0.08    &  1.98      &  23         &  0.51  \\
26b..... &  282   &  1.22    &  0.03    &  0.65      &  8          &  0.48  \\
26c..... &  323   &  1.28    &  0.02    &  0.49      &  5          &  0.48  \\
26d..... &  323   &  1.42    &  0.02    &  0.49      &  5          &  0.49  \\
26e..... &  323   &  1.31    &  0.02    &  0.49      &  5          &  0.49  \\
26f..... &  403   &  1.39    &  0.01    &  0.32      &  3          &  0.49  \\
26g..... &  403   &  1.37    &  0.01    &  0.32      &  3          &  0.48  \\
27a..... &  161   &  2.79    &  1.26    &  10.62     &  107        &  0.58  \\
27b..... &  202   &  2.37    &  0.81    &  6.80      &  90         &  0.54  \\
27c..... &  242   &  2.31    &  0.56    &  4.72      &  65         &  0.53  \\
27d..... &  282   &  2.62    &  0.41    &  3.47      &  39         &  0.55  \\
27e..... &  323   &  2.28    &  0.32    &  2.65      &  37         &  0.52  \\
27f..... &  403   &  2.79    &  0.20    &  1.70      &  17         &  0.55  \\
28a..... &  121   &  1.74    &  0.35    &  4.11      &  62         &  0.55  \\
28b..... &  202   &  2.05    &  0.13    &  1.48      &  18         &  0.53  \\
28c..... &  242   &  1.79    &  0.09    &  1.03      &  15         &  0.50  \\
28d..... &  282   &  2.22    &  0.06    &  0.75      &  8          &  0.52  \\
28e..... &  323   &  2.28    &  0.05    &  0.58      &  6          &  0.53  \\
28f..... &  323   &  1.99    &  0.05    &  0.58      &  7          &  0.51  \\
29a..... &  161   &  1.65    &  0.05    &  0.94      &  9          &  0.57  \\
29b..... &  161   &  1.62    &  0.05    &  0.94      &  10         &  0.57  \\
29c..... &  403   &  2.59    &  0.01    &  0.15      &  1          &  0.59  \\
29d..... &  403   &  2.02    &  0.01    &  0.15      &  1          &  0.56  \\
30a..... &  121   &  1.99    &  0.28    &  2.96      &  44         &  0.56  \\
30b..... &  121   &  1.59    &  0.28    &  2.96      &  61         &  0.54  \\
30c..... &  282   &  2.42    &  0.05    &  0.54      &  6          &  0.53  \\
30d..... &  363   &  2.05    &  0.03    &  0.33      &  5          &  0.51  \\
31a..... &  81    &  2.08    &  1.45    &  12.90     &  214        &  0.63  \\
31b..... &  161   &  1.74    &  0.36    &  3.22      &  69         &  0.51  \\
31c..... &  242   &  1.54    &  0.16    &  1.43      &  36         &  0.48  \\
31d..... &  242   &  1.79    &  0.16    &  1.43      &  30         &  0.50  \\
31e..... &  363   &  1.71    &  0.07    &  0.64      &  14         &  0.48  \\
31f..... &  363   &  2.08    &  0.07    &  0.64      &  11         &  0.50  \\
31g..... &  403   &  2.91    &  0.06    &  0.52      &  5          &  0.55  \\
32a..... &  161   &  1.54    &  0.08    &  1.80      &  19         &  0.52  \\
32b..... &  202   &  1.48    &  0.05    &  1.15      &  13         &  0.51  \\
32c..... &  323   &  1.28    &  0.02    &  0.45      &  6          &  0.48  \\
32d..... &  363   &  1.77    &  0.02    &  0.35      &  3          &  0.50  \\
32e..... &  363   &  1.97    &  0.02    &  0.35      &  3          &  0.52  \\
32f..... &  403   &  1.48    &  0.01    &  0.29      &  3          &  0.49  \\
33a..... &  81    &  1.88    &  0.24    &  3.59      &  42         &  0.65  \\
33b..... &  161   &  1.91    &  0.06    &  0.90      &  10         &  0.56  \\
33c..... &  202   &  1.82    &  0.04    &  0.57      &  7          &  0.54  \\
33d..... &  242   &  1.48    &  0.03    &  0.40      &  6          &  0.52  \\
33e..... &  363   &  2.17    &  0.01    &  0.18      &  2          &  0.54  \\
33f..... &  363   &  2.08    &  0.01    &  0.18      &  2          &  0.54  \\
33g..... &  403   &  1.48    &  0.01    &  0.14      &  2          &  0.51  \\
33h..... &  403   &  1.85    &  0.01    &  0.14      &  2          &  0.53  \\
34a..... &  161   &  1.82    &  0.11    &  1.38      &  20         &  0.53  \\
34b..... &  202   &  1.85    &  0.07    &  0.88      &  12         &  0.52  \\
34c..... &  242   &  1.74    &  0.05    &  0.61      &  9          &  0.50  \\
34d..... &  323   &  1.85    &  0.03    &  0.35      &  5          &  0.50  \\
34e..... &  403   &  1.45    &  0.02    &  0.22      &  4          &  0.48  \\
35a..... &  121   &  2.34    &  0.18    &  2.71      &  24         &  0.58  \\
35b..... &  161   &  1.77    &  0.10    &  1.52      &  20         &  0.53  \\
35c..... &  323   &  1.94    &  0.03    &  0.38      &  4          &  0.51  \\
35d..... &  323   &  2.17    &  0.03    &  0.38      &  4          &  0.52  \\
36a..... &  121   &  1.48    &  0.42    &  11.38     &  105        &  0.53  \\
36b..... &  161   &  1.25    &  0.23    &  6.40      &  73         &  0.49  \\
36c..... &  242   &  2.05    &  0.10    &  2.84      &  17         &  0.51  \\
36d..... &  282   &  1.19    &  0.08    &  2.09      &  25         &  0.46  \\
36e..... &  323   &  2.05    &  0.06    &  1.60      &  9          &  0.50  \\
36f..... &  323   &  1.88    &  0.06    &  1.60      &  11         &  0.49  \\
36g..... &  363   &  1.57    &  0.05    &  1.26      &  11         &  0.47  \\
37a..... &  40    &  1.65    &  0.74    &  13.33     &  164        &  0.95  \\
37b..... &  242   &  2.17    &  0.02    &  0.37      &  3          &  0.57  \\
37c..... &  282   &  1.82    &  0.02    &  0.27      &  3          &  0.55  \\
37d..... &  403   &  1.97    &  0.01    &  0.13      &  1          &  0.55  \\
38a..... &  161   &  1.31    &  0.06    &  1.47      &  18         &  0.53  \\
38b..... &  242   &  1.22    &  0.03    &  0.65      &  9          &  0.50  \\
38c..... &  282   &  1.28    &  0.02    &  0.48      &  6          &  0.50  \\
38d..... &  323   &  0.85    &  0.02    &  0.37      &  8          &  0.48  \\
39a..... &  81    &  0.99    &  0.20    &  5.39      &  86         &  0.63  \\
39b..... &  121   &  0.99    &  0.09    &  2.40      &  38         &  0.56  \\
39c..... &  242   &  0.88    &  0.02    &  0.60      &  11         &  0.51  \\
39d..... &  242   &  0.91    &  0.02    &  0.60      &  11         &  0.51  \\
39e..... &  363   &  0.97    &  0.01    &  0.27      &  4          &  0.50  \\
39f..... &  363   &  1.14    &  0.01    &  0.27      &  4          &  0.51  \\
39g..... &  403   &  1.57    &  0.01    &  0.22      &  2          &  0.52  \\
40a..... &  202   &  1.82    &  0.02    &  0.63      &  5          &  0.59  \\
40b..... &  282   &  1.45    &  0.01    &  0.32      &  3          &  0.56  \\
40c..... &  323   &  1.54    &  0.01    &  0.25      &  2          &  0.57  \\
40d..... &  323   &  1.25    &  0.01    &  0.25      &  3          &  0.55  \\
40e..... &  403   &  1.59    &  0.01    &  0.16      &  1          &  0.56  \\
40f..... &  403   &  1.25    &  0.01    &  0.16      &  2          &  0.55  \\
40g..... &  403   &  1.28    &  0.01    &  0.16      &  2          &  0.55  \\
41a..... &  81    &  0.95    &  0.10    &  4.55      &  50         &  0.75  \\
41b..... &  121   &  0.64    &  0.04    &  2.02      &  35         &  0.69  \\
41c..... &  161   &  0.66    &  0.03    &  1.14      &  19         &  0.66  \\
41d..... &  282   &  0.64    &  0.01    &  0.37      &  6          &  0.64  \\
41e..... &  282   &  0.55    &  0.01    &  0.37      &  8          &  0.64  \\
41f..... &  282   &  0.66    &  0.01    &  0.37      &  6          &  0.64  \\
41g..... &  323   &  1.06    &  0.01    &  0.28      &  3          &  0.65  \\
41h..... &  403   &  0.98    & 0.004    &  0.18      &  2          &  0.65  \\
42a..... &  40    &  1.58    &  0.52    &  22.60     &  142        &  0.99  \\
42b..... &  121   &  1.58    &  0.06    &  2.51      &  16         &  0.64  \\
42c..... &  161   &  1.58    &  0.03    &  1.41      &  9          &  0.62  \\
42d..... &  282   &  1.61    &  0.01    &  0.46      &  3          &  0.60  \\
42e..... &  323   &  1.35    &  0.01    &  0.35      &  3          &  0.59  \\
42f..... &  363   &  1.46    &  0.01    &  0.28      &  2          &  0.59  \\
42g..... &  363   &  1.12    &  0.01    &  0.28      &  3          &  0.58  \\
\enddata
\tablenotetext{a}{Fractional error}
\tablecaption{\label{tab:results_full} Candidate PDR results}

\end{deluxetable}

\clearpage


\begin{figure*}
  \centering
  \begin{tabular*}{\textwidth}{r p{1.5cm} l} 
    \includegraphics[width=0.35\textwidth]{fig3_01}
    & \centering{\scriptsize{\hspace{0.5cm}1\hspace{0.5cm}2}} &
    \includegraphics[width=0.35\textwidth]{fig3_02} \\
    \includegraphics[width=0.35\textwidth]{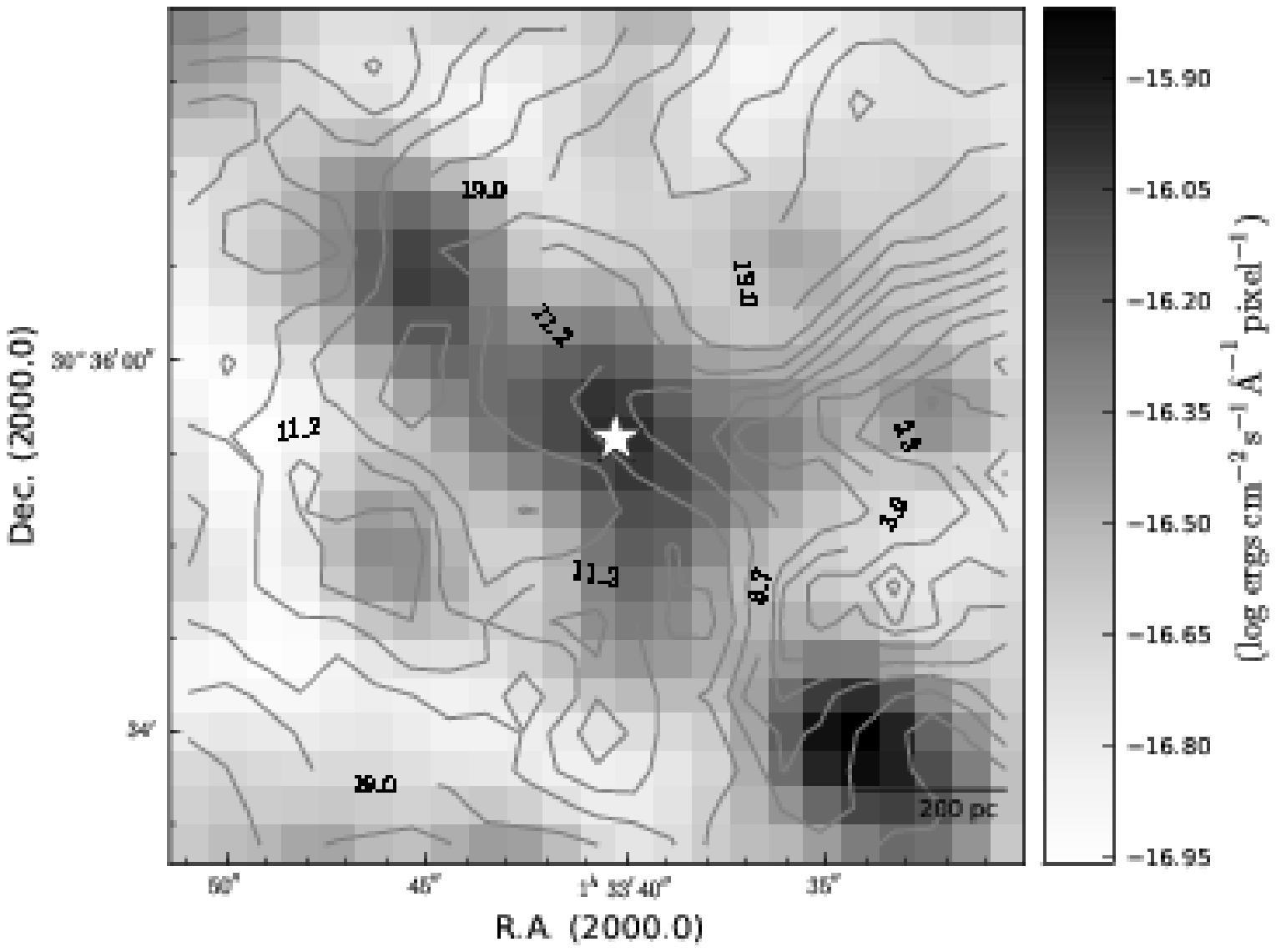}
    & \centering{\scriptsize{\hspace{0.5cm}3\hspace{0.5cm}4}} &
    \includegraphics[width=0.35\textwidth]{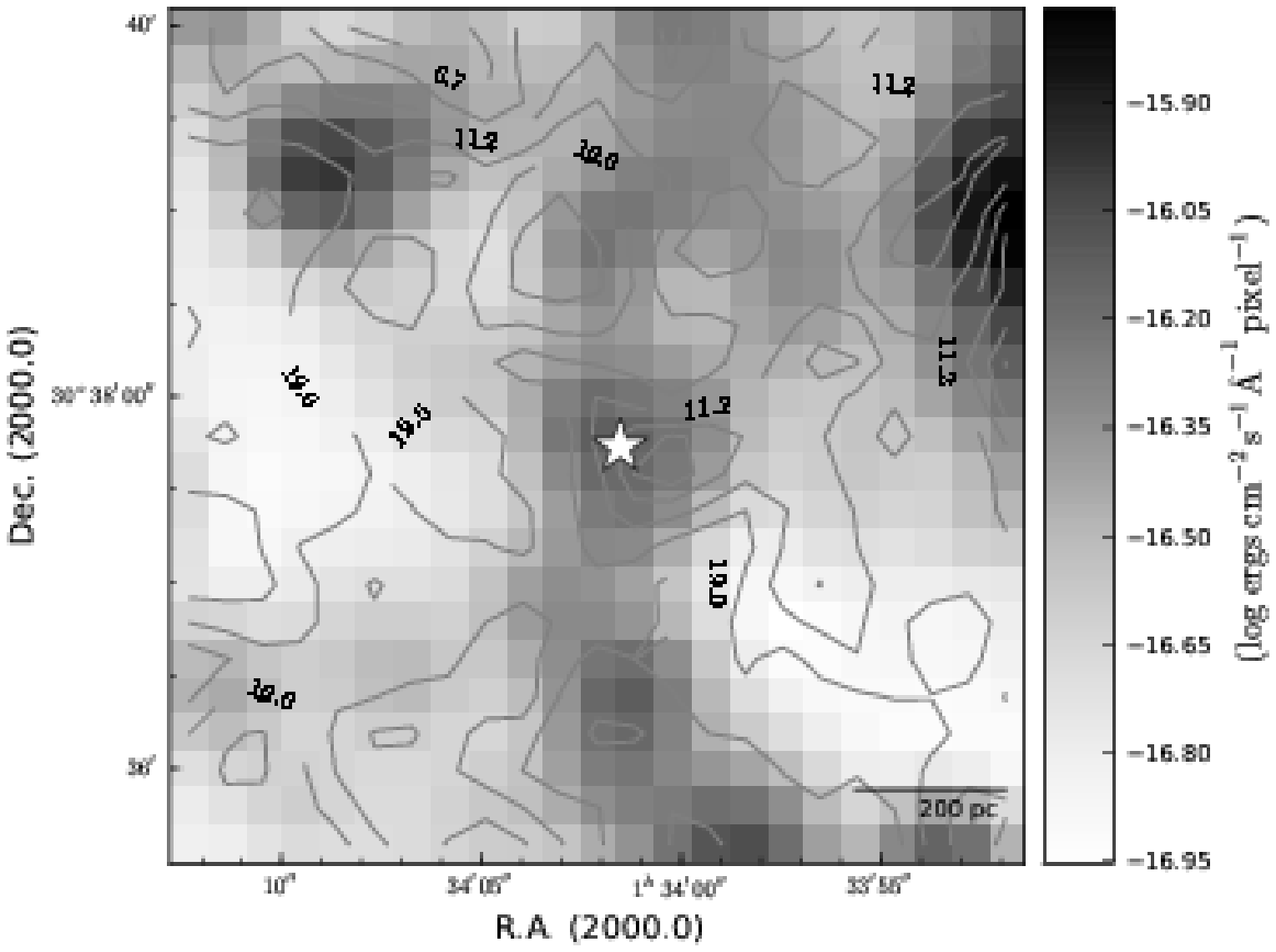} \\
    \includegraphics[width=0.35\textwidth]{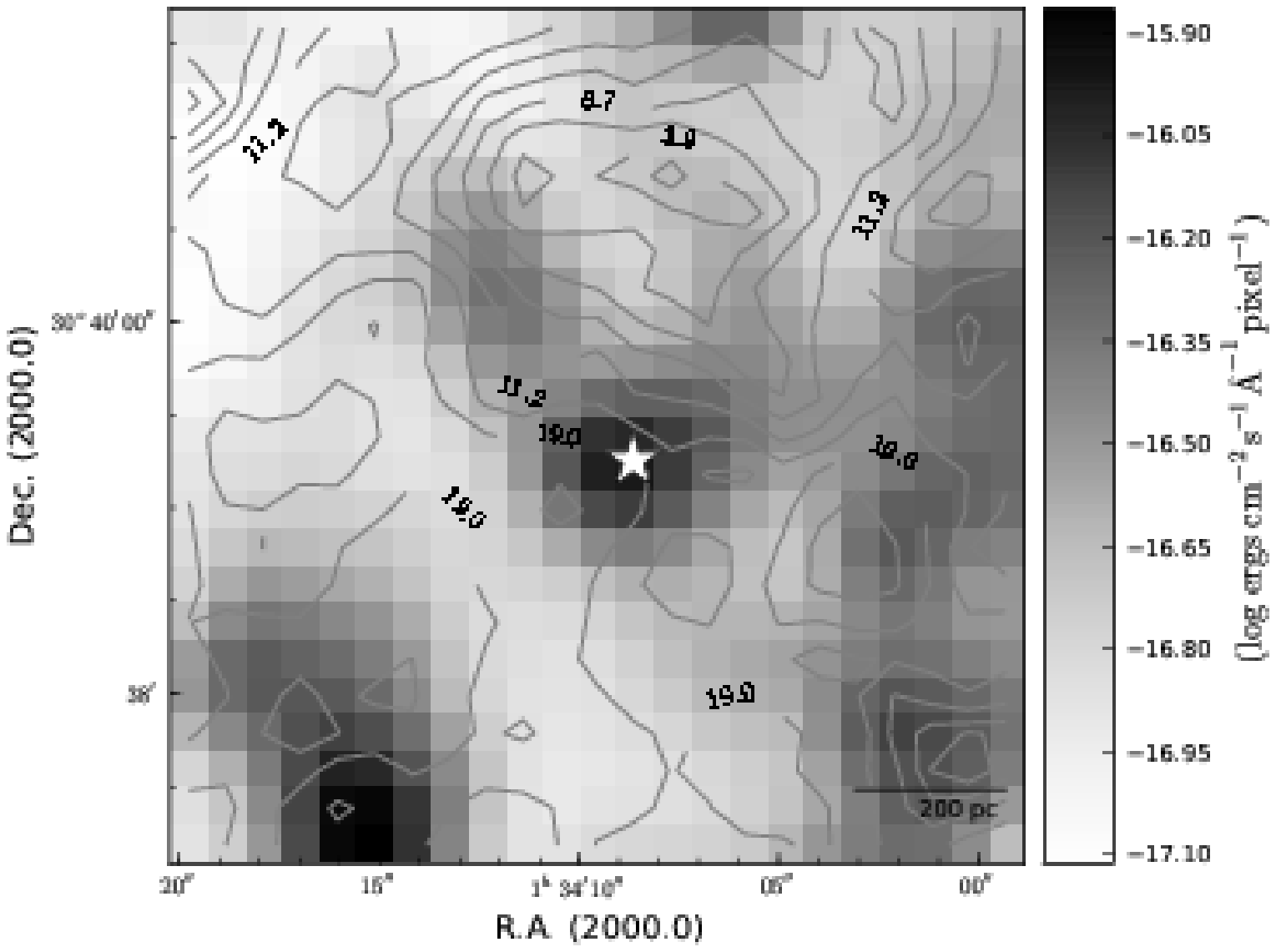}
    & \centering{\scriptsize{\hspace{0.5cm}5\hspace{0.5cm}6}} &
    \includegraphics[width=0.35\textwidth]{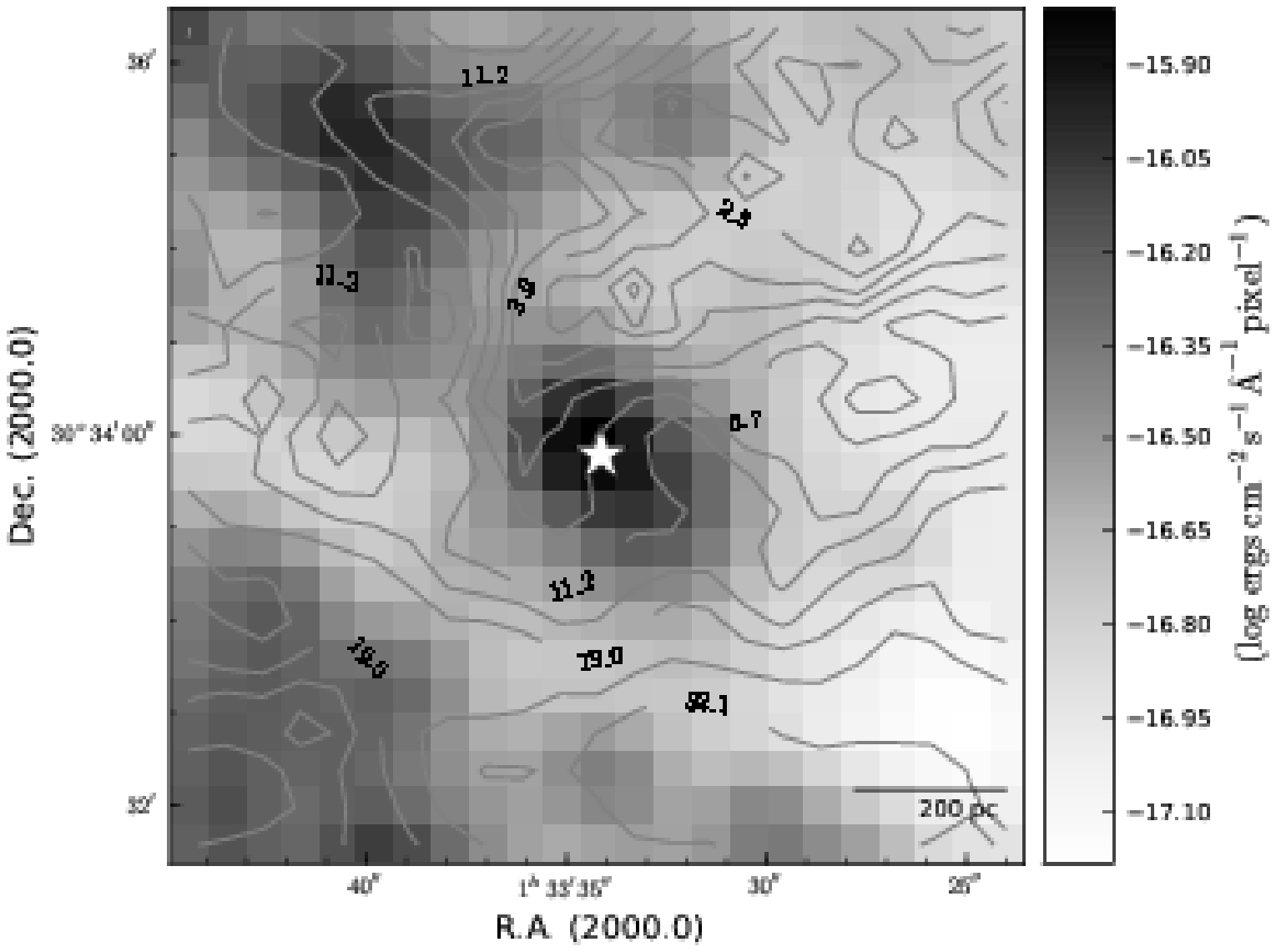} \\
    \includegraphics[width=0.35\textwidth]{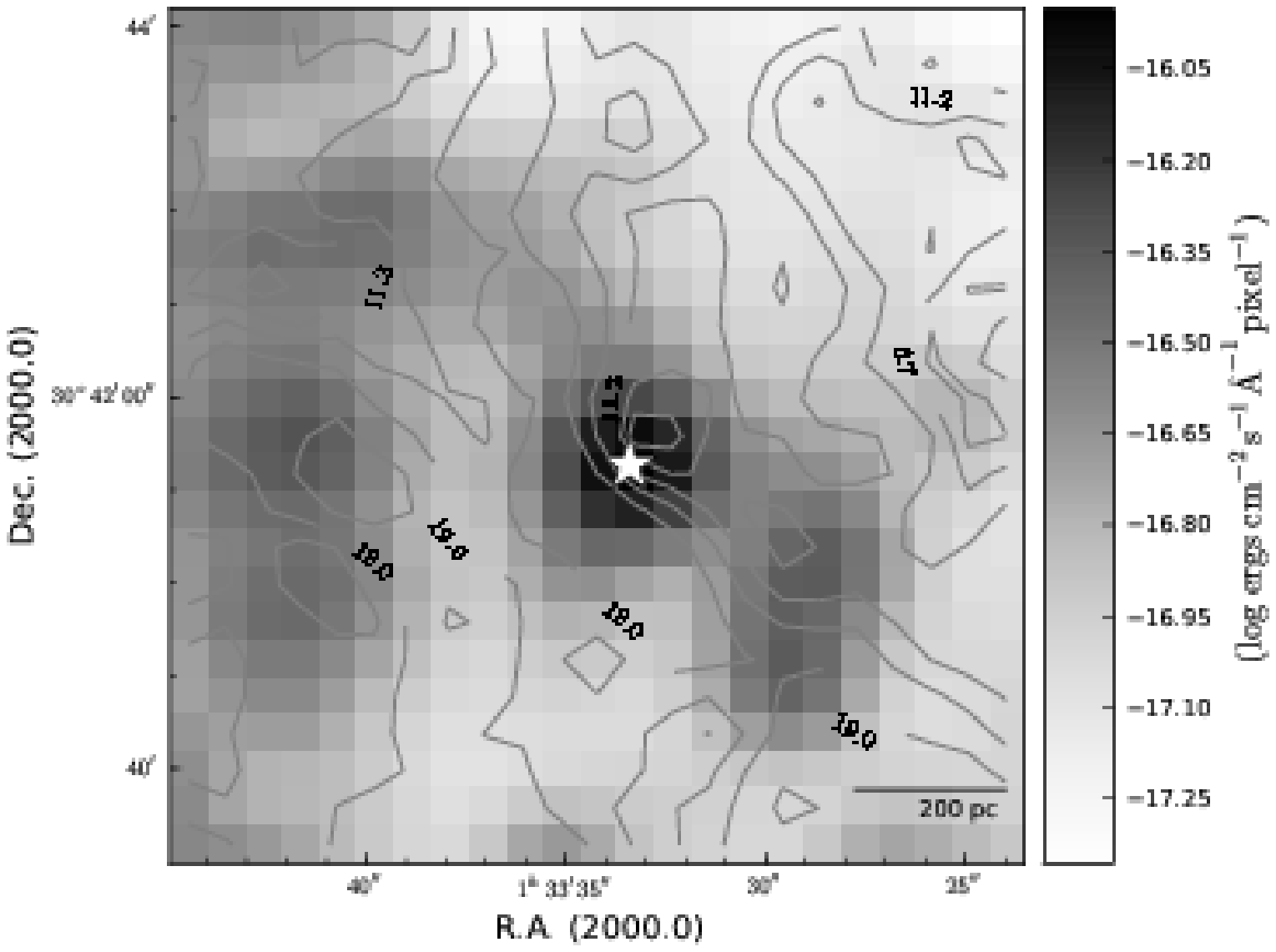}
    & \centering{\scriptsize{\hspace{0.5cm}7\hspace{0.5cm}8}} &
    \includegraphics[width=0.35\textwidth]{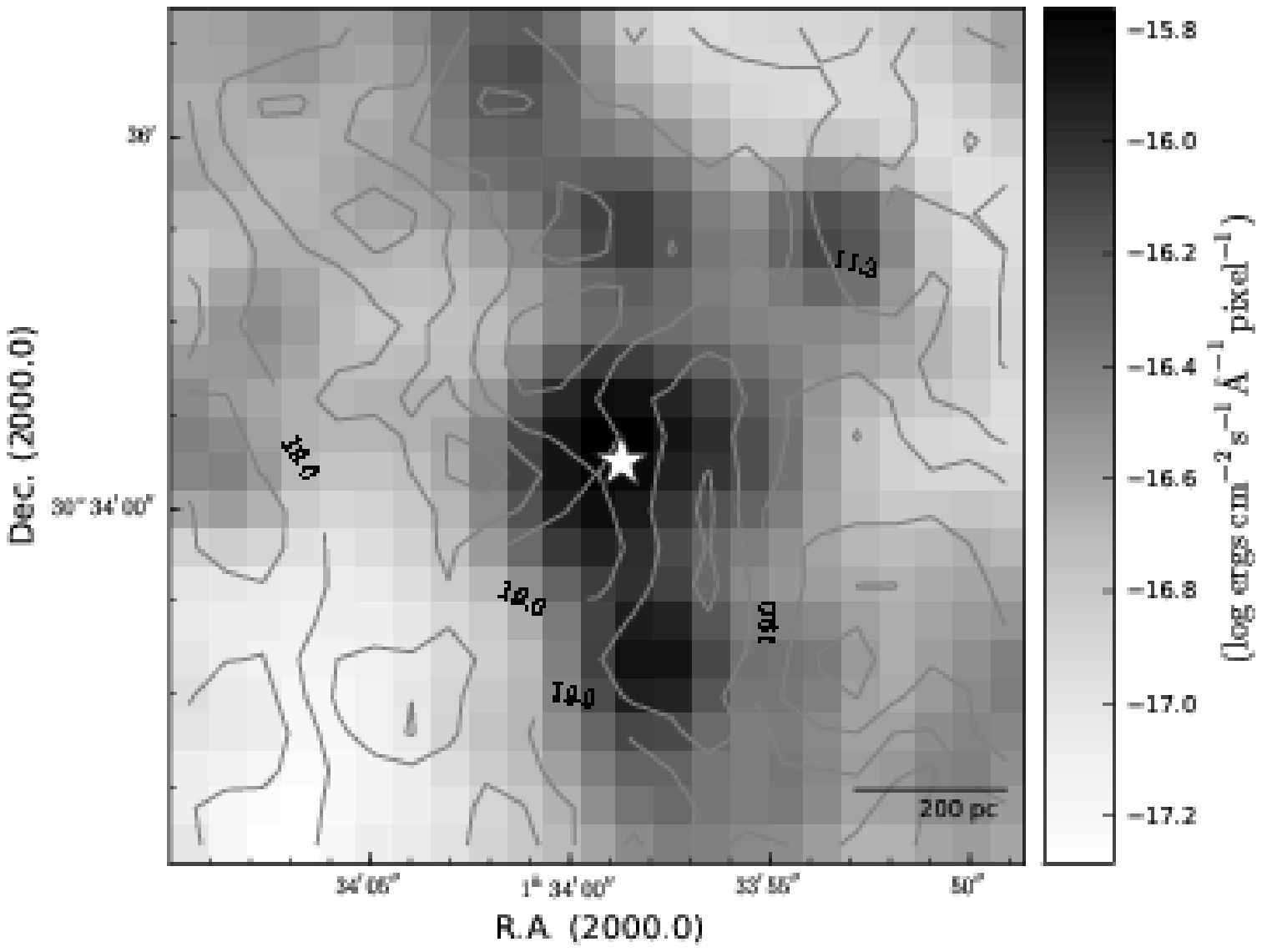} \\
  \end{tabular*}
	\caption[]{\label{fig:A_m33_1} M33 sources 1 -- 8, as described in Section \ref{sec:equiv} (from left to right, top to bottom). The \HI\ column density contours are plotted against a background of the FUV flux in gray scale. The fitted position of the central UV complex is marked by a star. The integrated FUV fluxes can be found in Table \ref{tab:sources_full}.}
\end{figure*}

\begin{figure*}
  \centering
  \begin{tabular*}{\textwidth}{l p{1.5cm} l} 
    \includegraphics[width=0.35\textwidth]{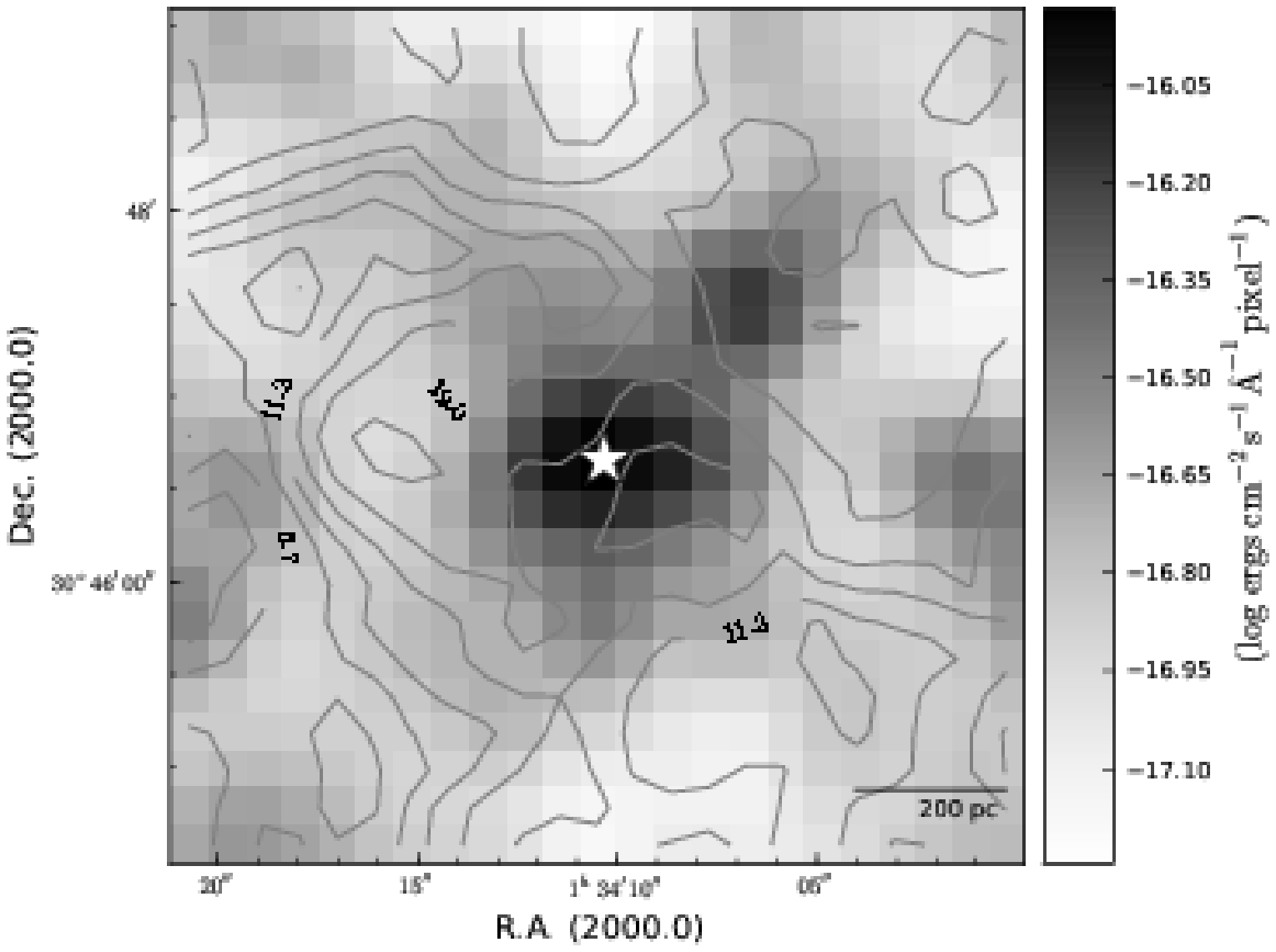}
    & \centering{\scriptsize{\hspace{0.5cm}\phn9\hspace{0.4cm}10}} &
    \includegraphics[width=0.36\textwidth]{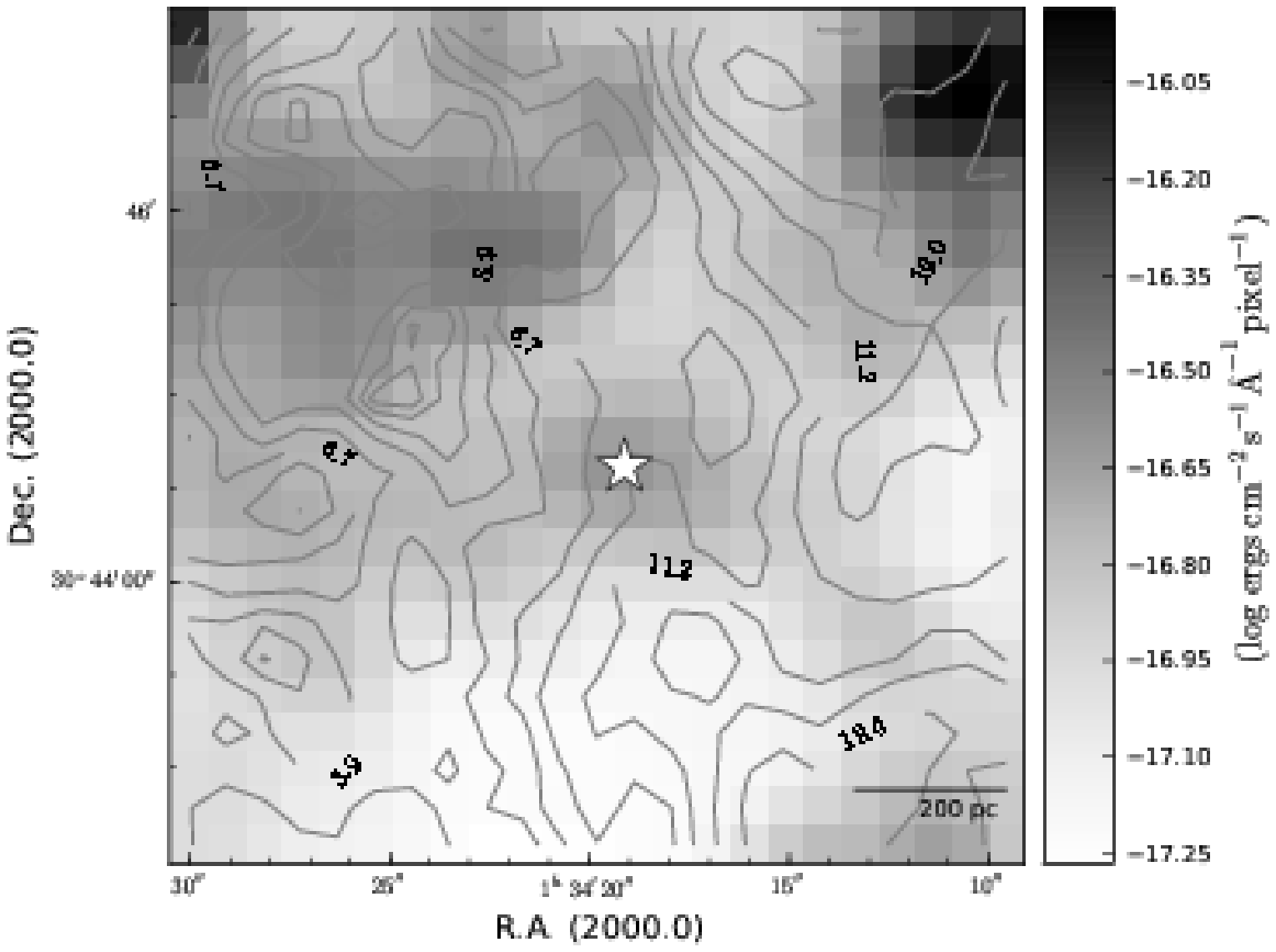} \\
    \includegraphics[width=0.37\textwidth]{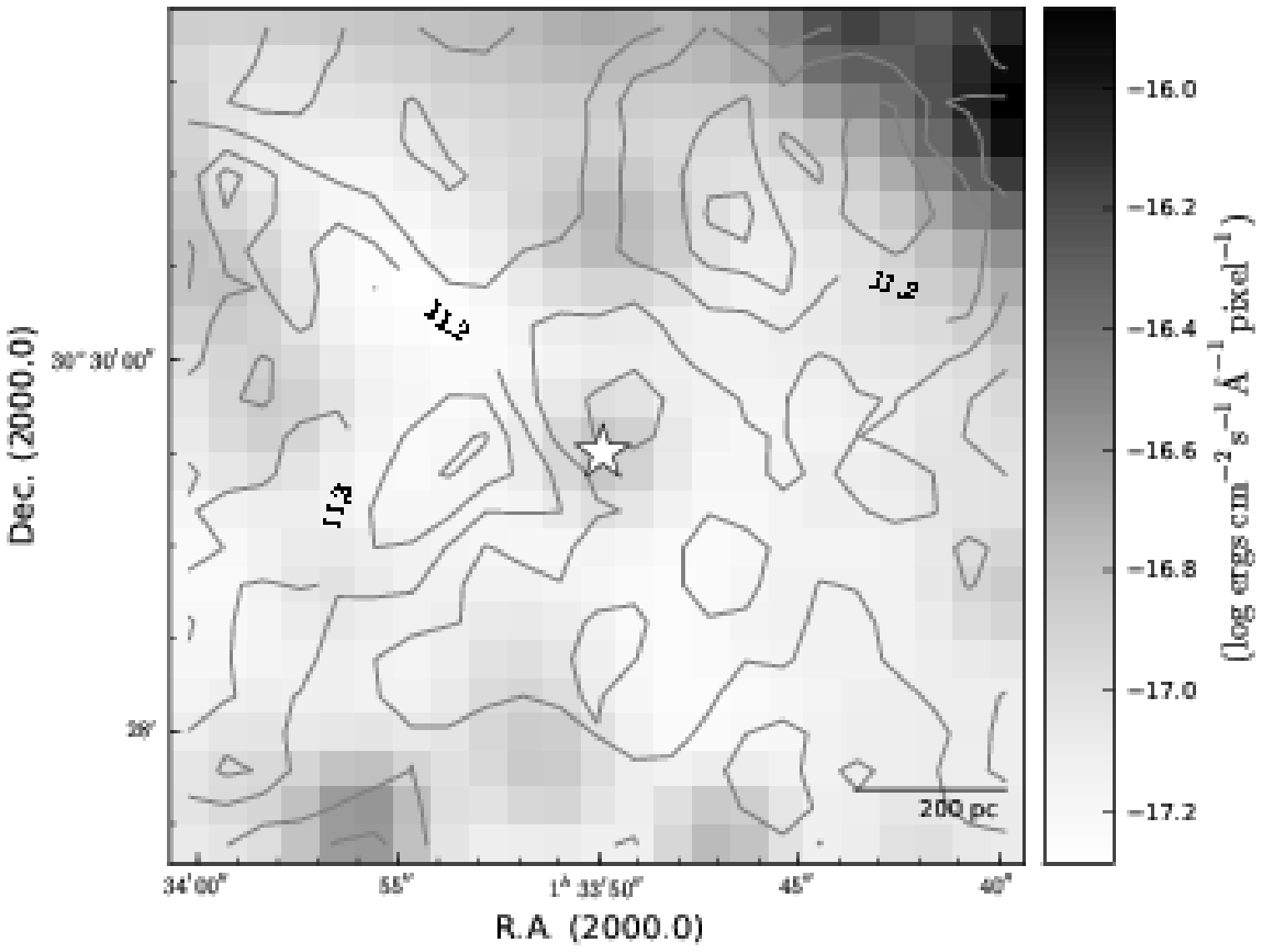}
    & \centering{\scriptsize{\hspace{0.5cm}11\hspace{0.4cm}12}} &
    \includegraphics[width=0.35\textwidth]{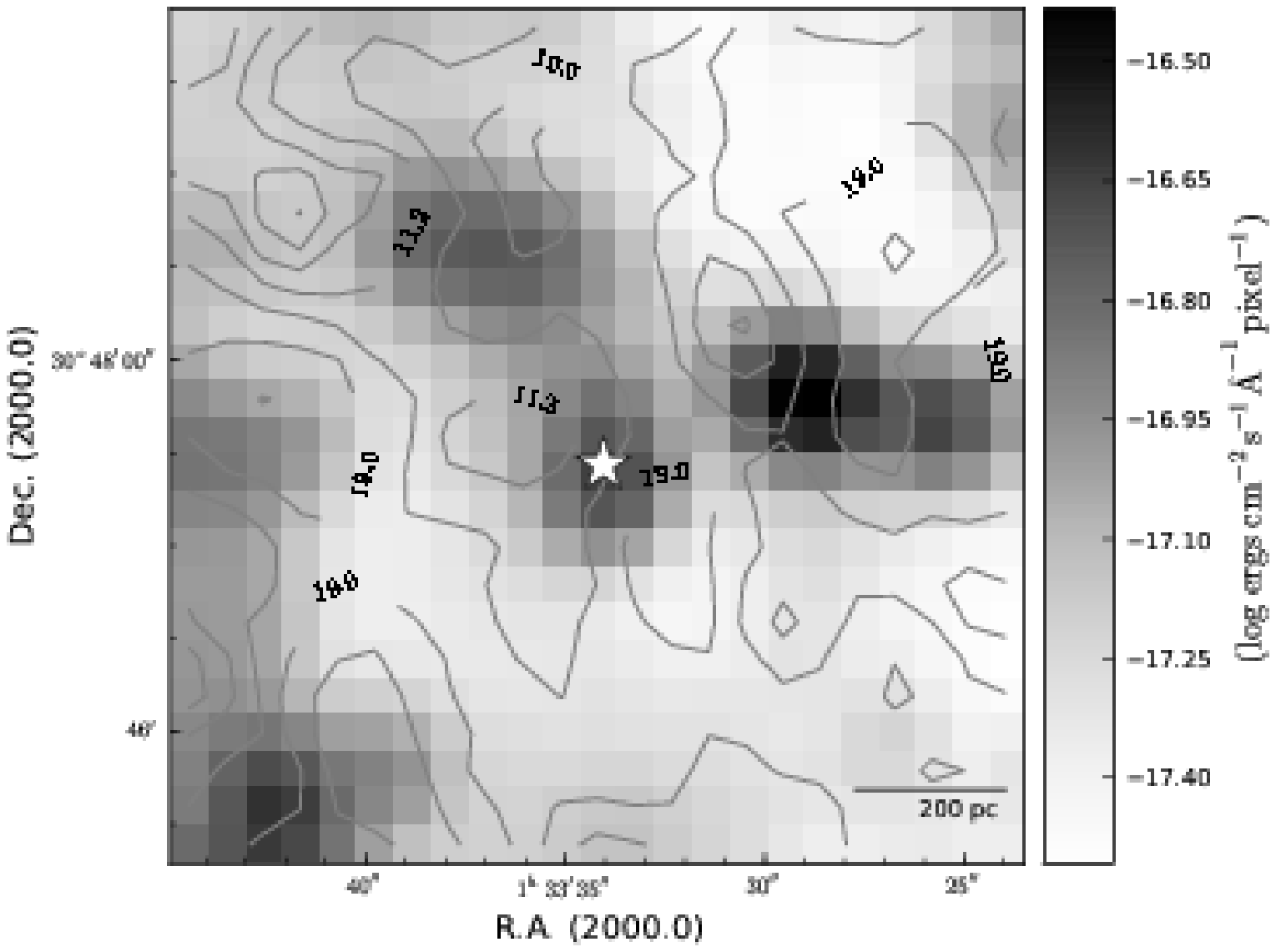} \\
    \includegraphics[width=0.35\textwidth]{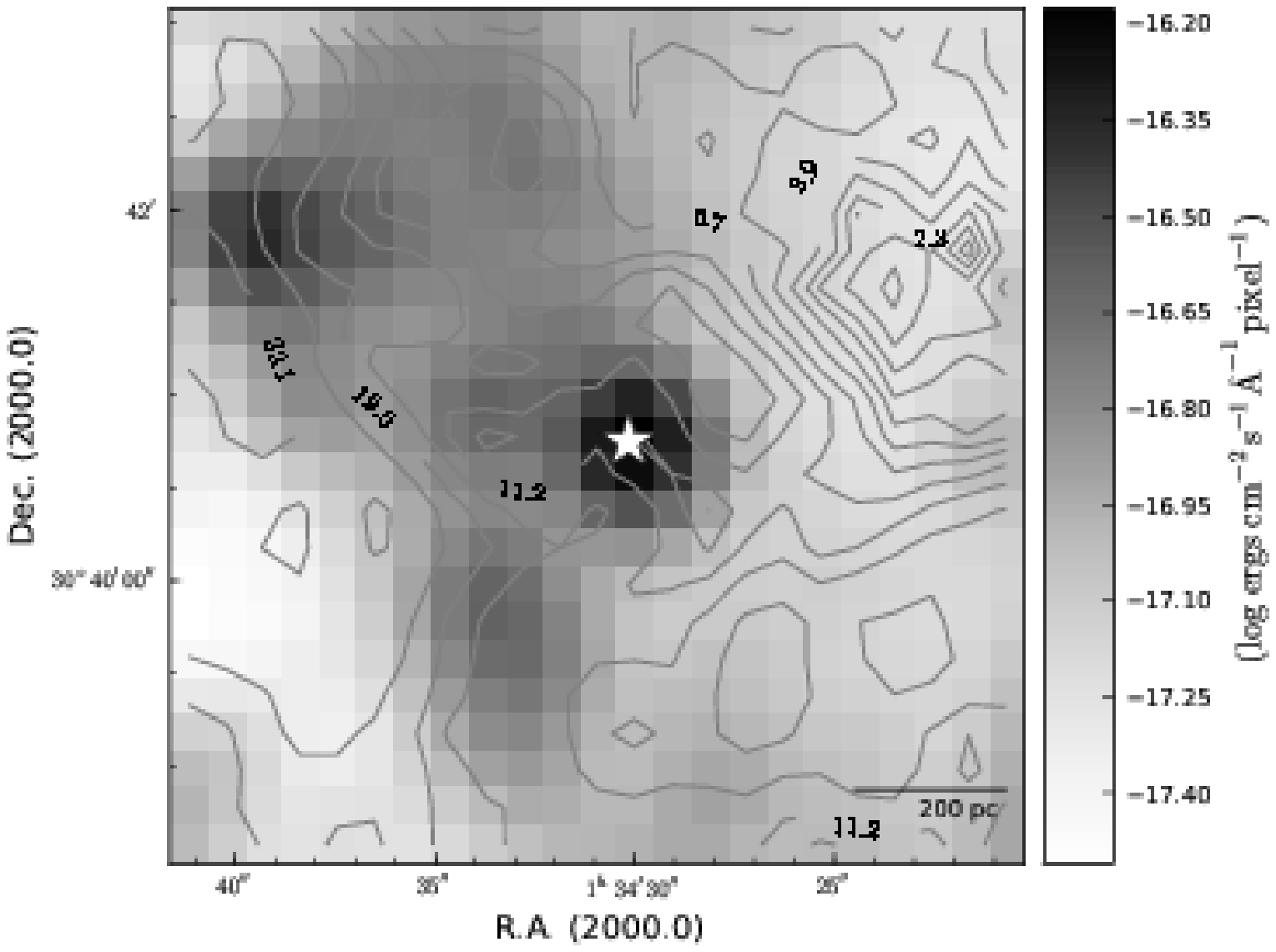}
    & \centering{\scriptsize{\hspace{0.5cm}13\hspace{0.4cm}14}} &
    \includegraphics[width=0.35\textwidth]{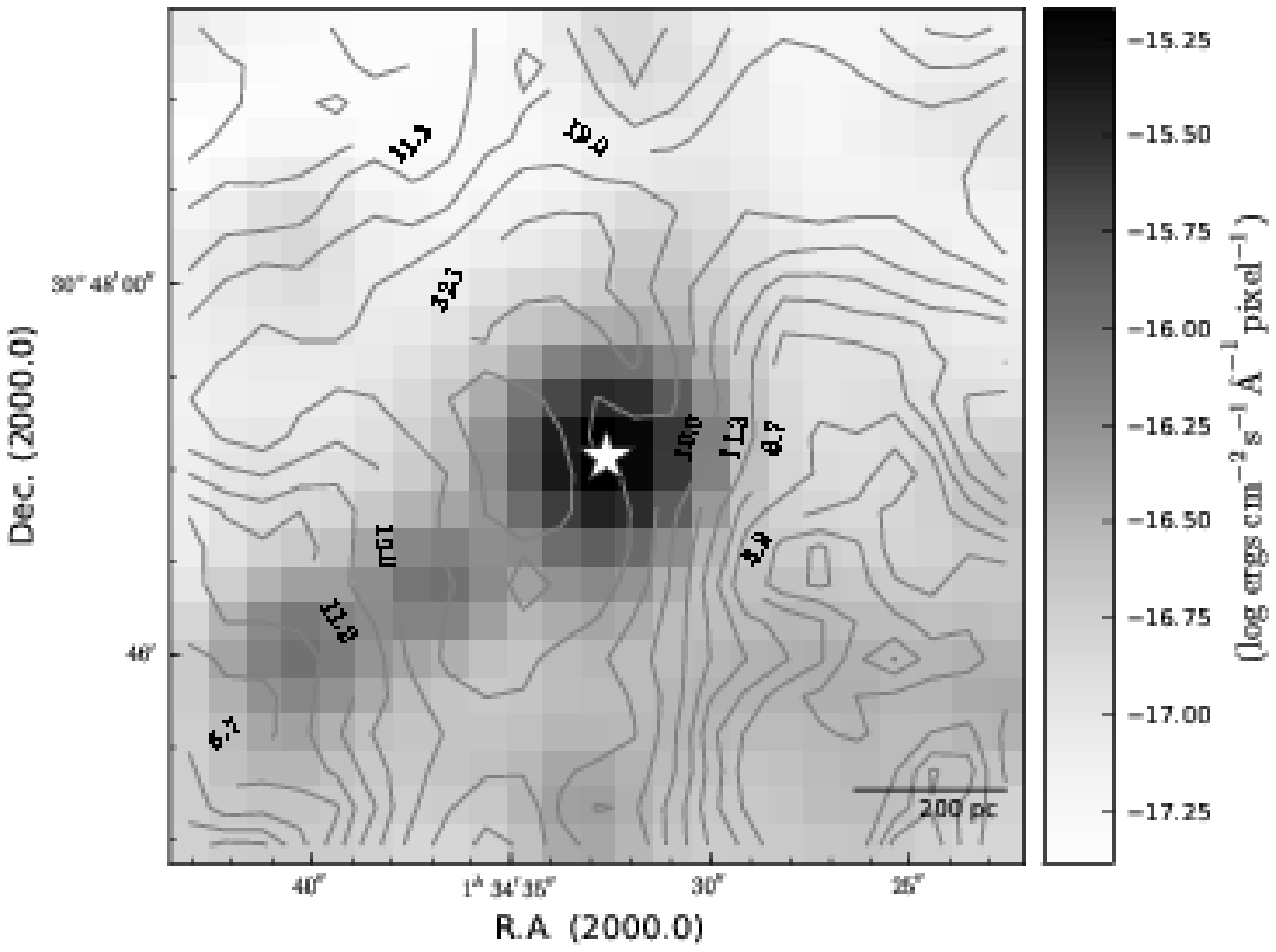} \\
    \includegraphics[width=0.35\textwidth]{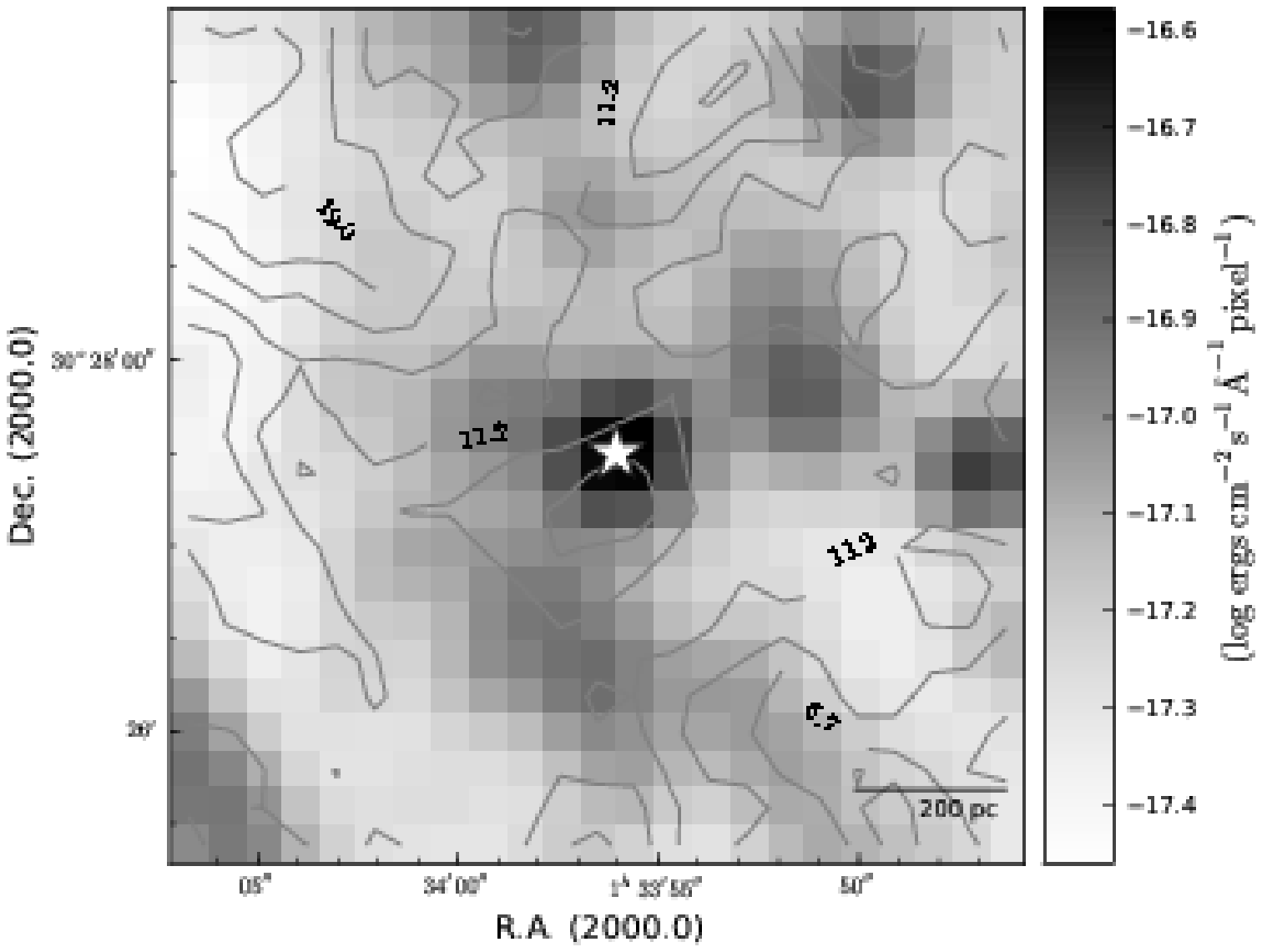}
    & \centering{\scriptsize{\hspace{0.5cm}15\hspace{0.4cm}16}} &
    \includegraphics[width=0.35\textwidth]{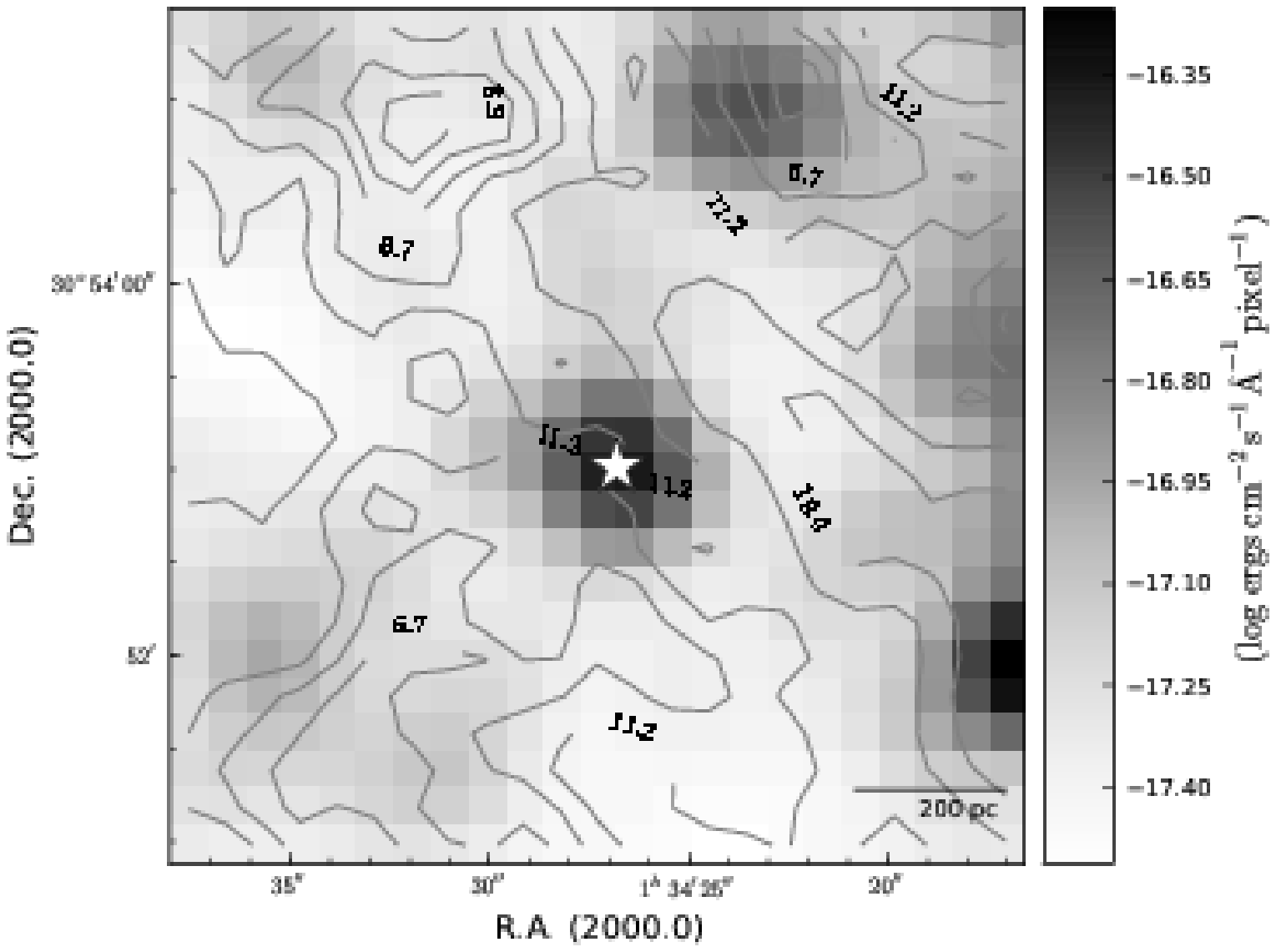} \\
  \end{tabular*}
  \caption[]{M33 sources 9 -- 16 (from left to right, top to bottom). See Figure \ref{fig:A_m33_1} for more details.}
\end{figure*}

\begin{figure*}
  \centering
  \begin{tabular*}{\textwidth}{l p{1.5cm} l} 
    \includegraphics[width=0.355\textwidth]{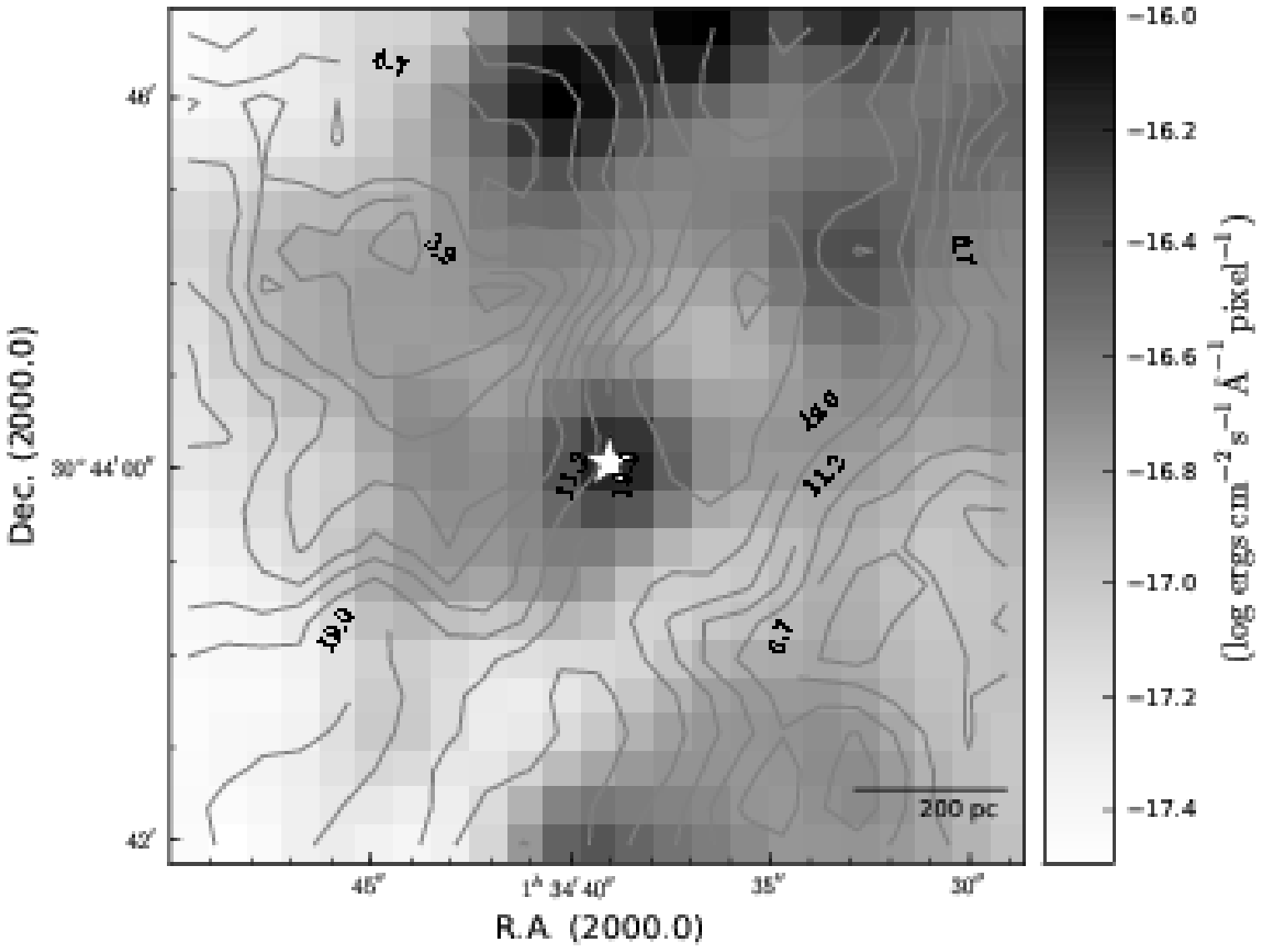}
    & \centering{\scriptsize{\hspace{0.5cm}17\hspace{0.4cm}18}} &
    \includegraphics[width=0.37\textwidth]{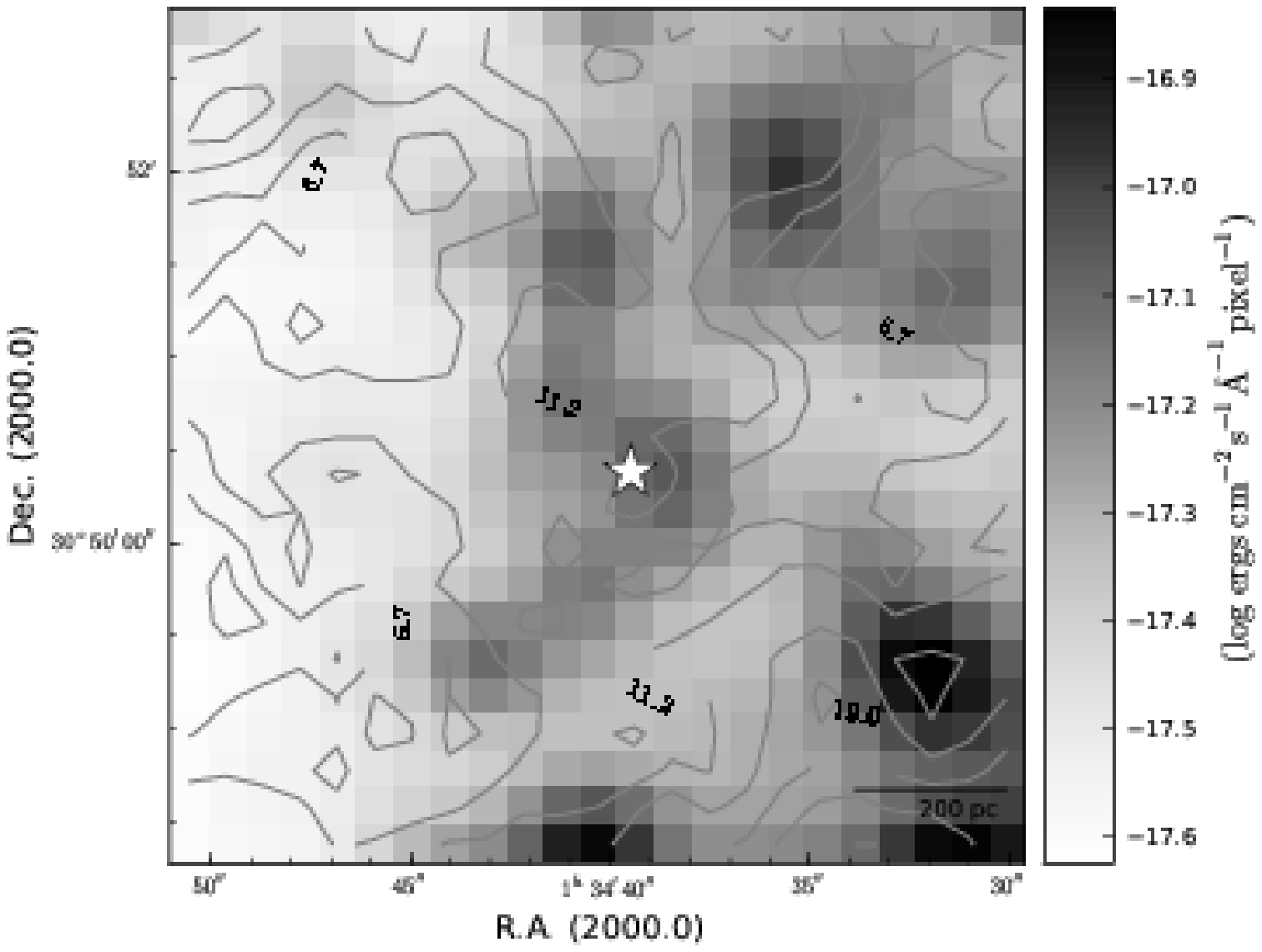} \\
    \includegraphics[width=0.35\textwidth]{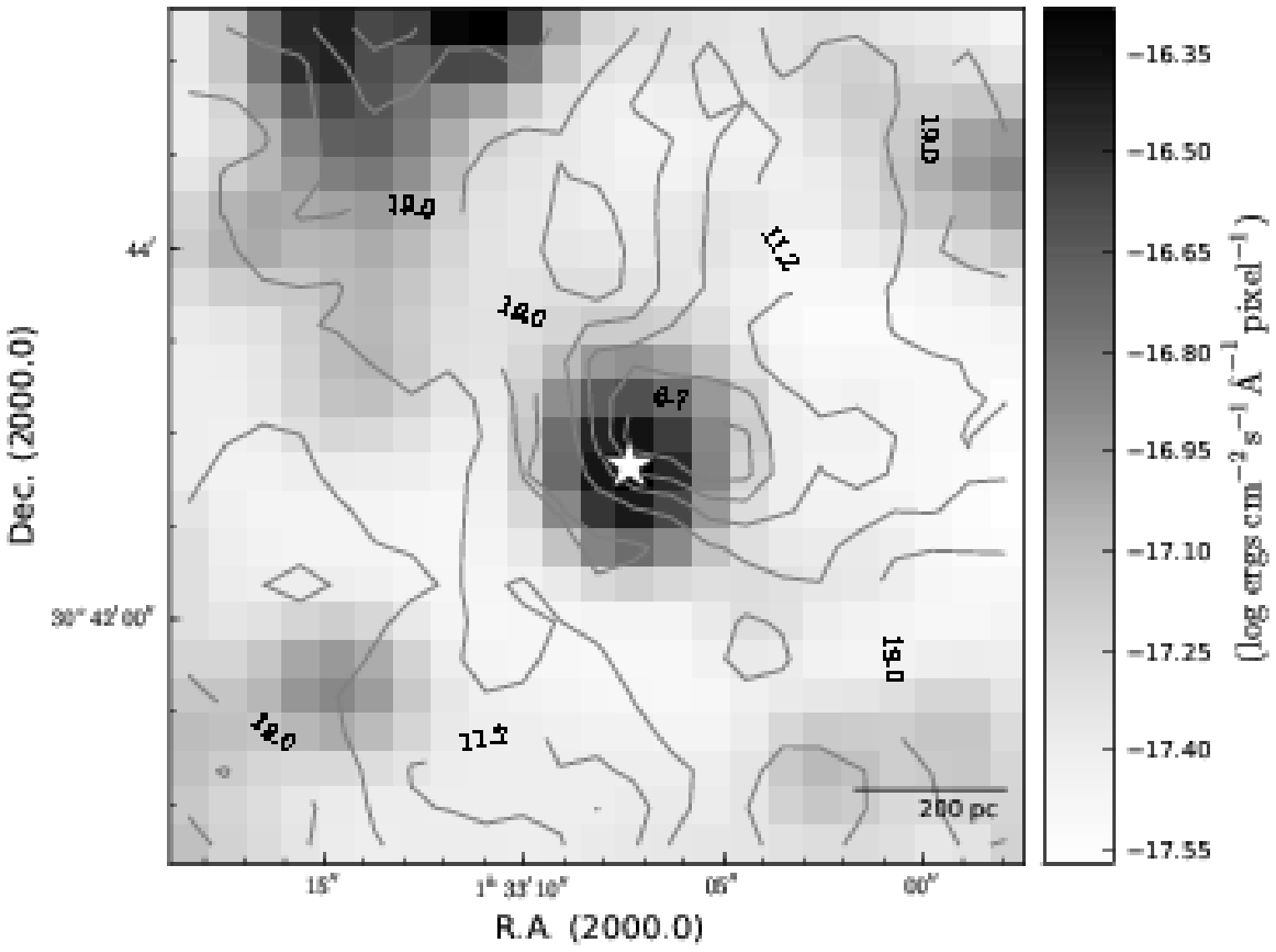}
    & \centering{\scriptsize{\hspace{0.5cm}19\hspace{0.4cm}20}} &
    \includegraphics[width=0.35\textwidth]{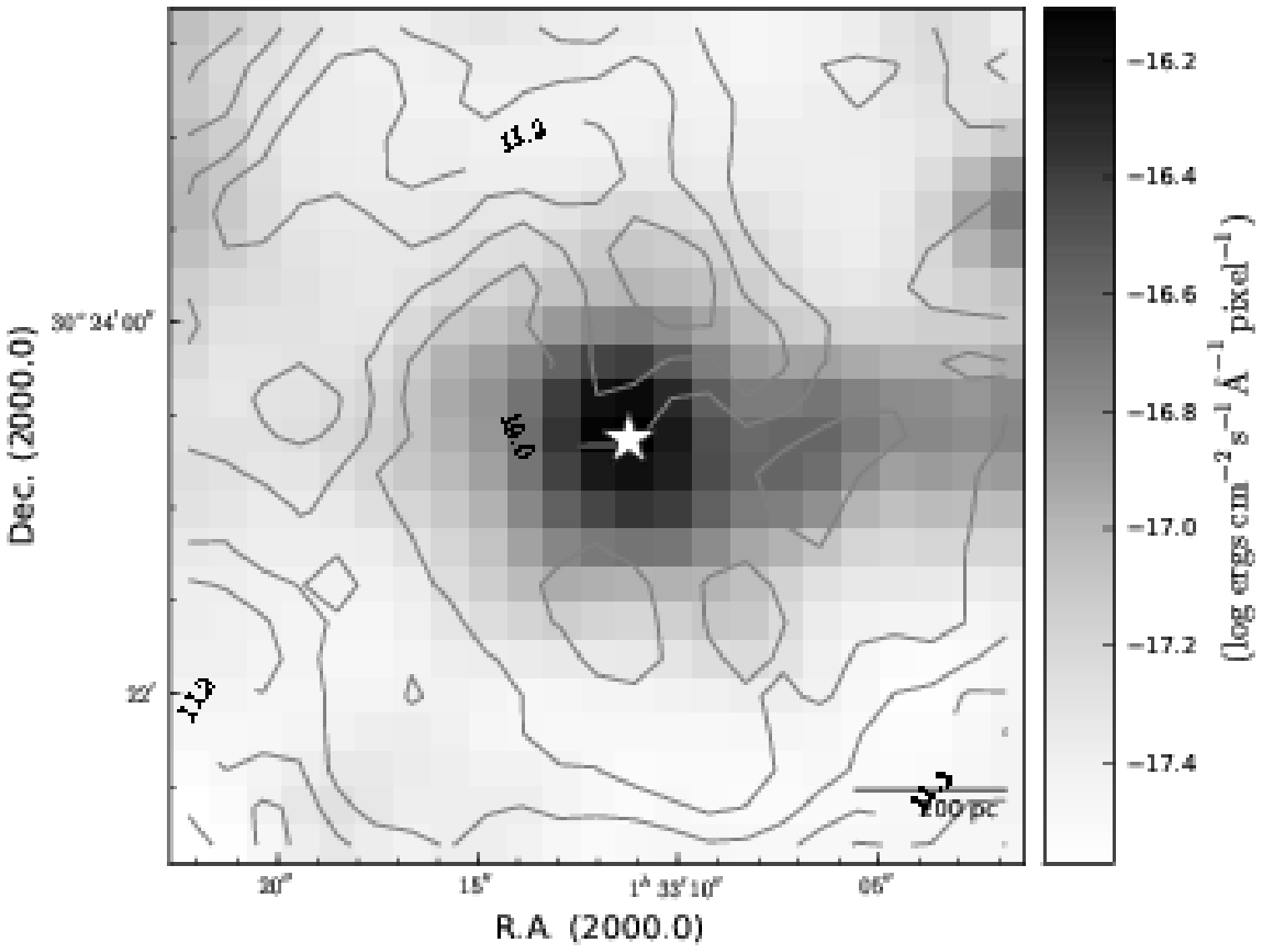} \\
    \includegraphics[width=0.35\textwidth]{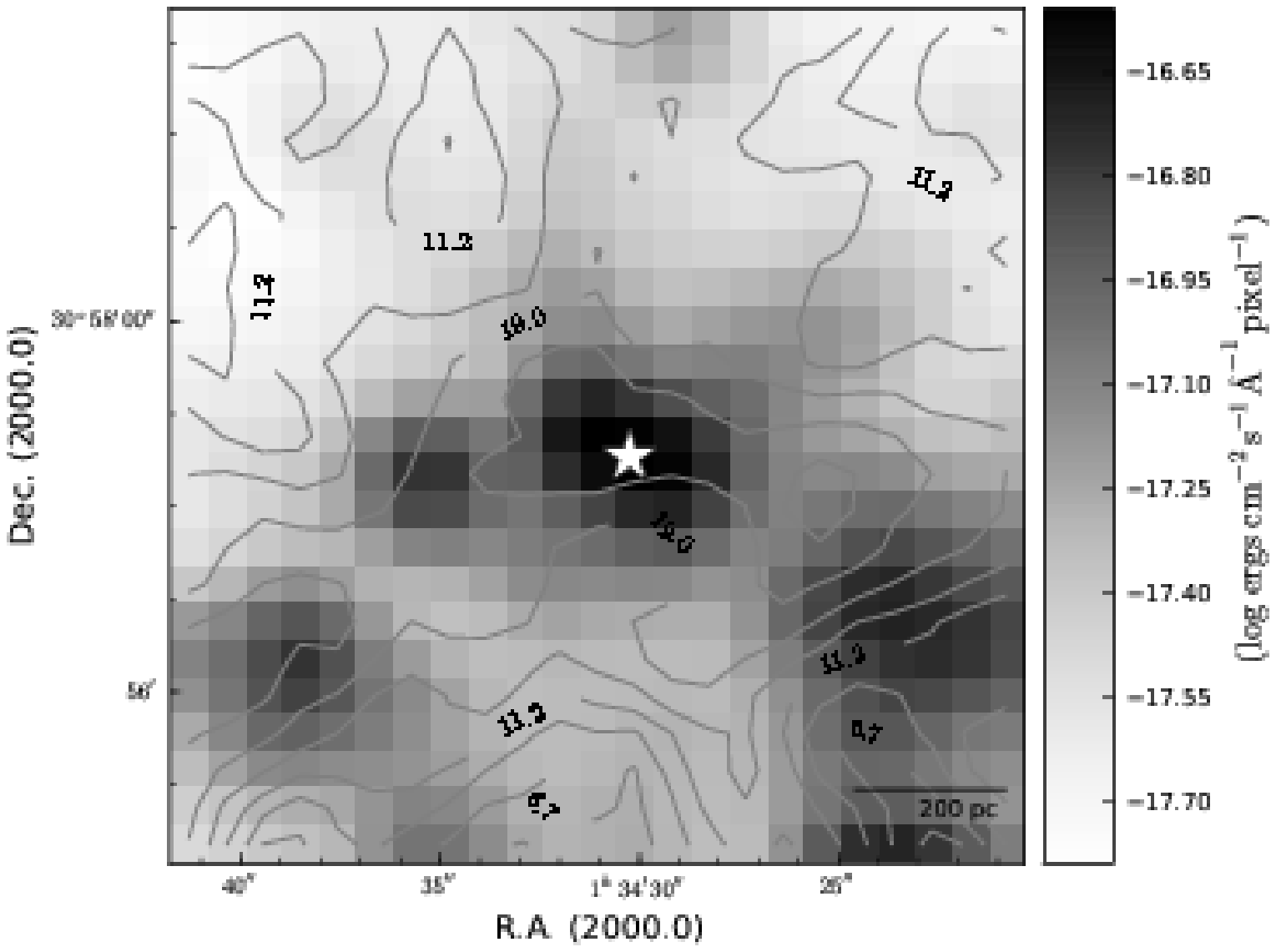}
    & \centering{\scriptsize{\hspace{0.5cm}21\hspace{0.4cm}22}} &
    \includegraphics[width=0.36\textwidth]{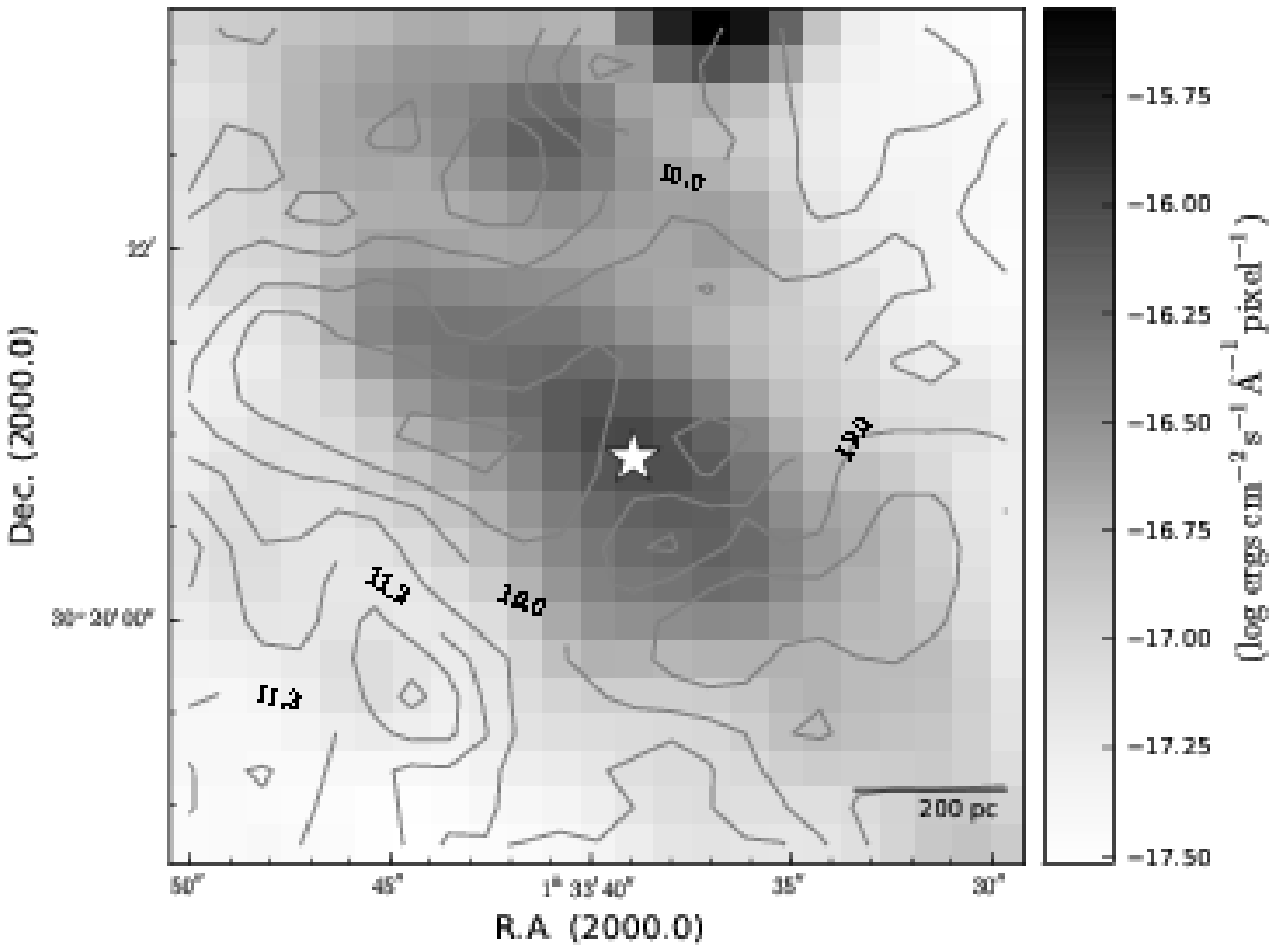} \\
    \includegraphics[width=0.36\textwidth]{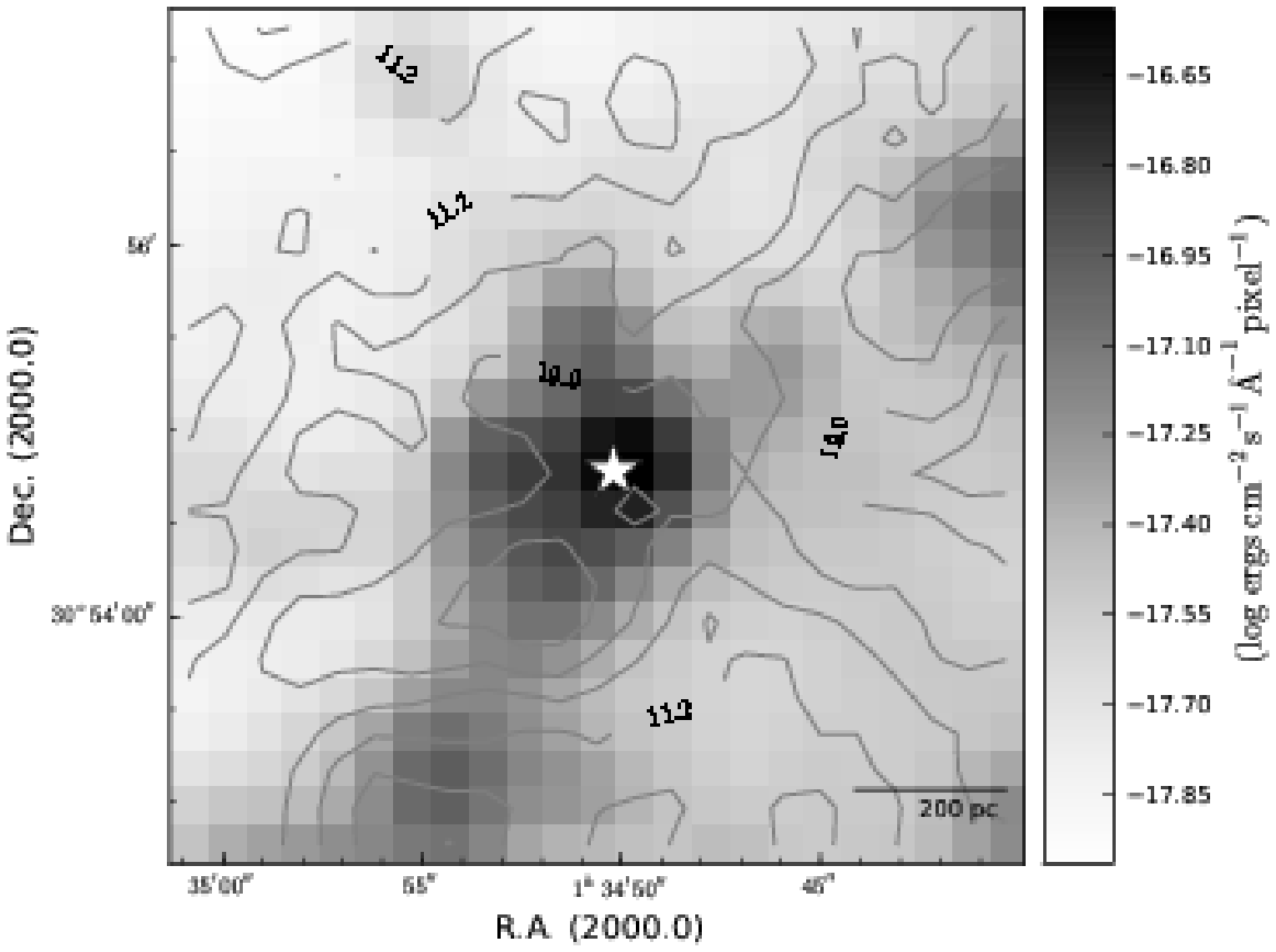}
    & \centering{\scriptsize{\hspace{0.5cm}23\hspace{0.4cm}24}} &
    \includegraphics[width=0.35\textwidth]{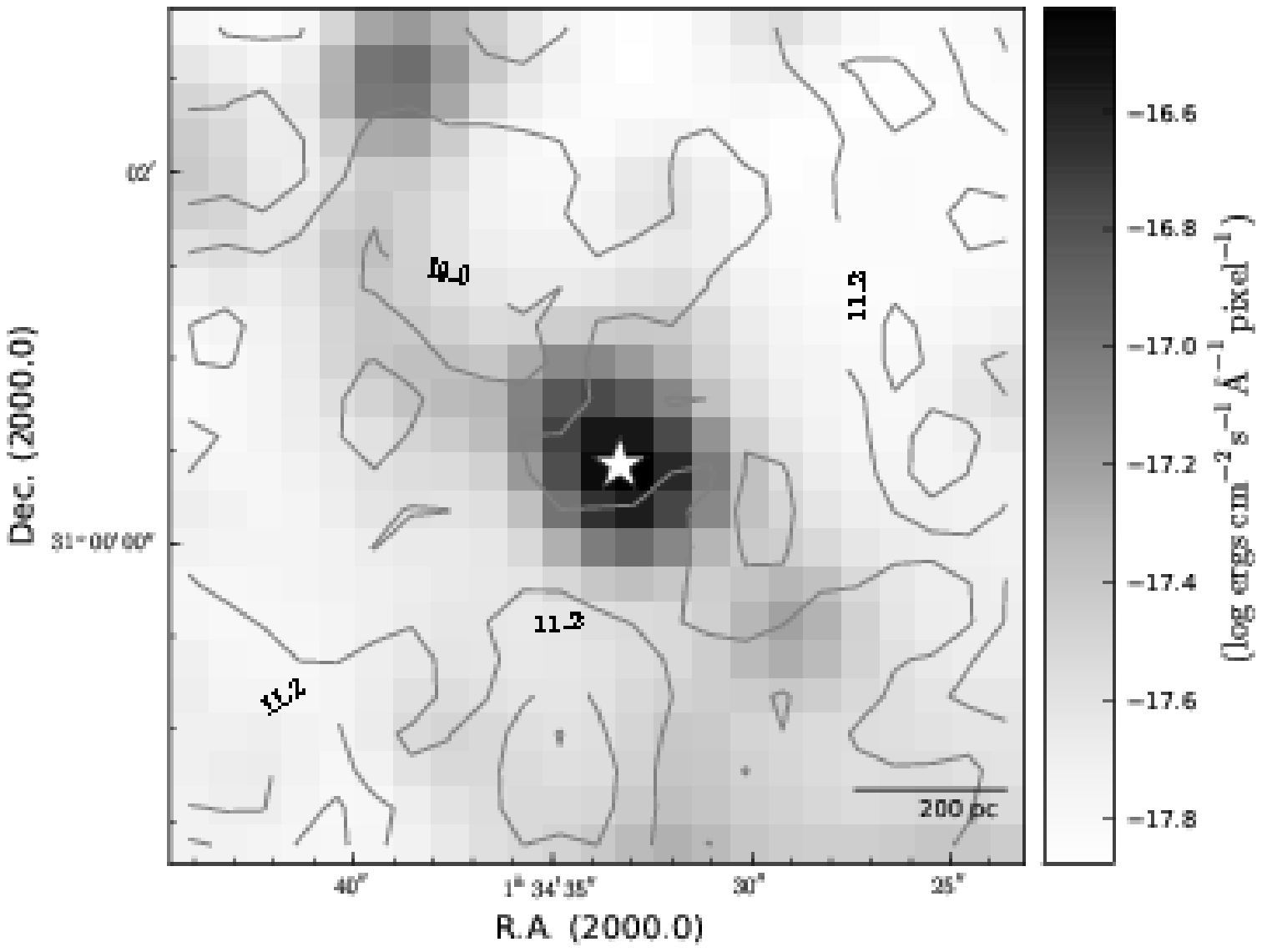} \\
  \end{tabular*}
  \caption[]{M33 sources 17 -- 24 (from left to right, top to bottom). See Figure \ref{fig:A_m33_1} for more details.}
\end{figure*}

\begin{figure*}
  \centering
  \begin{tabular*}{\textwidth}{r p{1.5cm} l} 
    \includegraphics[width=0.35\textwidth]{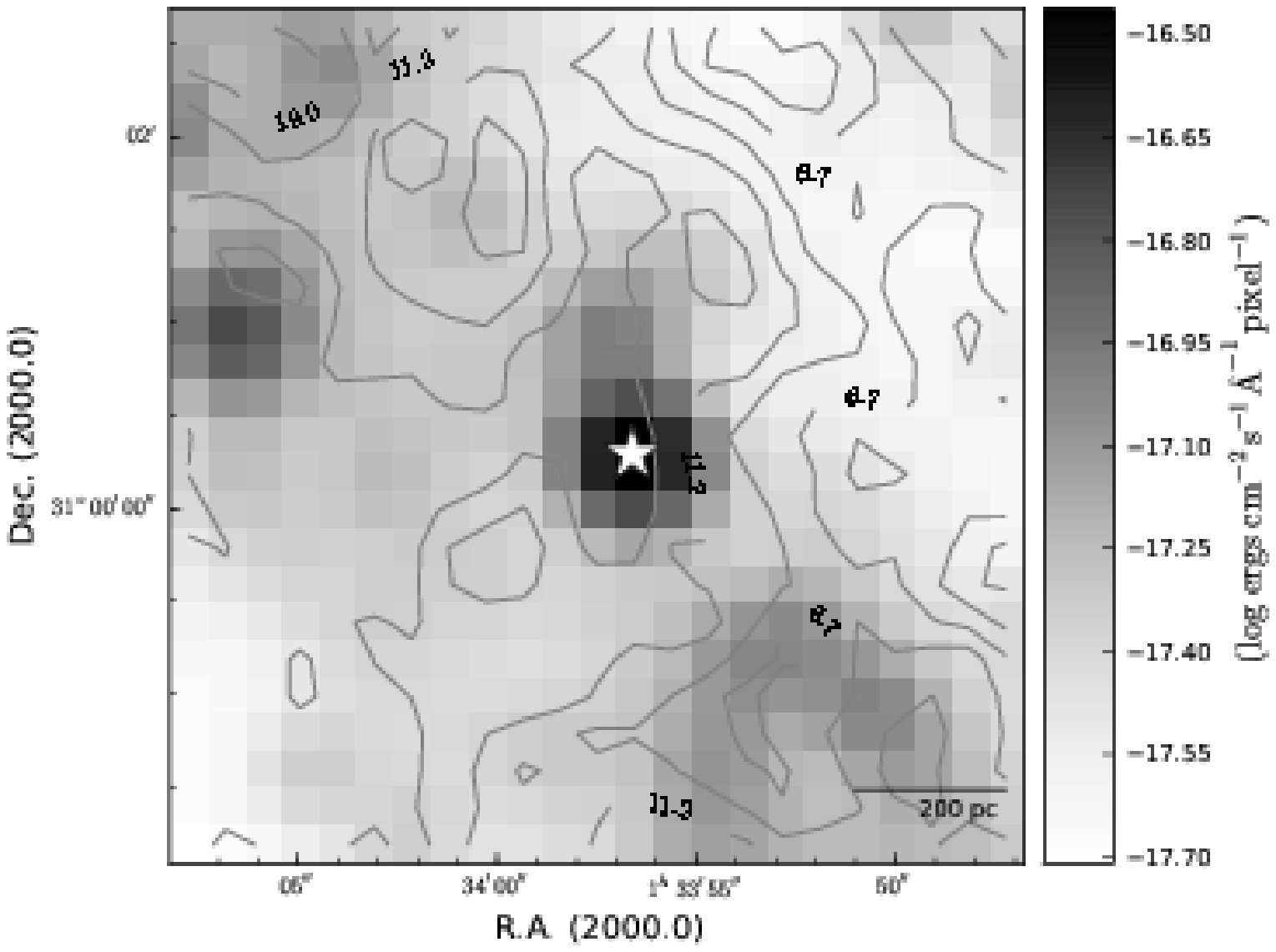}
    & \centering{\scriptsize{\hspace{0.5cm}25\hspace{0.4cm}26}} &
    \includegraphics[width=0.35\textwidth]{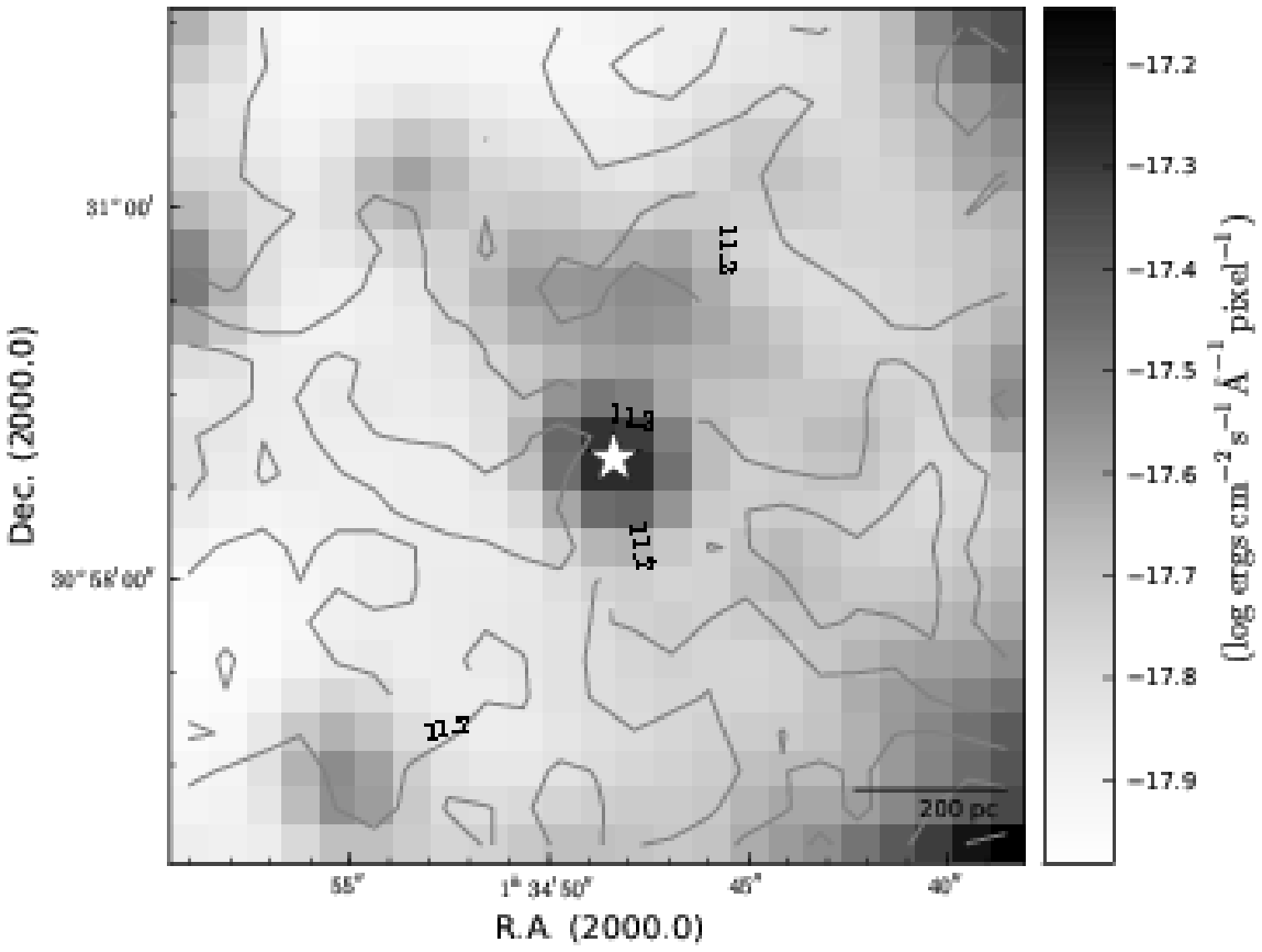} \\
    \includegraphics[width=0.35\textwidth]{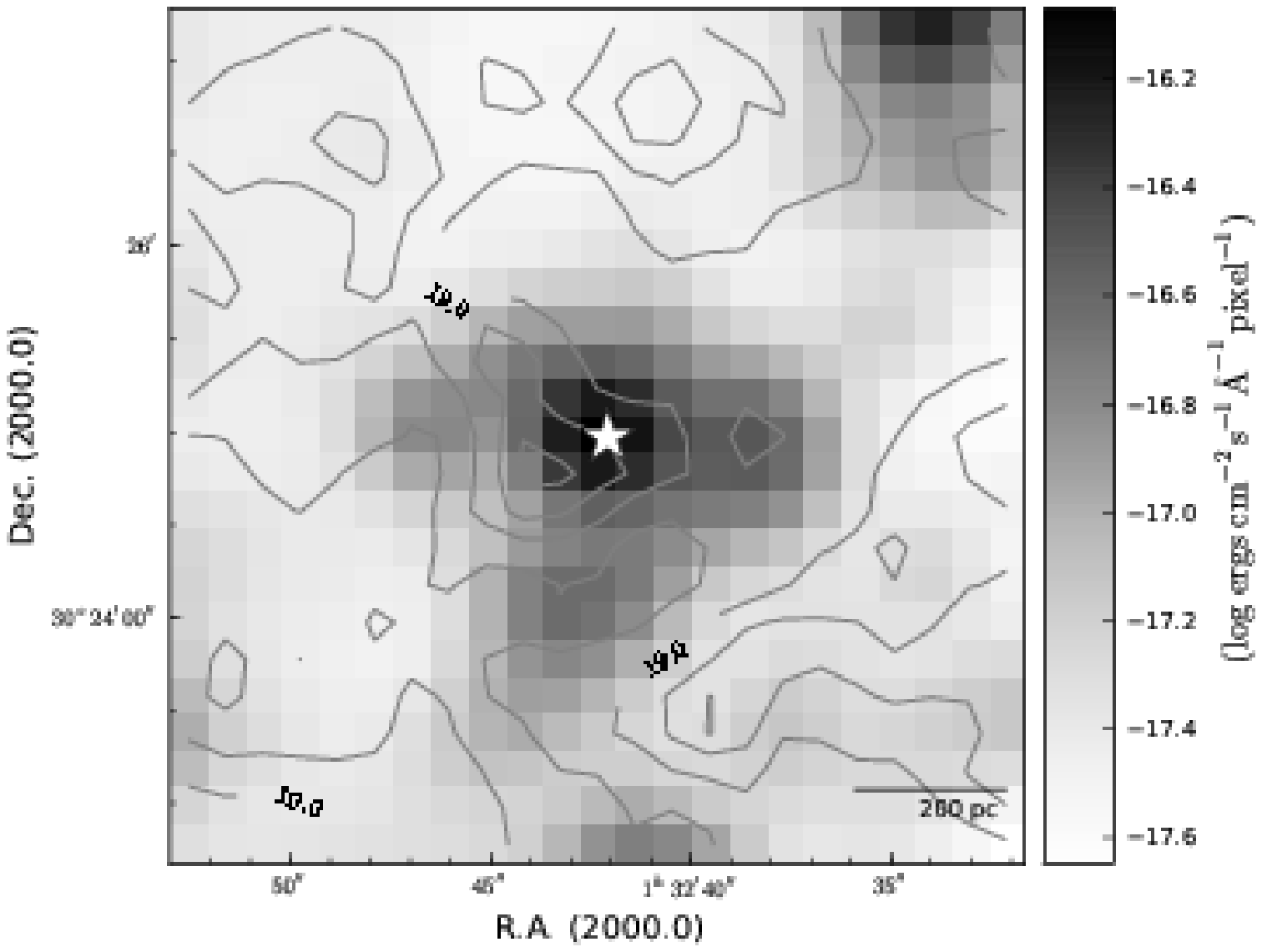}
    & \centering{\scriptsize{\hspace{0.5cm}27\hspace{0.4cm}28}} &
    \includegraphics[width=0.35\textwidth]{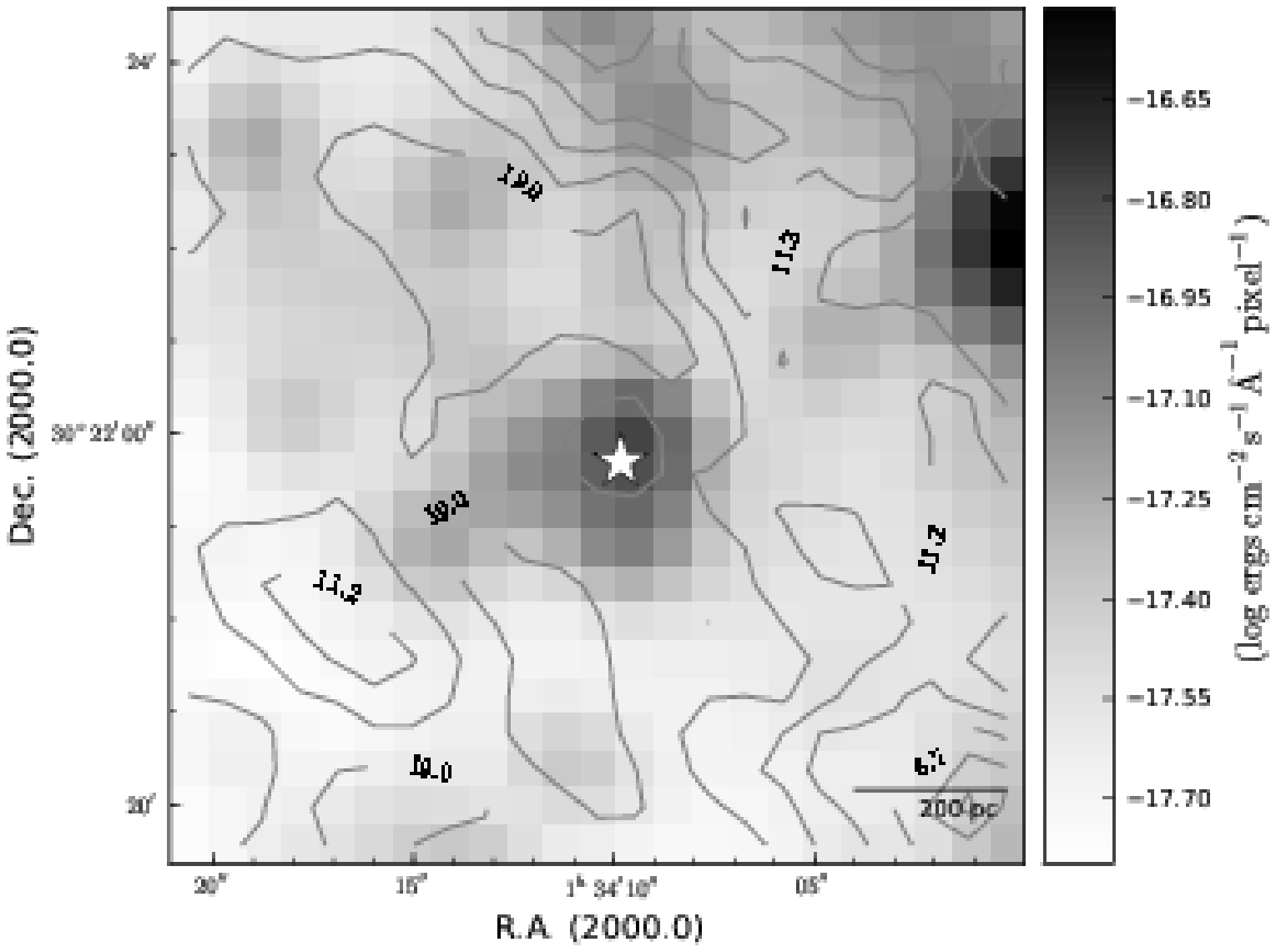} \\
    \includegraphics[width=0.35\textwidth]{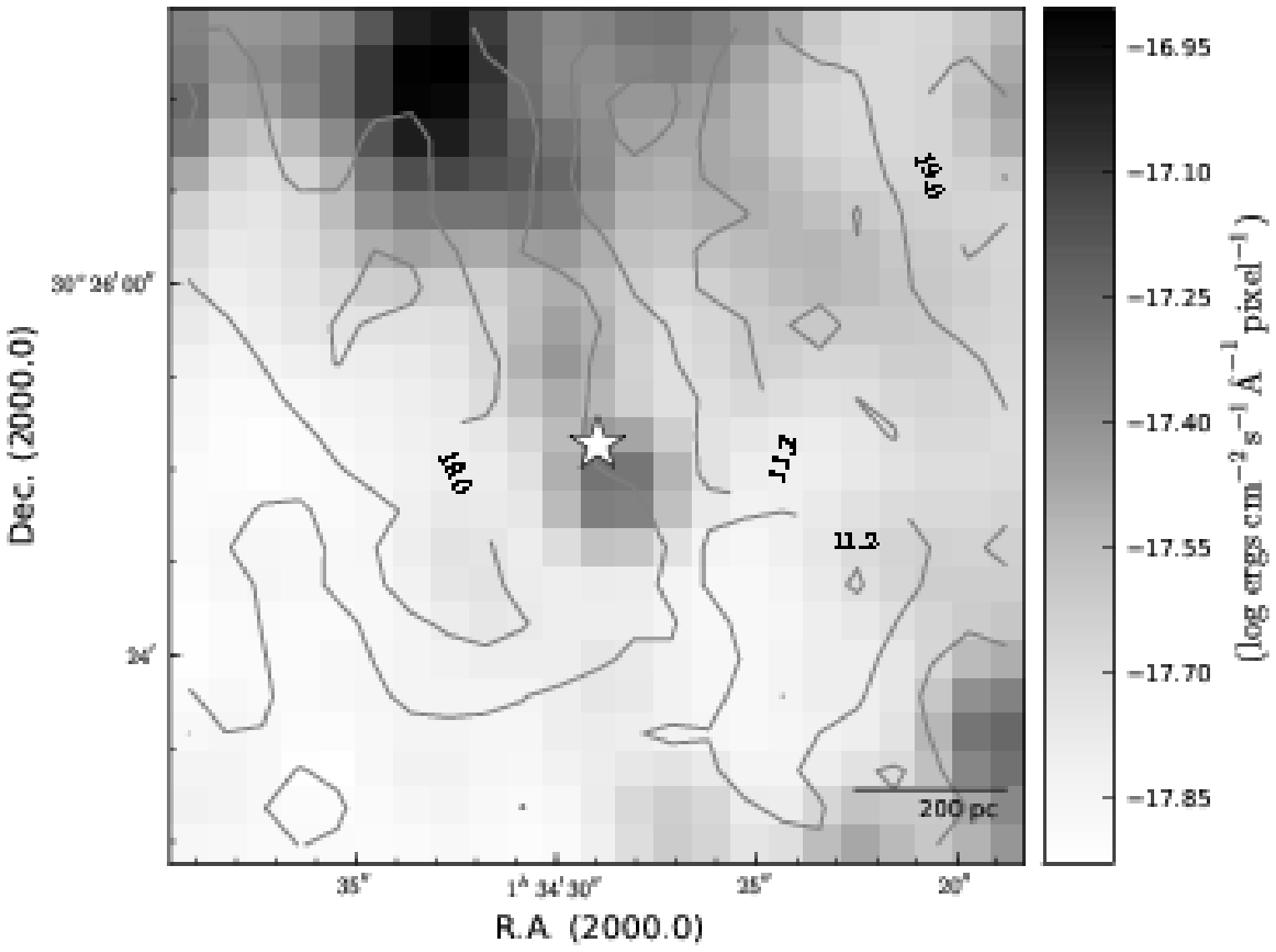}
    & \centering{\scriptsize{\hspace{0.5cm}29\hspace{0.4cm}30}} &
    \includegraphics[width=0.35\textwidth]{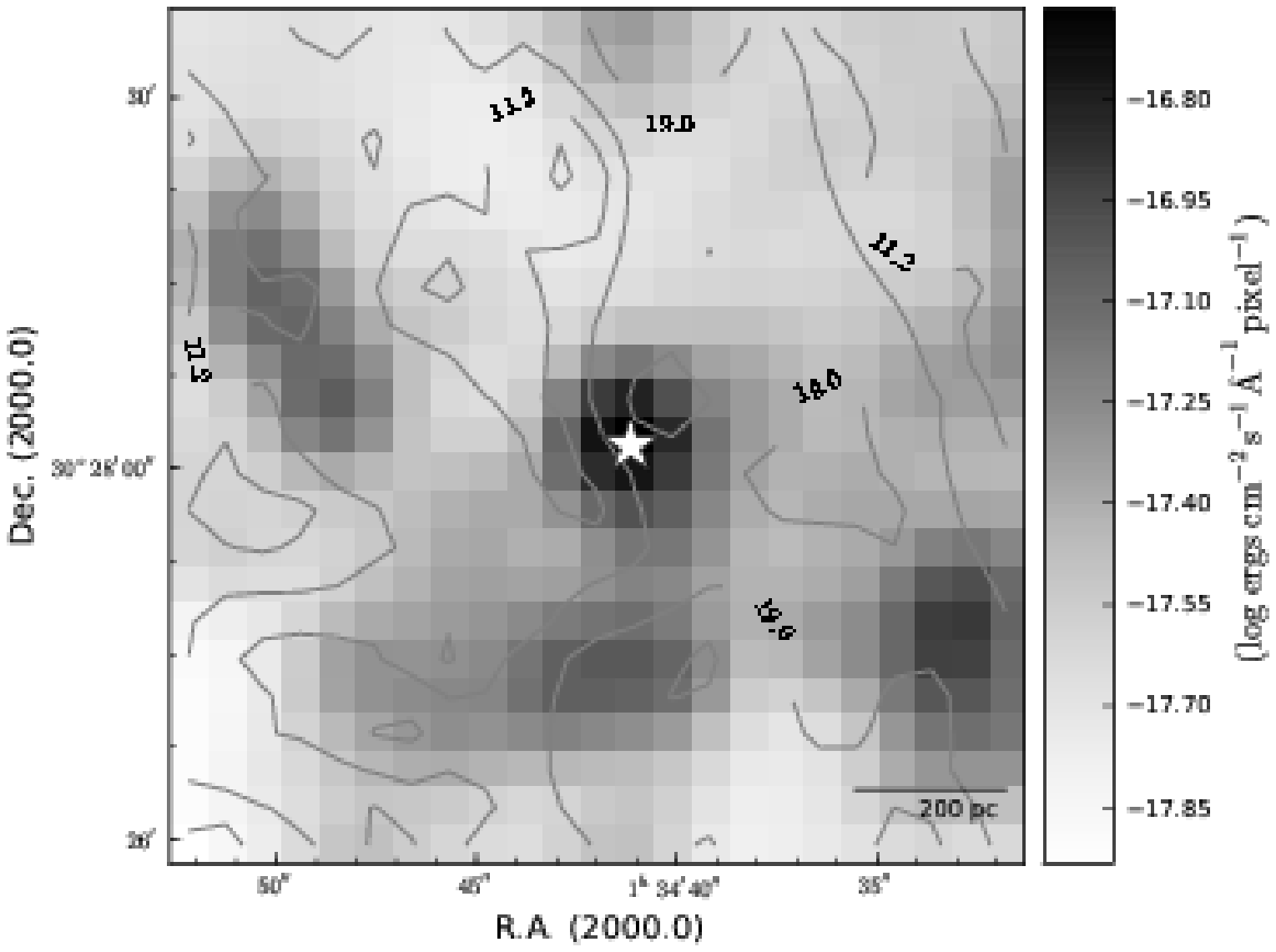} \\
    \includegraphics[width=0.35\textwidth]{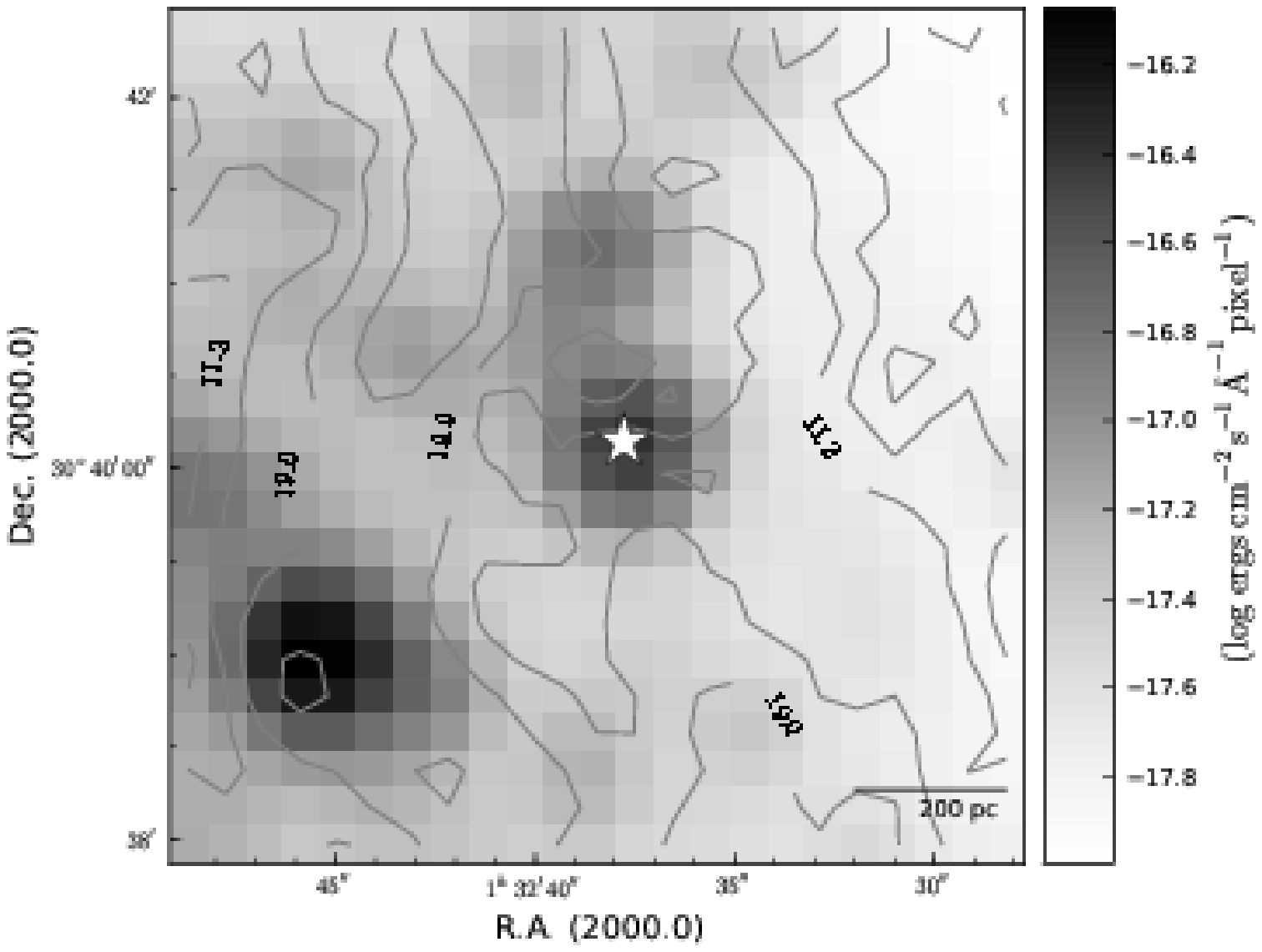}
    & \centering{\scriptsize{\hspace{0.5cm}31\hspace{0.4cm}32}} &
    \includegraphics[width=0.35\textwidth]{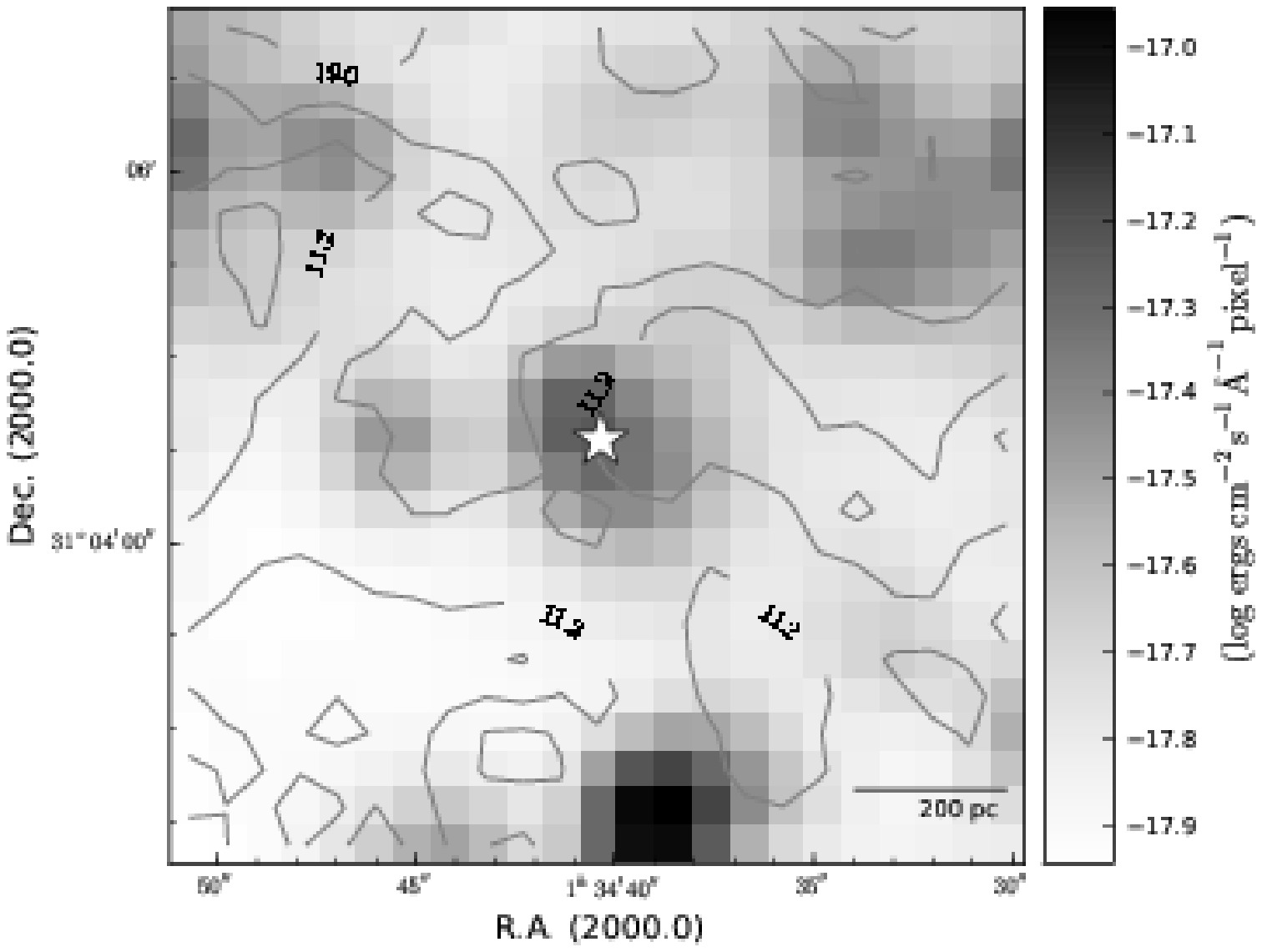} \\
  \end{tabular*}
  \caption[]{M33 sources 25 -- 32 (from left to right, top to bottom). See Figure \ref{fig:A_m33_1} for more details.}
\end{figure*}

\begin{figure*}
  \centering
  \begin{tabular*}{\textwidth}{r p{1.5cm} l} 
    \includegraphics[width=0.35\textwidth]{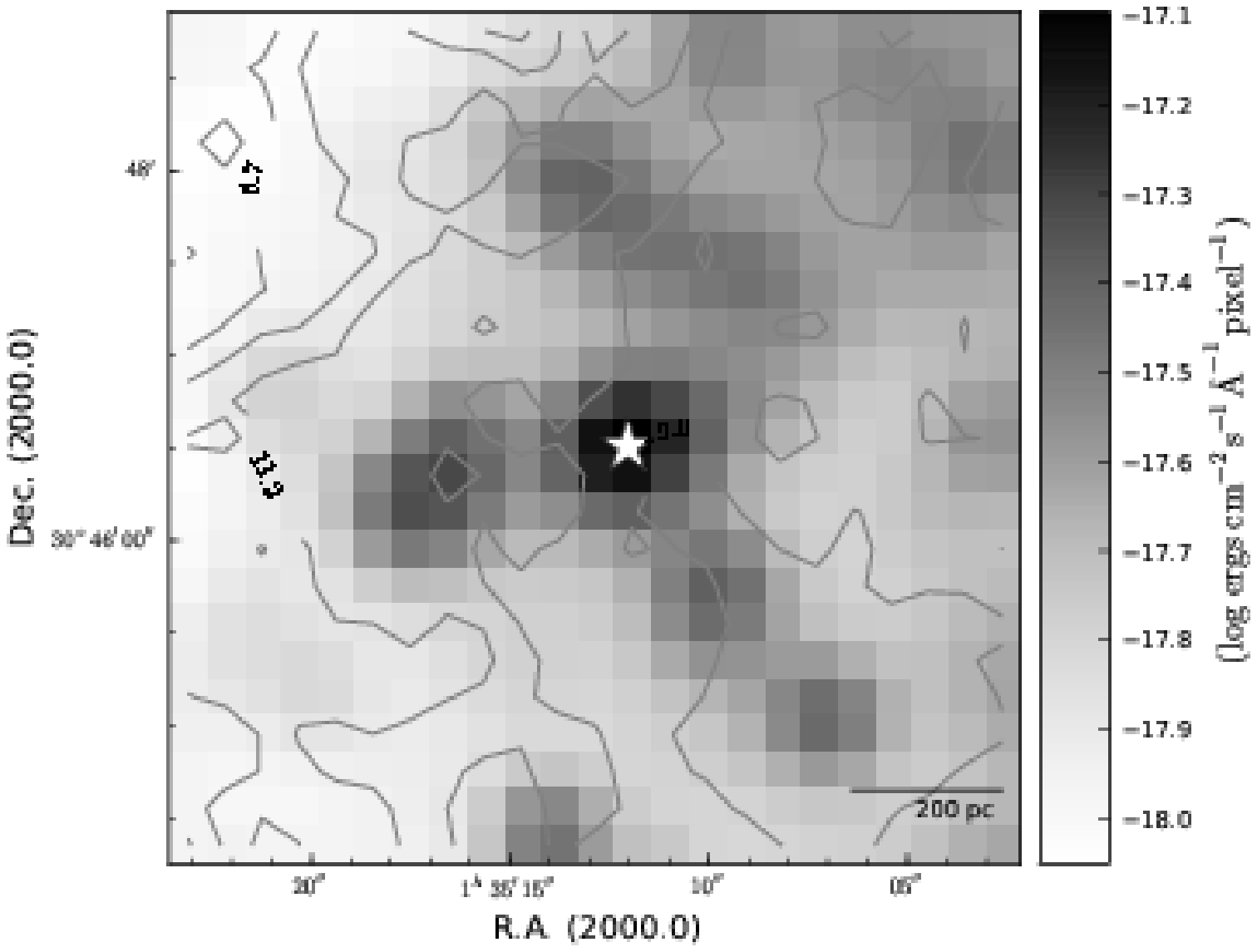}
    & \centering{\scriptsize{\hspace{0.5cm}33\hspace{0.4cm}34}} &
    \includegraphics[width=0.35\textwidth]{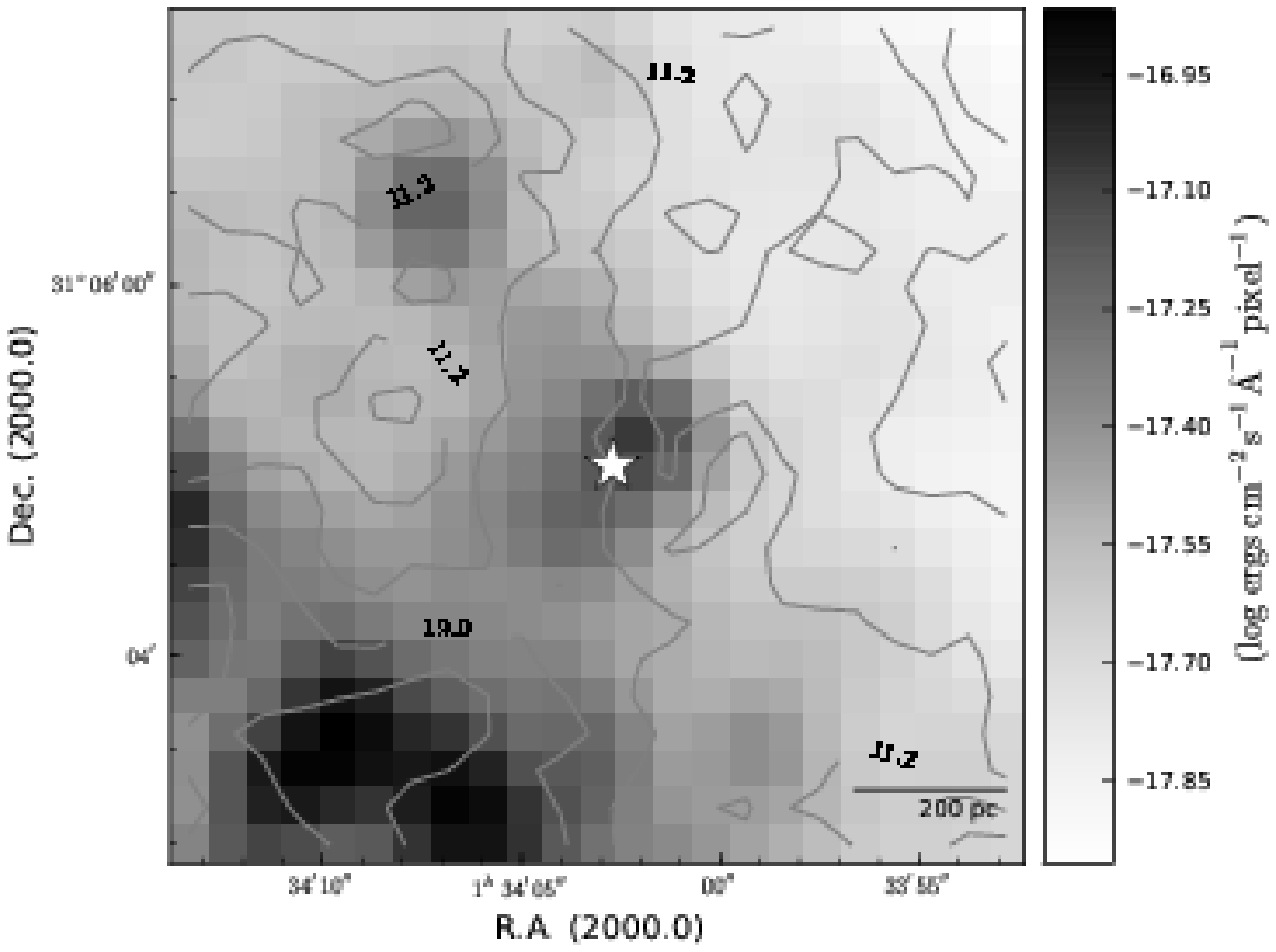} \\
    \includegraphics[width=0.35\textwidth]{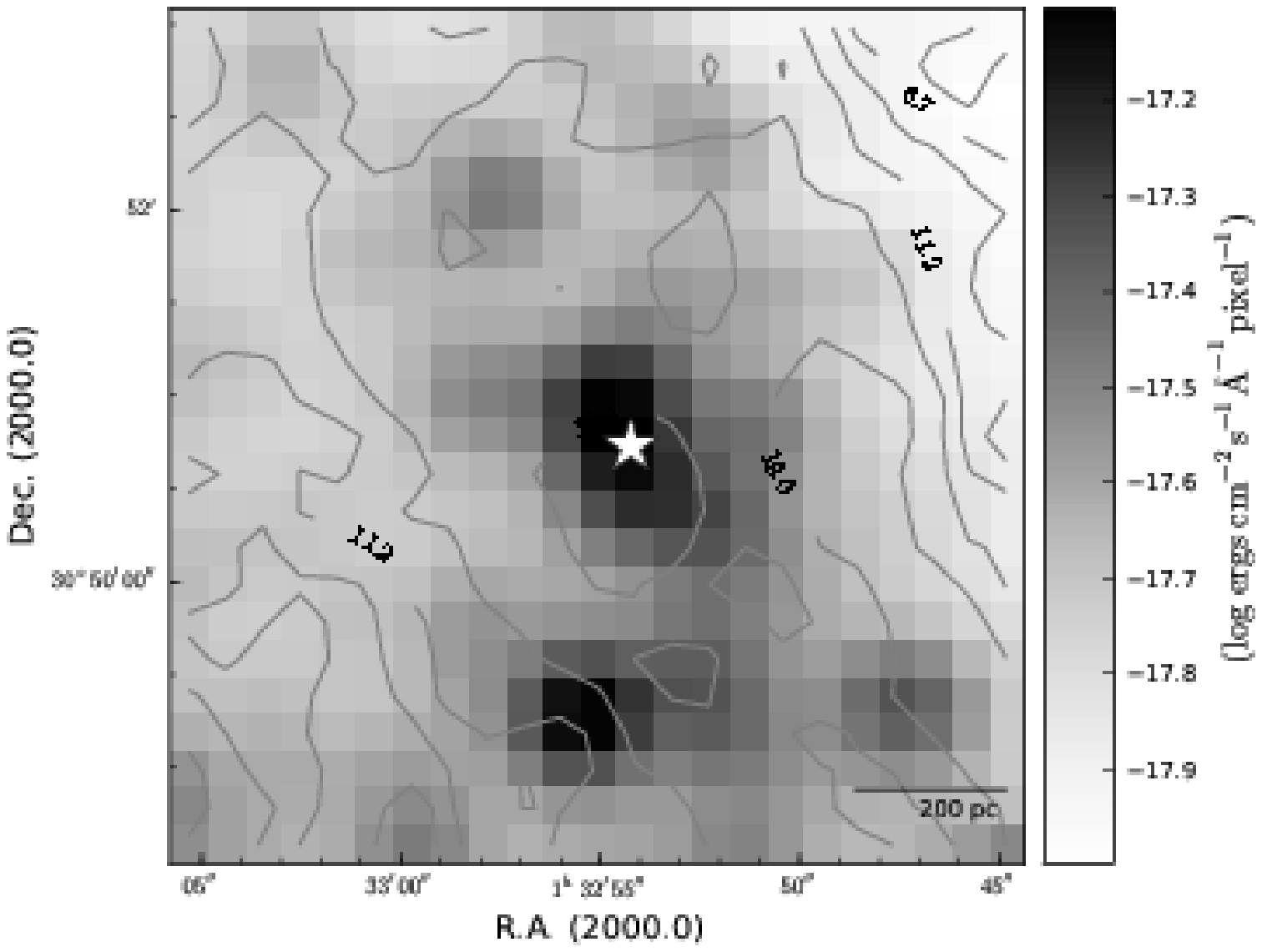}
    & \centering{\scriptsize{\hspace{0.5cm}35\hspace{0.4cm}36}} &
    \includegraphics[width=0.35\textwidth]{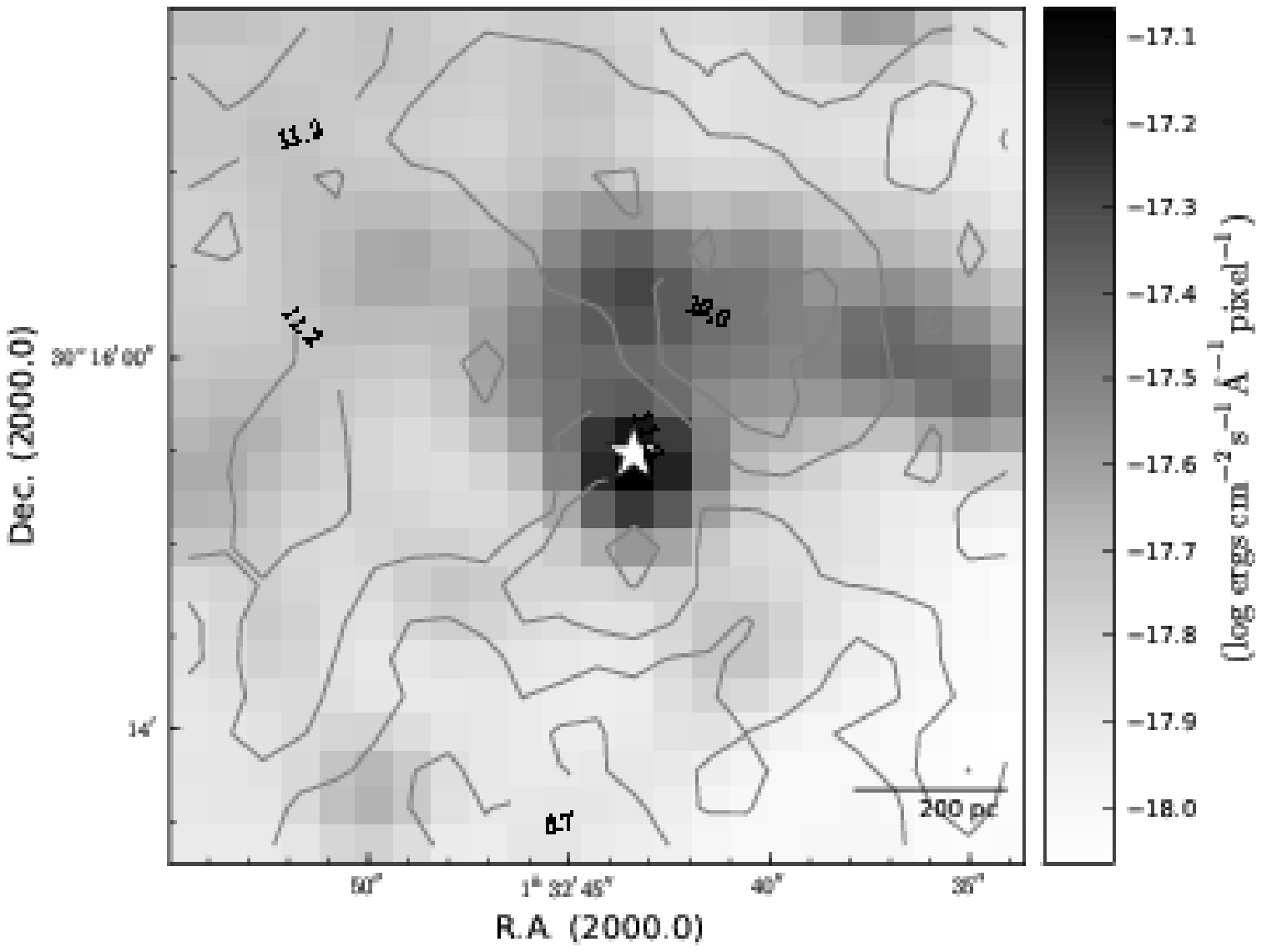} \\
    \includegraphics[width=0.35\textwidth]{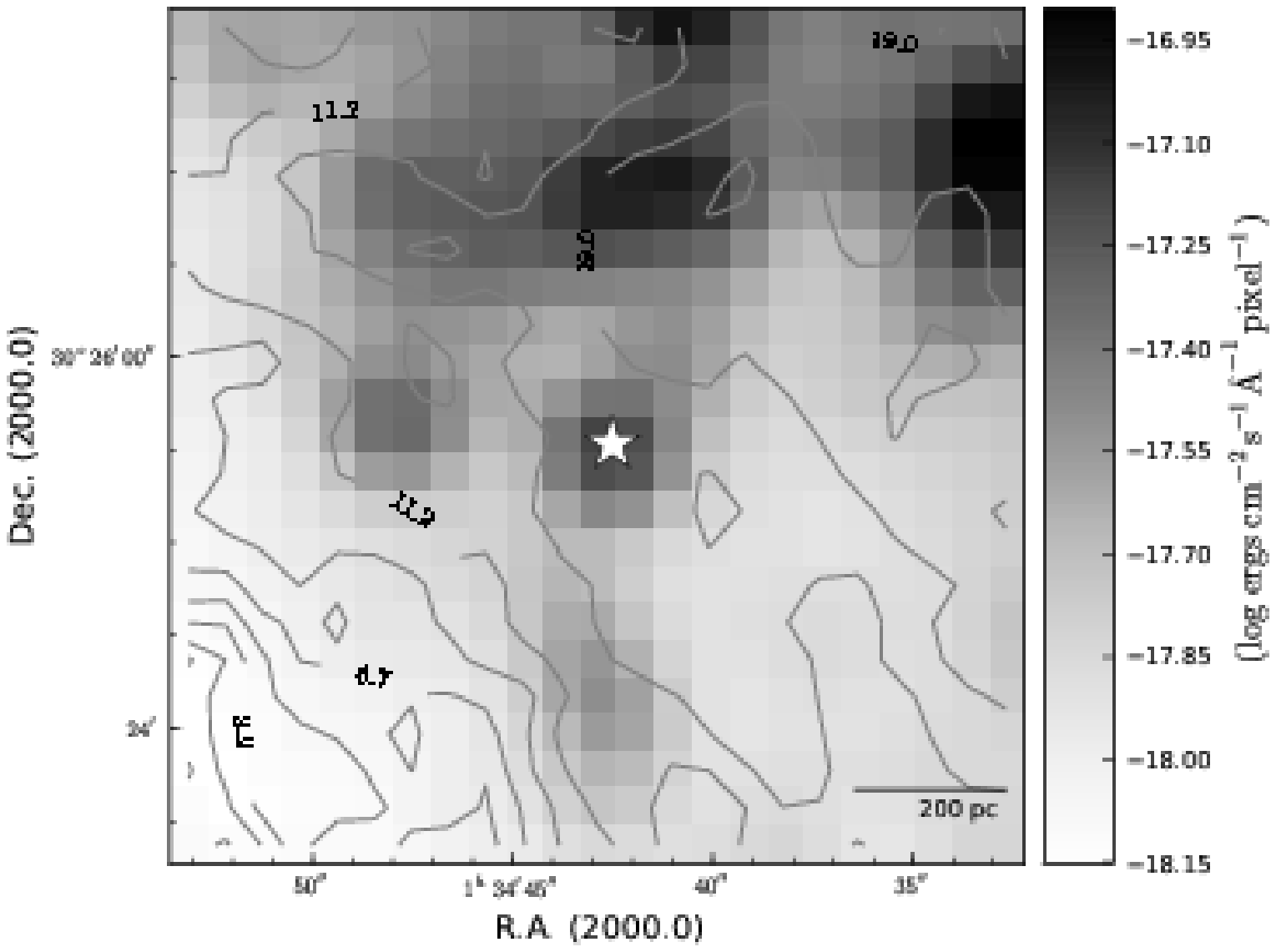}
    & \centering{\scriptsize{\hspace{0.5cm}37\hspace{0.4cm}38}} &
    \includegraphics[width=0.35\textwidth]{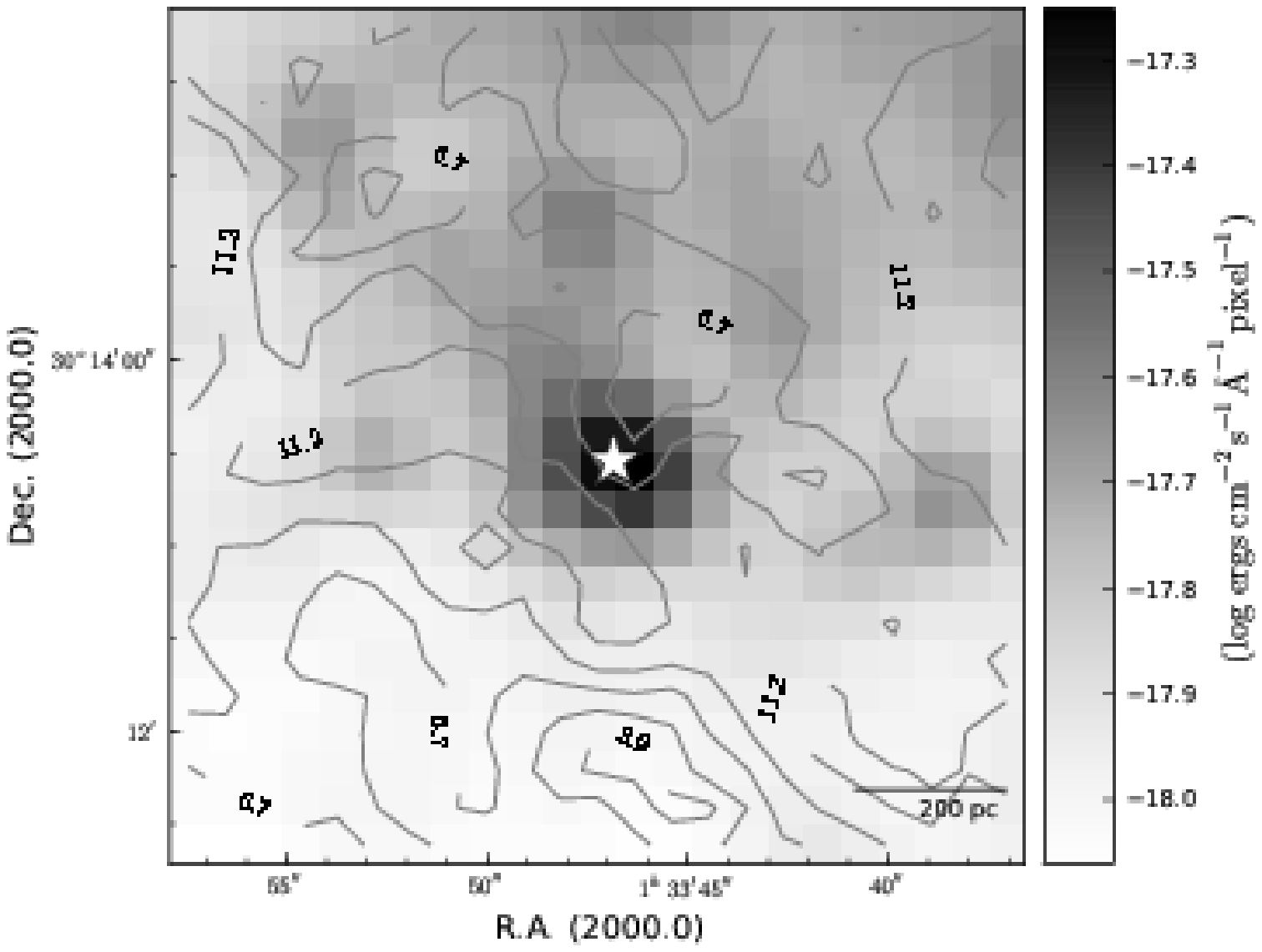} \\
    \includegraphics[width=0.35\textwidth]{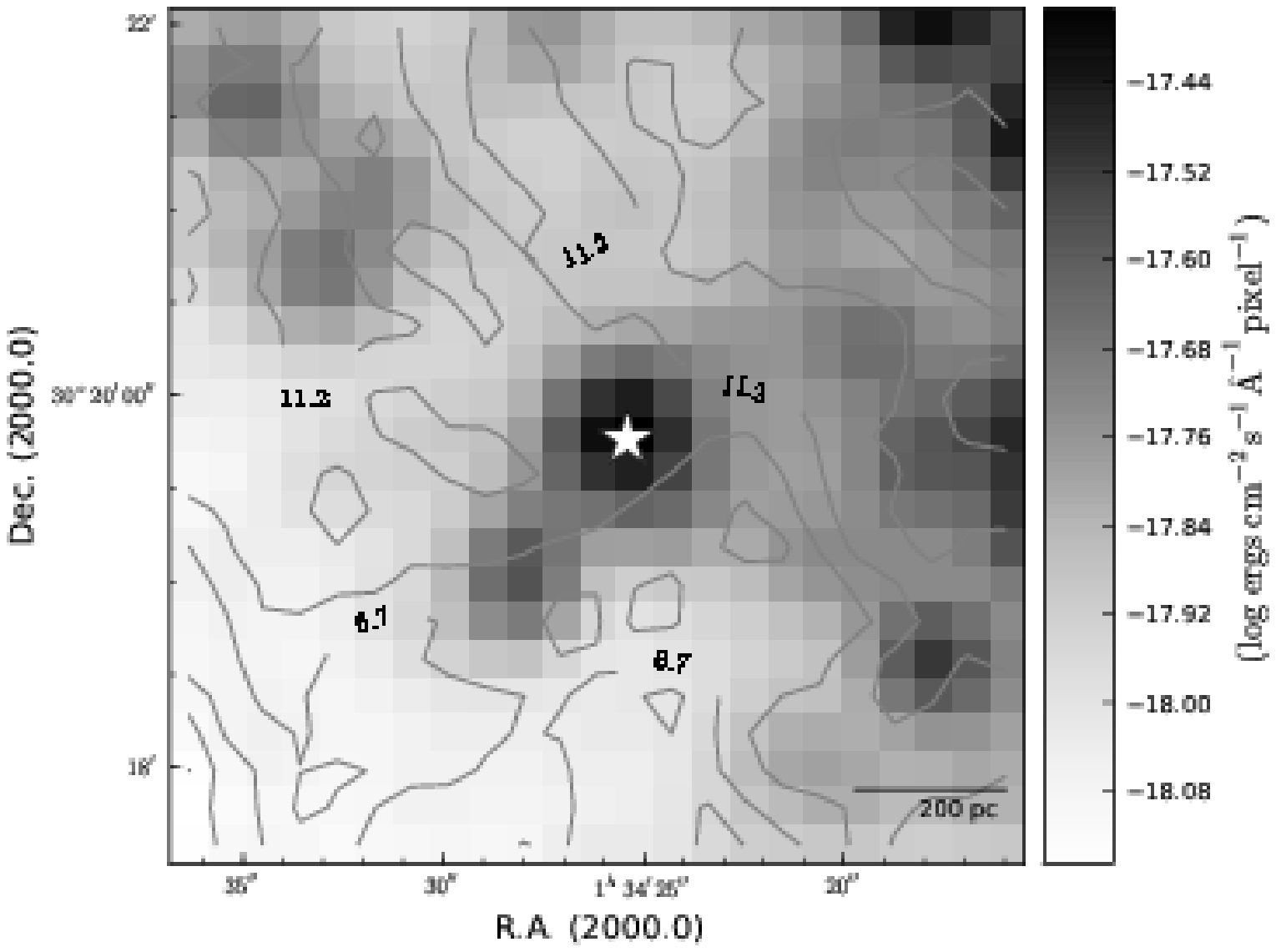}
    & \centering{\scriptsize{\hspace{0.5cm}39\hspace{0.4cm}40}} &
    \includegraphics[width=0.35\textwidth]{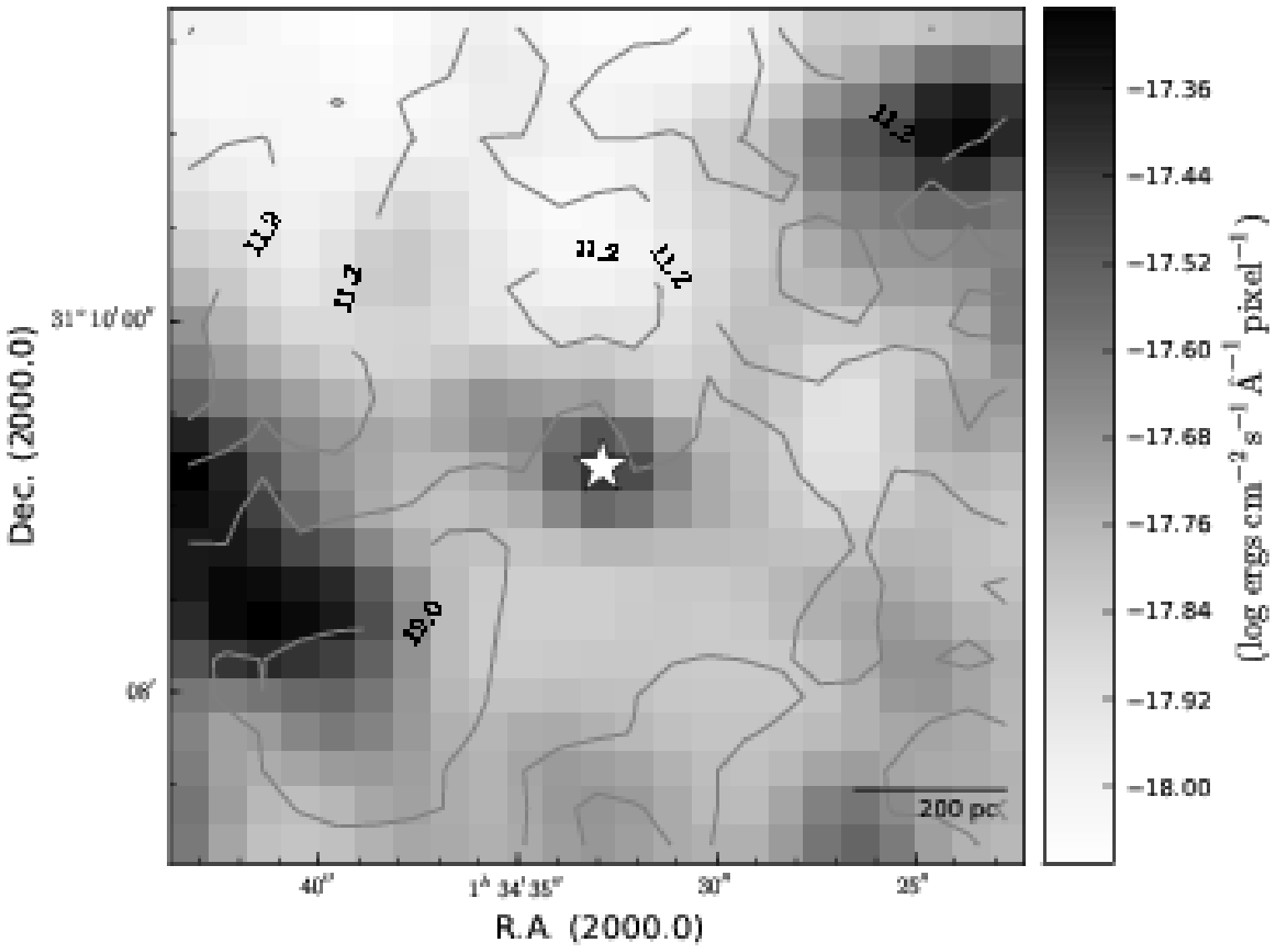} \\
  \end{tabular*}
  \caption[]{M33 sources 33 -- 40 (from left to right, top to bottom). See Figure \ref{fig:A_m33_1} for more details.}
\end{figure*}

\begin{figure*}
  \centering
  \begin{tabular*}{\textwidth}{r p{1.5cm} l} 
    \includegraphics[width=0.35\textwidth]{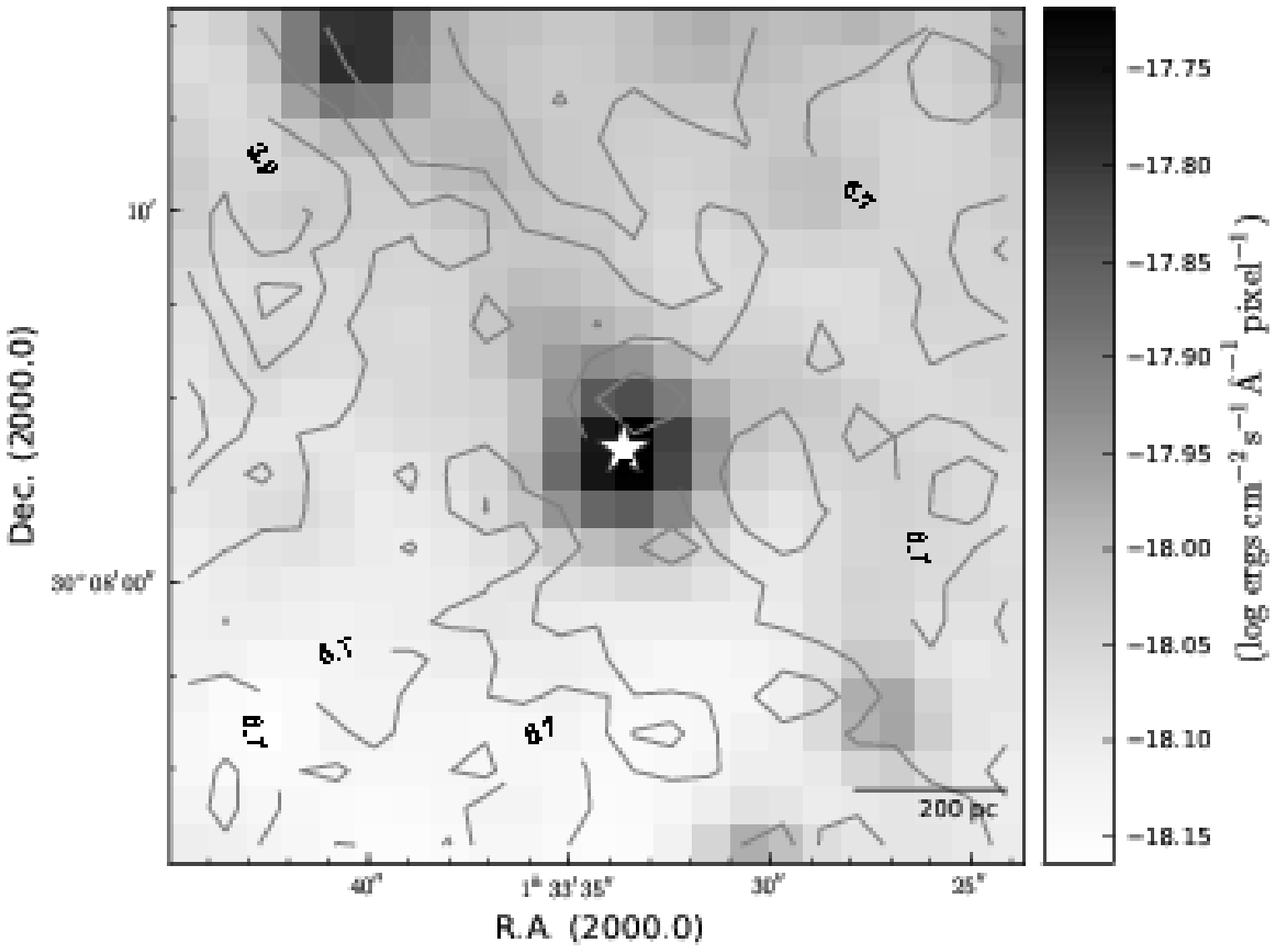}
    & \centering{\scriptsize{\hspace{0.5cm}41\hspace{0.4cm}42}} &
    \includegraphics[width=0.35\textwidth]{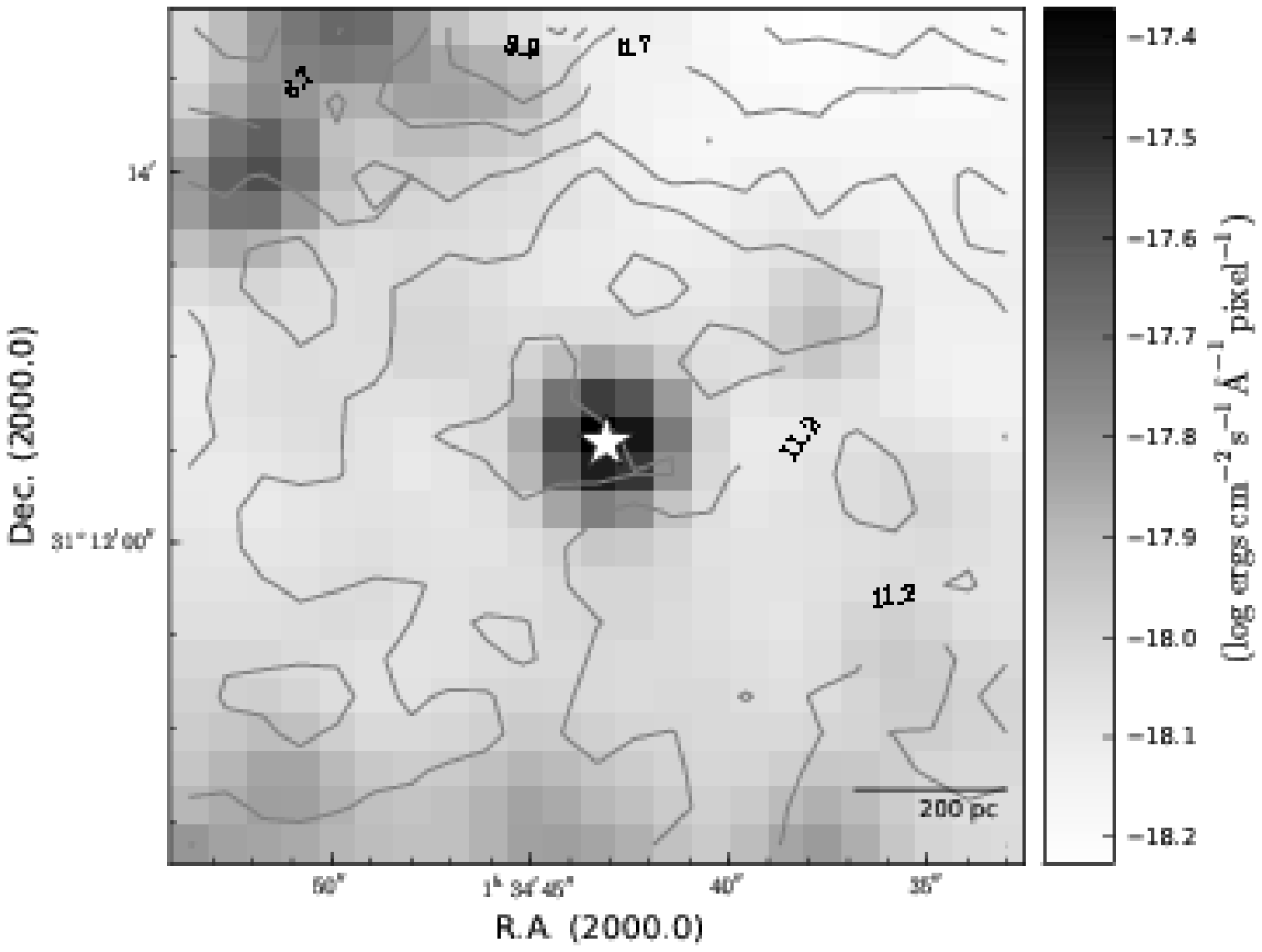} \\
  \end{tabular*}
  \caption[]{M33 sources 41 and 42. See Figure \ref{fig:A_m33_1} for more details.}
\end{figure*}
\clearpage



\begin{table*}
  \begin{tabular*}{\textwidth}{l p{6.5cm}}
    \hline\hline
    $R_{gal}$ (kpc) 		& 1.05 \\
    \dtg			& 0.65 \\
    FUV sources (J2000)         & \hmsdms{1}{34}{2.363}{30}{38}{42.30}$^a$,
                                  \hmsdms{1}{34}{2.175}{30}{38}{52.13}$^b$,
                                  \hmsdms{1}{34}{1.792}{30}{38}{20.99}$^c$,
                                  \hmsdms{1}{34}{1.754}{30}{39}{01.48}$^d$,
                                  \hmsdms{1}{34}{0.781}{30}{39}{08.93}$^e$,
                                  \hmsdms{1}{34}{0.127}{30}{39}{01.85}$^f$\\
    FUV fluxes ($10^{-15}$ \ecsa)	& $4.55^a$, $4.68^b$, $4.55^c$, $1.56^d$, $1.65^e$, $4.72^f$ \\
    $N_{bg} (\HIunits)$		& 0.96 \\
    $N_{HI} (\HIunits)$		& $3.37^{1:abcd}$, $4.10^{2:abc}$, $3.67^{3:abcdef}$, $4.63^{4:abcd}$, $3.64^{5:abcd}$ \\
    $G_0$ (cumulative)		& $1.69^1$, $0.32^2$, $0.97^3$, $0.48^4$, $0.37^{5}$ \\
    $G/G_{bg}$ range		& $0.01-1.77^1$, $0.03-0.17^2$, $0.01-0.23^3$, $0.01-0.16^4$, $0.02-0.09^{5}$\\
    n (derived, in \pccm)	& $15^1$, $1^2$, $6^3$, $1^4$, $3^{5}$ \\
    Fractional error range	& $0.35 - 0.45$ \\
    \hline
  \end{tabular*}
  \caption{\label{tab:cpsdp0087g} Detailed measurements of CPSDP 0087g}
\end{table*}

\clearpage


\begin{table*}
  \begin{tabular*}{\textwidth}{l p{6.5cm}}
    \hline\hline
    $R_{gal}$ (kpc)		 	& 1.63 \\ 
    \dtg				& 0.38 \\
    FUV sources (J2000)                 & \hmsdms{1}{34}{8.422}{30}{39}{03.93}$^a$,
                                          \hmsdms{1}{34}{8.874}{30}{39}{11.48}$^b$,
                                          \hmsdms{1}{34}{9.746}{30}{39}{12.35}$^c$,
                                          \hmsdms{1}{34}{9.543}{30}{39}{18.92}$^d$,
                                          \hmsdms{1}{34}{7.112}{30}{39}{23.94}$^e$\\
    FUV fluxes ($10^{-15}$ \ecsa)	& $6.03^a$, $6.73^b$, $9.61^c$, $3.72^d$, $10.9^e$ \\
    $N_{bg} (\HIunits)$		& 0.96 \\
    $N_{HI} (\HIunits)$		& $4.56^{1:abcd}$, $4.41^{2:e}$, $3.58^{3:abcd}$, $3.38^{4:abcde}$, $3.22^{5:abcd}$ \\
    $G_0$ (cumulative)		& $0.87^1$, $0.65^2$, $0.45^3$, $7.20^4$, $3.97^5$ \\
    $G/G_{bg}$ range		& $0.02-0.18^1$, $0.57^2$, $0.01-0.09^3$, $0.13-1.65^4$, $0.05-0.51^5$\\
    n (derived, in \pccm)	& $18^1$,   $15^2$,   $16^3$,   $295^4$,  $180^5$ \\
    Fractional error range	& $0.26 - 0.36$ \\
    \hline
  \end{tabular*}
  \caption{\label{tab:cpsdpz204} Detailed measurements of CPSDP Z204}
\end{table*}

\clearpage


\begin{table*}
  \begin{tabular*}{\textwidth}{l p{6.5cm}}
    \hline\hline
    $R_{gal}$ (kpc)			& 1.88 \\ 
    \dtg				& 0.59 \\
    FUV sources (J2000)                 & \hmsdms{1}{33}{34.134}{30}{41}{39.68}$^a$,
                                          \hmsdms{1}{33}{33.491}{30}{41}{35.03}$^b$,
                                          \hmsdms{1}{33}{32.937}{30}{41}{38.15}$^c$,
                                          \hmsdms{1}{33}{32.785}{30}{41}{46.67}$^d$,
                                          \hmsdms{1}{33}{32.208}{30}{41}{32.86}$^e$,
                                          \hmsdms{1}{33}{33.401}{30}{41}{37.88}$^\dagger$\\
    FUV fluxes ($10^{-15}$ \ecsa)	& $2.75^a$, $59.8^b$, $4.74^c$, $3.79^d$, $1.76^e$, $63.0^\dagger$\\
    $N_{bg} (\HIunits)$			& 0.93 \\
    $N_{HI} (\HIunits)$			& $1.19^{1:abcde}$, $3.79^{2:abc}$, $3.90^{3:abcd}$, $4.18^{4:ab}$, $2.48^\dagger$\\
    $G_0$ (cumulative)			& $1.10^1$, $1.40^2$, $1.33^3$, $0.54^4$, $1.79^\dagger$\\
     $G/G_{bg}$ range			& $0.01-1.29^1$; $0.01-2.29^2$; $0.01-2.01^3$; $0.01, 0.89^4$; $3.56^\dagger$\\  
    n (derived, in \pccm)		& $105^1$, $12^2$, $10^3$, $3^4$, $45^\dagger$\\
    Fractional error range		& 0.21-0.38, $0.28^\dagger$\\
    \hline
  \end{tabular*}
  \caption{\label{tab:ngc595} Detailed measurements of NGC 595}
\end{table*}

\clearpage


\begin{table*}
  \begin{tabular*}{\textwidth}{l p{6.5cm}}
    \hline\hline
    $R_{gal}$ (kpc)		 	& 2.49 \\  
    \dtg				& 0.40 \\
    FUV sources (J2000)                 & \hmsdms{1}{33}{58.768}{30}{49}{13.37}$^a$,
                                          \hmsdms{1}{33}{57.689}{30}{49}{01.45}$^b$,
                                          \hmsdms{1}{33}{56.577}{30}{49}{07.47}$^c$,
                                          \hmsdms{1}{33}{58.480}{30}{48}{44.72}$^d$,
                                          \hmsdms{1}{33}{57.651}{30}{48}{49.38}$^e$,
                                          \hmsdms{1}{33}{56.593}{30}{48}{55.91}$^f$\\
    FUV fluxes ($10^{-15}$ \ecsa)	& $2.26^a$, $1.11^b$, $0.79^c$, $6.38^d$, $1.73^e$, $1.08^f$\\
    $N_{bg} (\HIunits)$		& 0.77 \\
    $N_{HI} (\HIunits)$		& \parbox[t]{6.5cm}{\raggedright$2.60^{1:abcd}$, $1.78^{2:abcdef}$, $2.77^{3:abcd}$, $1.89^{4:abcdef}$, $1.74^{5:abcef}$, $1.46^{6:cf}$}\\
    $G_0$ (cumulative)		& $0.38^1$, $1.74^2$, $0.13^3$, $1.14^4$, $0.16^5$, $0.04^6$\\
    $G/G_{bg}$ range		& $0.03-0.40^1$; $0.05-1.69^2$; $0.01-0.11^3$; $0.02-1.02^4$; $0.02-0.20^5$; $0.02,0.06^6$\\
    n (derived, in \pccm)	& $23^1$,   $195^2$,   $7^3$,   $116^4$,  $19^5$, $5^6$\\
    Fractional error range	& $0.20 - 0.43$\\
    \hline
  \end{tabular*}
  \caption{\label{tab:bclmp0695} Detailed measurements of BCLMP 0695}
\end{table*}

\clearpage


\begin{table*}
  \begin{tabular*}{\textwidth}{l p{6.5cm}}
    \hline\hline
    $R_{gal}$ (kpc)			& 3.47 \\ 
    \dtg				& 0.63 \\
    FUV sources (J2000)			& \hmsdms{1}{34}{32.42}{30}{47}{05.2}$^a$,
					  \hmsdms{1}{34}{33.30}{30}{47}{07.3}$^b$,
					  \hmsdms{1}{34}{33.19}{30}{46}{58.7}$^c$,
					  \hmsdms{1}{34}{31.98}{30}{46}{53.0}$^d$,
					  \hmsdms{1}{34}{31.64}{30}{47}{19.9}$^e$,
					  \hmsdms{1}{34}{32.44}{30}{47}{05.2}$^\dagger$\\
    FUV fluxes ($10^{-15}$ \ecsa)	& $406^a$, $15.3^b$, $22.6^c$, $3.47^d$, $8.54^e$, $914^\dagger$\\
    $N_{bg} (\HIunits)$			& 0.97 \\
    $N_{HI} (\HIunits)$			& $4.19^{1:abcd}$, $2.63^{2:ade}$, $2.98^{3:abcd}$, $4.64^{4:abcd}$, $1.93^\dagger$\\
    $G_0$ (cumulative)			& $7.40^1$, $6.06^2$, $10.53^3$, $6.93^4$, $15.21^\dagger$\\
    $G/G_{bg}$ range			& $0.01-4.45^1$; $<0.01-4.01^2$; $0.01-6.27^3$; \mbox{$<0.01-4.45^4$}; $6.49^\dagger$\\  
    n (derived, in \pccm)		& $34^1$, $109^2$, $139^3$, $22^4$, $539^\dagger$\\
    Fractional error range		& 0.30-0.44, $0.26^\dagger$\\
    \hline
  \end{tabular*}
  \caption{\label{tab:ngc604} Detailed measurements of NGC 604}
\end{table*}

\clearpage


\begin{table*}
  \begin{tabular*}{\textwidth}{l p{6.5cm}}
    \hline\hline
    $R_{gal}$ (kpc)		 	& 4.36 \\  
    \dtg				& 0.36 \\
    FUV	source (J2000)			& \hmsdms{1}{33}{07.349}{30}{42}{49.34}\\
    FUV flux ($10^{-15}$ \ecsa)	& $33.2$\\
    $N_{bg} (\HIunits)$			& 1.37 \\
    $N_{HI} (\HIunits)$			& $1.07$\\
    $G_0$				& $0.44$\\
    $G/G_{bg}$ 				& $3.55$\\
    n (derived, in \pccm)		& $125$ \\
    Fractional error 			& $0.19$\\
    \hline
  \end{tabular*}
  \caption{\label{tab:bclmp0288} Detailed measurements of BCLMP 0288}
\end{table*}

\clearpage


\begin{table*}
  \begin{tabular*}{\textwidth}{l p{6.5cm}}
    \hline\hline
    $R_{gal}$ (kpc)		 	& 4.55 \\  
    \dtg				& $0.21^{67}$, $0.22^{1235\dagger}$, $0.4^{4}$ \\
    FUV sources (J2000)                 & \hmsdms{1}{33}{12.566}{30}{23}{39.21}$^a$,
                                          \hmsdms{1}{33}{10.670}{30}{23}{38.13}$^b$,
                                          \hmsdms{1}{33}{11.541}{30}{23}{27.84}$^c$,
                                          \hmsdms{1}{33}{12.608}{30}{23}{08.57}$^d$,
                                          \hmsdms{1}{33}{10.353}{30}{23}{04.95}$^e$,
                                          \hmsdms{1}{33}{08.072}{30}{23}{10.32}$^f$,
                                          \hmsdms{1}{33}{07.235}{30}{23}{15.18}$^g$,
                                          \hmsdms{1}{33}{06.760}{30}{23}{10.58}$^h$,
                                          \hmsdms{1}{33}{07.304}{30}{23}{01.77}$^i$,
                                          \hmsdms{1}{33}{06.435}{30}{23}{30.40}$^j$,
                                          \hmsdms{1}{33}{11.217}{30}{23}{21.93}$^\dagger$\\
    FUV fluxes ($10^{-15}$ \ecsa)	& $2.58^a$, $2.29^b$, $16.9^c$, $4.95^d$, $11.0^e$, $1.69^f$, $0.91^g$, $2.34^h$, $1.36^i$, $4.35^j$, $61.7^\dagger$\\
    $N_{bg} (\HIunits)$			& $1.37$\\
    $N_{HI} (\HIunits)$			& $3.10^{1:acd}$, $2.54^{2:acd}$, $2.93^{3:a-e}$, $2.94^{4:befghi}$, $3.77^{5:bcefi}$, $3.21^{6:bfghij}$, $4.17^{7:f-j}$, $1.48^\dagger$ \\
    $G_0$ (cumulative)			& $0.59^1$, $0.98^2$, $1.25^3$, $0.23^4$, $0.87^5$, $0.51^6$, $0.97^7$, $0.74^\dagger$\\
    $G/G_{bg}$ range			& $0.11-0.38^1$; $0.10-0.92^2$, $0.02-0.53^3$, $0.01-0.15^4$, $0.03-0.86^5$, $0.02-0.73^6$, $0.08-1.26^7$, $2.30^\dagger$\\
    n (derived, in \pccm)		& $93^1$, $207^2$, $217^3$, $11^4$, $102^5$, $86^6$, $109^7$, $319^\dagger$\\
    Fractional error range 		& $0.18 - 0.29$\\
    \hline
  \end{tabular*}
  \caption{\label{tab:bclmp0256} Detailed measurements of BCLMP 0256}
\end{table*}

\clearpage


\begin{table*}
  \begin{tabular*}{\textwidth}{l p{6.5cm}}
    \hline\hline
    $R_{gal}$ (kpc)		 	& 5.60 \\  
    \dtg				& 0.33 \\
    FUV sources (J2000)                 & \hmsdms{1}{34}{31.961}{31}{1}{09.22}$^a$,
                                          \hmsdms{1}{34}{33.158}{31}{0}{21.19}$^b$,
                                          \hmsdms{1}{34}{33.368}{31}{0}{30.29}$^c$,
                                          \hmsdms{1}{34}{34.374}{31}{0}{26.79}$^d$,
                                          \hmsdms{1}{34}{34.767}{31}{0}{44.28}$^e$,
                                          \hmsdms{1}{34}{33.785}{31}{0}{45.55}$^f$,
                                          \hmsdms{1}{34}{32.966}{31}{0}{00.25}$^g$,
                                          \hmsdms{1}{34}{33.331}{31}{0}{25.62}$^\dagger$\\
    FUV fluxes ($10^{-15}$ \ecsa)	& $0.68^a$, $5.86^b$, $1.69^c$, $1.11^d$, $3.08^e$, $0.83^f$, $0.46^g$, $15.7^\dagger$\\
    $N_{bg} (\HIunits)$		& 0.71 \\
    $N_{HI} (\HIunits)$		& $1.24^{1:abcfg}$, $1.77^{2:bcdefg}$, $1.91^{3:bcdeg}$, $2.24^{4:af}$, $1.14^{5:bcdefg}$, $1.99^{6:cdef}$, $1.04^\dagger$\\
    $G_0$ (cumulative)		& $0.13^1$, $0.61^2$, $0.13^3$, $0.1^4$, $1.04^5$, $0.23^6$, $0.055^\dagger$\\
    $G/G_{bg}$ range		& $0.02-0.08^1$; $0.04-1.22^2$; $0.01-0.21^3$; $0.03,1.45^4$; $0.06-1.42^5$; $0.01-0.86^6$; $0.23^\dagger$\\
    n (derived, in \pccm)	& $35^1$,   $102^2$,   $20^3$,   $12^4$,  $310^5$, $32^6$, $18^\dagger$ \\
    Fractional error range	& $0.19 - 0.27$, $0.18^\dagger$ \\
    \hline
  \end{tabular*}
  \caption{\label{tab:bclmp0650} Detailed measurements of BCLMP 0650}
\end{table*}

\clearpage


\begin{table*}
  \begin{tabular*}{\textwidth}{l p{6.5cm}}
    \hline\hline
    $R_{gal}$ (kpc)		 	& 5.90 \\  
    \dtg				& 0.32 \\
    FUV sources (J2000)                 & \hmsdms{1}{32}{41.199}{30}{25}{57.90}$^a$,
                                          \hmsdms{1}{32}{42.991}{30}{25}{36.90}$^b$,
                                          \hmsdms{1}{32}{41.256}{30}{25}{40.42}$^c$,
                                          \hmsdms{1}{32}{44.007}{30}{25}{21.28}$^d$,
                                          \hmsdms{1}{32}{42.159}{30}{25}{22.11}$^e$,
                                          \hmsdms{1}{32}{40.696}{30}{25}{28.71}$^f$,
                                          \hmsdms{1}{32}{41.433}{30}{25}{07.25}$^g$,
                                          \hmsdms{1}{32}{42.409}{30}{24}{56.60}$^h$,
                                          \hmsdms{1}{32}{46.560}{30}{25}{00.72}$^i$,
                                          \hmsdms{1}{32}{42.053}{30}{24}{57.91}$^\dagger$\\
    FUV fluxes ($10^{-15}$ \ecsa)	& $0.24^a$, $0.52^b$, $0.55^c$, $0.31^d$, $0.56^e$, $1.05^f$, $1.66^g$, $36.1^h$, $38.2^\dagger$\\
    $N_{bg} (\HIunits)$			& $1.55$\\
    $N_{HI} (\HIunits)$			& $2.32^{1:a-h}$, $3.43^{2:i}$, $1.15^\dagger$ \\
    $G_0$ (cumulative)			& $0.42^1$, $0.43^2$, $0.25^\dagger$\\
    $G/G_{bg}$ range			& $0.01-0.84^1$; $2.80^2$, $0.87^\dagger$\\
    n (derived, in \pccm)		& $48^1$, $24^2$, $77^\dagger$\\
    Fractional error range 		& $0.19 - 0.26$\\
    \hline
  \end{tabular*}
  \caption{\label{tab:bclmp0269} Detailed measurements of BCLMP 0269}
\end{table*}

\clearpage


\begin{table*}
  \begin{tabular*}{\textwidth}{l p{6.5cm}}
    \hline\hline
    $R_{gal}$ 			& 8.63 \\  
    \dtg			& 0.27 \\
    FUV center (J2000)		& \hmsdms{1}{34}{43.099}{31}{12}{33.25}$^a$\\
    FUV flux ($10^{-15}$ \ecsa)	& $1.44^a$ \\
    $N_{bg} (\HIunits)$		& 1.0 \\
    $N_{HI} (\HIunits)$		& $1.49^1$, $1.56^2$, $1.47^3$, $1.37^4$. $1.73^5$, $2.34^6$, $1.60^7$, $1.59^8$, {\it 0.51}\\
    $G_0$ 			& $0.77^1$, $0.14^2$, $0.14^3$, $0.09^4$, $0.06^5$, $0.04^6$, $0.04^7$, $0.03^8$, {\it 0.07} \\
    $G/G_{bg}$			& $18.63^1$, $3.43^2$, $3.42^3$, $2.07^4$, $1.39^5$, $0.99^6$, $0.99^7$, $0.66^8$, {\it 1.68} \\
    n (derived, in \pccm)	& $227^1$, $39^2$, $43^3$, $28^4$, $14^5$, $7^6$, $11^7$, $7^8$, {\it $71$} \\
    Fractional error range	& $0.19 - 0.31$, {\it 0.21} \\
    \hline
  \end{tabular*}
  \caption{\label{tab:r43} Detailed measurements of Region 42}
\end{table*}

\clearpage

\end{document}